\documentclass{aa}  
\usepackage[dvipsnames]{xcolor}
\usepackage{graphicx}
\usepackage{caption}
\usepackage{subcaption}
\usepackage{dirtytalk}
\usepackage{natbib}
\usepackage{stfloats}
\usepackage[breaklinks=true,colorlinks=true,citecolor=blue]{hyperref}
\bibpunct{(}{)}{;}{a}{}{,}

\usepackage{comment}

\usepackage{txfonts}

\begin{document}

   \title{CHEX-MATE: pressure profiles of 6 galaxy clusters as seen by SPT and {\em Planck}}
    \titlerunning{Pressure profiles of 6 galaxy clusters as seen by SPT and {\em Planck}}
   \subtitle{}

   \author{F. Oppizzi,
          \inst{1,2}
          F. De Luca,
          \inst{1,2}
          H. Bourdin
          \inst{1,2},
          P. Mazzotta,
          \inst{1,2}
          S. Ettori,
          \inst{3,4}
          F. Gastaldello,
          \inst{5}
          S. Kay,
          \inst{6}
          L. Lovisari,
          \inst{3,7},
          B.J. Maughan,
          \inst{8}
          E. Pointecouteau,
          \inst{9}
          G. W. Pratt,
          \inst{10}
          M. Rossetti,
          \inst{5}
          J. Sayers,
          \inst{11}
          \and
          M. Sereno
          \inst{3,4}
          \
          }

   \institute{Universit\`a degli studi di Roma `Tor Vergata',  Via della ricerca              scientifica, 1, 00133, Roma, Italy
   \and
   INFN, Sezione di Roma `Tor Vergata', Via della Ricerca Scientifica, 1, 00133, Roma, Italy
   \and
    INAF, Osservatorio di Astrofisica e Scienza dello Spazio, Via Pietro Gobetti 93/3, 40129 Bologna, Italy 
    \and
    INFN, Sezione di Bologna, Viale Berti Pichat 6/2, 40127 Bologna, Italy
    \and
     INAF – Istituto di Astrofisica Spaziale e Fisica Cosmica di Milano, Via A. Corti 12, 20133 Milano, Italy
    \and
    Jodrell Bank Centre for Astrophysics, Department of Physics and Astronomy, School of Natural Sciences, Alan Turing Building, Room 3.116, University of Manchester, Manchester M13 9PL, UK
    \and
    Center for Astrophysics, Harvard \& Smithsonian, 60 Garden Street, Cambridge, MA 02138, USA 
    \and
    HH Wills Physics Laboratory, University of Bristol, Tyndall Ave, Bristol BS8 1TL, UK
    \and
    IRAP, Université de Toulouse, CNRS, CNES, UPS, 9 av du colonel Roche, BP 44346, 31028 Toulouse Cedex 4, France 
    \and
    AIM, CEA, CNRS, Université Paris-Saclay, Université Paris Diderot, Sorbonne Paris Cité, 91191 Gif-sur-Yvette, France
    \and
     California Institute of Technology, Pasadena, CA 91125, USA 
     }
    \authorrunning{The CHEX-MATE collaboration}
   \date{}

 
  \abstract
     {Pressure profiles are sensitive probes of the thermodynamic conditions and the internal structure of galaxy clusters.
   The intra-cluster gas resides in hydrostatic equilibrium within the Dark Matter gravitational potential.
   However, this equilibrium may be perturbed, e.g. as a consequence of thermal energy losses, feedback and non-thermal pressure supports.
   Accurate measures of the gas pressure over the cosmic times are crucial to constrain the cluster evolution as well as the contribution of astrophysical processes.
   }
   {In this work we presented a novel algorithm to derive the pressure profiles of galaxy clusters from the Sunyaev-Zeldovich (SZ) signal measured on a combination of {\em Planck} and South Pole Telescope (SPT) observations.
   The synergy of the two instruments made it possible to track the profiles on a wide range of spatial scales.
   We exploited the sensitivity to the larger scales of the {\it Planck} High-Frequency Instrument to observe the faint peripheries, and the higher spatial resolution of SPT to solve the innermost regions. 
   }
   {We developed a two-step pipeline to take advantage of the specifications of each instrument. We first performed a component separation on the two data-sets separately to remove the background (CMB) and foreground (galactic emission) contaminants. Then we jointly fitted a parametric pressure profile model on a combination of {\it Planck} and SPT data.}
     {We validated our technique on a sample of 6 CHEX-MATE clusters detected by SPT. We compare the results of the SZ analysis with profiles derived from X-ray observations with XMM-\textit{Newton}. We find an excellent agreement between these two independent probes of the gas pressure structure.}
    {}

  
   \maketitle

%

\section{Introduction}
\begin{table*}
\caption{Main properties of the galaxy clusters sample}     
\centering          
\begin{tabular}{c c c c c c c }     
\hline\hline       
   SPT ID               & {\em Planck} ID    & R.A.           & Dec.            & ${\rm M}_{500}$           & $r_{500}$ & $z$    \\  
                        &                    &                &                 & $[10^{14} {\rm M}_\odot]$ & [arcmin]  &      \\ \hline 
    SPT-CLJ0232-4421    & PSZ2 G259.98-63.43 & $38.0767 ^\circ$ & $-44.3464^\circ$  & 7.45                       & 4.87      & 0.28 \\
    SPT-CLJ0438-5419    & PSZ2 G262.73-40.92 & $69.5732 ^\circ$ & $-54.3225^\circ$  & 7.46                       & 3.57      & 0.42 \\
    SPT-CLJ0645-5413    & PSZ2 G263.68-22.55 & $101.3711^\circ$ & $-54.2272^\circ$  & 7.96                       & 7.89      & 0.16 \\
    SPT-CLJ0549-6205    & PSZ2 G271.18-30.95 & $87.3287 ^\circ$ & $-62.0874^\circ$  & 7.37                       & 3.93      & 0.37 \\
    SPT-CLJ0254-5857    & PSZ2 G277.76-51.74 & $43.5692 ^\circ$ & $-58.9491^\circ$  & 8.65                       & 3.65      & 0.44 \\
    SPT-CLJ2344-4243    & PSZ2 G339.63-69.34 & $356.1824^\circ$ & $-42.7197^\circ$  & 8.05                       & 2.84      & 0.60 \\
\hline       

\end{tabular}
\label{table:sample}
\tablefoot{The cluster centre coordinates correspond to the X-ray peaks (see sec.~\ref{sec:Xray bck} for further details), the masses and the redshifts come from the {\it Planck}  PSZ2 catalogue and have been used to derive $r_{500}$.}
\end{table*}

Galaxy clusters are the most massive virialized objects in the Universe. 
They evolve from the largest gravitational overdensities in the primordial field, and they probe the evolution of the Universe at the largest scales \citep{2012ARA&A..50..353K,2005RvMP...77..207V}.
Deep into the cluster potential well, a large amount of baryonic matter lies in the form of hot, ionised gas.
In the standard picture, the gas pressure balances the gravitational potential, driven by the Dark Matter (DM) halo.
Gas pressure acts as the connection between the large scale evolution that drives the growth of DM halos and the baryonic processes active within.

The tight correlation with the background density field links the clusters thermodynamic properties to the underlying Cosmology, making them a powerful probe of the Universe structure and evolution.
The hydrostatic equilibrium links the intra-cluster gas pressure to the total mass in a predictable way; if this condition holds, the simplest model of spherical collapse of scale-invariant perturbations predicts the pressure profiles to be self-similar \citep{1995ApJ...449..460K,1998ARA&A..36..599B,2006ApJ...650..128K}.
This motivates the use of scaling relations linking the gas pressure with the cluster mass.

The self-similar behaviour is well respected at intermediate radii $r\sim r_{500}$\footnote{$r_{\Delta}$ is the radius enclosing a mean overdensity of $\Delta$ times the critical density of the Universe.} \citep{2010A&A...517A..92A,2013A&A...550A.131P,2014ApJ...794...67M,2017ApJ...843...72B}, while significant deviations are expected in the innermost regions and the outskirts.
The former derive from the energetic non-gravitational processes taking place into the core, {\it e.g.} AGN feedback and cooling processes.
In the peripheries instead, the infall of material into the virialized region induces the breakdown of the perfect hydrostatic equilibrium condition \citep{2009ApJ...705.1129L}.
Different processes related to the mass accretion rates are responsible for these deviations reflecting non-thermal pressure support, {\it e.g.} inhomogeneities and anisotropies in the gas motion and the matter distribution surrounding the cluster, turbulence and shocks \citep{2015ApJ...806...68L,2014ApJ...792...25N}.
Numerical simulations suggest that clusters experiencing higher mass accretion present steeper profiles at large radii \citep{2012ApJ...758...75B,2014MNRAS.440.3645M}.
Furthermore, the complex physics of virialization is expected to affect differently the collisional baryons and the collisionless DM, leading to further deviations from the self-similarity in the outskirts \citep{2015ApJ...806...68L}.
Thus, this effect may depend on redshift following the mass accretion history, as some measurements suggest \citep{2014ApJ...794...67M}.

The Sunyaev-Zeldovic effect (SZ) in the Cosmic Microwave Background (CMB) is a direct, weakly biased tracer of the gas pressure \citep{1972CoASP...4..173S,1999PhR...310...97B}.
Free electrons interact with the background radiation via inverse Compton scattering, leaving a characteristic imprint in the CMB spectrum.
SZ depends only on the pressure integrated along the line of sight. 
The intensity of the SZ signal above the primordial microwave background, {\it i.e.} the Compton $y$ parameter, does not depend on the redshift.
At the same time, the gas emits intense X-ray bremsstrahlung radiation, providing a second independent probe of the thermodynamic conditions of the gas.
X-ray emission depends quadratically on the gas density, while the SZ dependence is linear. 
These properties togheter makes X-ray and SZ complementary to probe the clusters structure at different density regimes.

In this work, we present a pipeline to extract clusters pressure profiles from a combination of {\em Planck} and the South Pole Telescope (SPT) data.
Currently, these two instruments represent the state of the art of cosmological surveys at millimetre and sub-millimetre wavelengths.
The {\em Planck} satellite, launched by the European Space Agency in 2009, delivered maps of the full-sky at frequencies from 30 to 857 GHz with sub-Jansky sensitivity and a maximum resolution of $\sim5$ arcmin (see \cite{2020A&A...641A...1P} for a review of the main results of the {\em Planck} team).
SPT is a 10-meter telescope located at the Amundsen-Scott station at the South Pole \citep{2011PASP..123..568C}.
Having a resolution $<1.7$arcmin, it can solve the inner regions of distant galaxy clusters, inaccessible by space observations \citep{2015ApJS..216...27B}.
The high sensitivity of {\em Planck} and the SPT high resolution makes the two instruments complementary in constraining the clusters structure from the core to the faint peripheries.

Thanks to the wide sky coverage and the independence on redshift, Millimetric surveys are precious tools for cluster Cosmology.
The improvement in resolution and sensitivity of ground-based observations allows us to refine the results from {\em Planck} alone \citep{2016A&A...594A..27P,
2013A&A...550A.131P}.
In literature, some works exist building clusters catalogues from the combination of {\em Planck} with SPT \citep{2021A&A...647A.106M,2021arXiv211203606S}.
Other works \citep{2019A&A...632A..47A,2021A&A...651A..73P} show how to combine {\em Planck} observation with data from the Atacama Cosmology Telescope  and measure the pressure profile on a sample of clusters.
Here, we present for the first time pressure profiles of individual clusters obtained from SPT and {\em Planck}.

We compare the results of our pipeline and the profiles derived from X-ray data.
We analyse six SPT clusters selected from the sample of the \say{Cluster HEritage project with XMM-\textit{Newton}-Mass Assembly and Thermodynamics at the Endpoint of structure formation} (CHEX-MATE) \citep{CHEX-MATE}.
X-ray data represent a powerful benchmark for our technique.
We do not only compare the profiles derived from two completely independent analyses  but, thanks to the higher resolution of XMM-\textit{Newton} data, we can investigate the impact of sub-structures on the relation between the two.

The paper is organised as follows: in section \ref{sec:data} we describe the data-sets and the clusters sample selected for the analysis; in section \ref{sec:meth} we outline the data reduction pipelines for the three data-sets; in section \ref{sec:res} we present our results; section \ref{sec:con} is dedicated to the conclusion.

In our analysis we assume a $\Lambda CDM$ cosmology with $H_{0}=70{\rm Kms^{-1}Mpc^{-1}}$, $\Omega_{M}=0.3$, and $\Omega_{\Lambda}=0.7$.

\section{Data-sets and Cluster Sample}
\label{sec:data}
\subsection{The CHEX-MATE sample}
We study a sample of 6 galaxy clusters common to the SPT-SZ catalogue \citep{2015ApJS..216...27B}  and the CHEX-MATE clusters sample. The CHEX-MATE project is an X-ray follow up of 118 clusters detected by {\em Planck} and present in the PSZ2 catalogue with the ESA satellite XMM-{\it Newton}.
The CHEX-MATE sample is designed to be minimally biased and signal-to-noise-limited ($S/N>6.5$).
In addition to millimetric and X-ray data-sets, multi-wavelength observations in optical and radio are available for most targets.
Its purpose is the study of the most massive objects that have formed and the statistical characterisation of the local clusters population. 
For this reason, the CHEX-MATE sample is further divided into two sub-sets: the Tier-1, collecting the most recently formed objects, with $0.05<z<0.2$ and $2\times10^{14}{\rm M_{\odot}}<{\rm M_{500}}<9\times10^{14}{\rm M_\odot})$, and the Tier-2, assembling some of the most massive clusters in the Universe with $z<0.6$ and ${\rm M_{500}}>7.25\times10^{14}{\rm M_{\odot}}$. In particular, the six clusters studied in this paper are a sub-set of the Tier-2 sub-sample. We refer the reader to the introductory study \citep{CHEX-MATE} for a more detailed description of the cluster selection, the observational strategy and the future outcomes of the collaboration.

This selection provides us with reliable X-ray counterparts as a benchmark for the pressure profiles derived with our technique from SPT and {\em Planck}.
Our sample spans a redshift range from $z=0.16$ to $z=0.60$, and have angular dimension from $r_{500}=2.8 \ {\rm arcmin}$ to $r_{500}=7.9$ arcmin. 
The main properties of the clusters are summarised in Table \ref{table:sample}. 
We report the SPT and the PSZ2 identifiers, the coordinates of the X-ray peak in RA and Dec, the redshifts, the ${\rm M}_{500}$ and the angular dimensions $r_{500}$. 
The redshifts and the masses come from the the PSZ2 union catalogue \citep{2016A&A...594A..27P}, where the masses are computed assuming the best-fit Y-M scaling relation of \cite{2010A&A...517A..92A} as a prior. 
We derive $r_{500}$ as a function of $z$ and ${\rm M}_{500}$ for the given Cosmology.

\subsection{{\em Planck} data}
We use the data from the second public data release (PR2) of the {\em Planck} High-Frequency Instrument (HFI), from the full 30-month mission \citep{2016A&A...594A...8P}. The dataset comprise 6 full sky maps in HEALPIX format with Nside=2048, ({\it i.e.} with pixel angular size of 1.72 arcmin)  with nominal frequencies $100$, $143$, $217$, $353$, $545$ and $857 GHz$, with resolution 9.66, 7.22, 4.90, 4.92, 4.67, 4.22 arcmin FWHM Gaussian, respectively.
The Planck team detected more than $1000$ galaxy clusters through the SZ effect with identified counterpart \citep{2016A&A...594A..27P}.

\subsection{SPT data}

The SPT observes the microwave sky in three frequency bands centred at 95, 150 and 220 GHz with 1.7, 1.2, and 1.0 arcmin resolution, respectively.
The SPT-SZ survey consists of three maps derived from the data collected from 2008 (2009 for the 150GHz channel) to 2011 over an area of 2540 $\deg^2$ located in the southern hemisphere, from 20h to 7h in right ascension and from $65^\circ$ to $40^\circ$ in declination \citep{2018ApJS..239...10C}. 

In this area, the SPT team identified 677 clusters candidates, 516 with optical and/or infrared counterpart \citep{2015ApJS..216...27B}. The full sample spans a redshift range from $z=0.047$ to $z=1.7$; it is nearly mass limited independently of redshift, with median mass  ${\rm M_{500}}\sim3.5\times10^{14}{\rm M}_\odot h^{-1}_{70}$.

In this work, we use the public SPT data\footnote{\url{https://pole.uchicago.edu/public/data/chown18/index.html}} that are convolved with a common Gaussian beam with 1.75 arcmin FWHM.
The SPT collaboration release the data in HEALPIX format with resolution parameter Nside = 8192 corresponding to a pixel angular size of 0.43 arcmin.
Notice that, despite the SPT team also provides combined {\em Planck} and SPT maps, in this work we use the "SPT Only" products and we process the two data-sets independently.

\subsection{X-ray data}
\begin{table*}
\caption{List of the XMM-\textit{Newton} observations for the galaxy clusters of this work.}     
\centering          
\begin{tabular}{ c c c c }
\hline\hline       
Cluster name &  & ObsIDs & \\ 
\hline 
PSZ2 G259.98-63.43 & 0042340301 & 0827350201 & \\ 

PSZ2 G262.73-40.92 & 0656201601 & 0827360501 & \\ 

PSZ2 G263.68-22.55 & 0201901201 & 0201903401 & 0404910401 \\ 

PSZ2 G271.18-30.95 & 0656201301 & 0827050701 &  \\ 

PSZ2 G277.76-51.74 & 0656200301 & 0674380301 &  \\ 

PSZ2 G339.63-69.34 & 0693661801 & 0722700101 & 0722700201 \\ 
\hline 
\end{tabular} 
\label{table:obsid}
\end{table*}
The X-ray study of our galaxy cluster sample is based on a joint analysis of the observations performed with the three cameras, MOS1, MOS2, and PN, of the European Photon Imaging Camera (EPIC) of the XMM-\textit{Newton} space telescope. In particular, all the public data available on the XMM-\textit{Newton} Science Archive\footnote{\url{http://nxsa.esac.esa.int/nxsa-web/\#home}} are collected and prepared for the analysis detailed in Sec.~\ref{sec:Xana}. In Table~\ref{table:obsid}, we list all the XMM-\textit{Newton} pointings we use for our sample, with the OBSid that identifies the observations in the XMM-Newton Science Archive.

\section{Methods}
\label{sec:meth}
\subsection{Millimetric data}
Our pipeline consists of two steps. First, we perform a component separation to remove the background and foreground contaminants on the SPT and {\em Planck} observations separately. Then, we combine the two datasets in a joint-fit to derive the pressure profiles.
\subsubsection{Data processing of {\it Planck} observations}
\label{sec:planckback}
To obtain images of each cluster, we project the full sky {\em Planck} maps on smaller tiles of $512\times512$ pixels into the tangential plane, with the resolution of 1 arcmin/pixel (corresponding to $\sim8.5^\circ\times8.5^\circ$) around the XMM-\textit{Newton} cluster centre.
Each image is then processed with the technique developed in \cite{2017ApJ...843...72B} to isolate the cluster signal from the background and foreground components, namely the CMB and the dust emission from our Galaxy and the cluster itself.
In the following, we summarise the data reduction pipeline and refer to \cite{2017ApJ...843...72B} for additional details.

We first perform a high-pass filter convolving the maps with a third-order B-spline kernel (as defined by \cite{Curry1966OnPF}) to remove the large scale fluctuations. 
The largest scales are more contaminated by CMB anisotropies and Galactic Thermal Dust (GTD). 
At the same time, we do not expect to find significant contribution from the cluster at those scales.

We then build the spatial templates of the GTD with a wavelet reconstruction of the 857GHz channel.
Since this frequency is strongly dust dominated we do not expect to find any contribution from the other sky components in the 857GHz maps.
We expand the map with an isotropic undecimated wavelet transform and we apply a thresholding on the coefficients with a $3\sigma$ threshold to clean the template from the noise.
We model the dust spectral energy distribution (SED) as a two-temperature grey body as proposed in \citet{2015ApJ...798...88M}, with the frequency scaling:
\begin{equation}
    s_{GTD}(\nu)=\left[ \frac{f_1q_1}{q_2}\left(\frac{\nu}{\nu_0}\right)^{\beta_{d,1}}B_\nu(T_1)+(1-f_1)\left(\frac{\nu}{\nu_0}\right)^{\beta_{d,2}}B_\nu(T_2)\right],
\end{equation}
where $B_\nu$ is the Planck function, $T_1$ and $T_2$ are the \say{cold} and \say{hot} component temperatures, $\beta_{d,1}$ and $\beta_{d,2}$ the respective power law indices, $f_1$ the cold component fraction and $q_1$ and $q_2$ are the ratios of far infrared emission to optical absorption. In our fit we left free the overall amplitude, the cold component fraction $f_1$ and spectral index $\beta_{d,1}$, while the other parameters remain fixed to their all sky values: $q_1/q_2=8.22$ and $\beta_{d,2}=2.82$, $T_2$ derives from {\it Planck} and IRAS all sky maps and $T_1=f(T_2,q_1/q_2,\beta_{d,1},\beta_{d,2})$ \citep{2015ApJ...798...88M,1999ApJ...524..867F,2014A&A...571A..11P,2014A&A...571A...1P,2016A&A...586A.132P}.

We construct the CMB template as the difference between the 217 GHz map and the GTD template, de-noised with the same wavelet procedure used on the 857 GHz map to obtain the dust template.
These templates are then rescaled, with a convolution with the proper beam, to match the resolution of each channel.

To obtain the the cluster templates, we work under the assumption that the pressure follows the spherically symmetric profile proposed in \cite{2007ApJ...668....1N}:
\begin{equation}
    P(r)=P_{0}\times \frac{P_{500}}{x^\gamma(1+x^\alpha)^{(\beta-\gamma)/\alpha}},
    \label{eq:pressure}
\end{equation}
where $x=c_{500}r/r_{500}$, $r_{500}$ is the radius where the density is 500 times the critical density and $c_{500}$ is the concentration parameter, $\alpha,\beta$ and $\gamma$ are the slopes at $r\sim r_{500}/c_{500}$,  $r\gg r_{500}/c_{500}$ and $r\ll r_{500}/c_{500}$, respectively. Following the self-similar model, $P_{500}$ is given by:
\begin{equation}
    {\rm P_{500}}=1.65\times10^{-3}E(z)^\frac{8}{3}\times\left[ \frac{{\rm M_{500}}}{3\times10^{14}{\rm M_{\odot}}}\right]^\frac{2}{3}h_{70}^2\frac{\rm keV}{\rm cm^{3}},
\end{equation}
where $E(z)=\sqrt{\Omega_M(1+z)^3+\Omega_\Lambda}$ is the normalised expansion rate and the numerical coefficient is set as in \cite{2010A&A...517A..92A}.
To obtain the SZ signal, we first project the 3D profile in Eq. \ref{eq:pressure} integrating along the line of sight and then we convert in SZ brightness $I_{SZ,c}$ for each frequency channel $c$:
\begin{equation}
    I_{SZ,c}=s_{SZ,c}\frac{\sigma_T}{m_e c^2}\int {\rm d}l\ P(r).
    \label{eq:2dtemp}
\end{equation}
The $s_{SZ,c}$ coefficients define the frequency scaling and derive from the non-relativistic Kompaneets equation, adjusted for the channel spectral response $R_c(\nu)$:
\begin{equation}
    s_{SZ,c}=\int {\rm d} \nu \ R_c(\nu) x(\nu)\left[ \frac{e^{x(\nu)}+1}{e^{x(\nu)}-1}-4\right]
    \label{eq:SZcoeff}
\end{equation}
where $x(\nu)=h\nu/kT$, $R_c(\nu)$ is given by the HFI model \citep{2016A&A...594A...7P}.

Furthermore, we add a correction term to take into account contamination by the thermal dust emission from the cluster (CTD), that has been observed in {\em Planck} data \citep{2016A&A...596A.104P,2016A&A...594A..23P}.
The SED of this component $s_{CTD}(\nu)$ is modelled as a grey body with spectral index $\beta_{d,CTD}=1.5$ and temperature $T=19.2/(1+z){\rm K}$, as in \cite{2016A&A...596A.104P}. We compute the spatial template as the projection along the line of sight of a Navarro-Frenk-White profile \citep{1997ApJ...490..493N} with concentration parameter $c_{500}=1$.
These cluster templates, SZ and dust contributions, are further convolved with the HFI beam and the same high-pass filter applied to the {\em Planck} maps.

The whole model, GTD, CMB, CTD and cluster SZ signal is then fitted to the frequency maps to obtain the channel by channel amplitudes of the templates of the contaminants components marginalized over the cluster signal. 
The difference between the background and foreground templates are the cleaned cluster maps that we will use in the joint fit with SPT data processed as described in the next section.
\begin{figure*}
    \centering
    \begin{subfigure}{\textwidth}
        \includegraphics[width=1\textwidth]{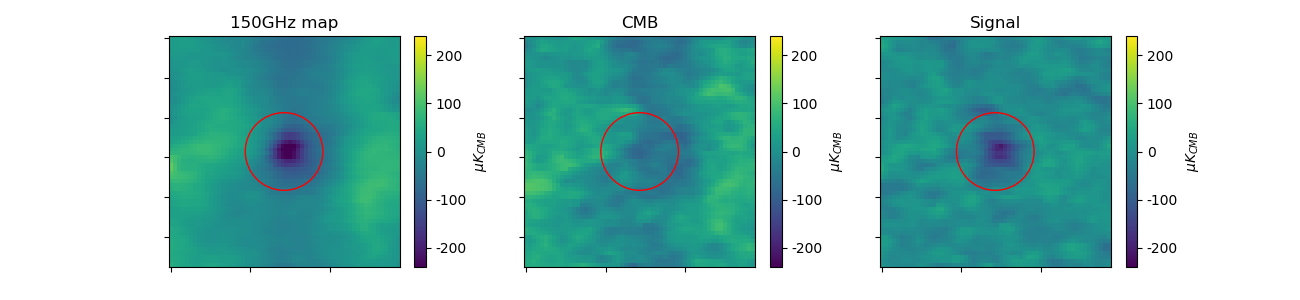}
        \caption{PSZ2 G259.98-63.43}
    \end{subfigure}
    \begin{subfigure}{\textwidth}
        \includegraphics[width=\textwidth]{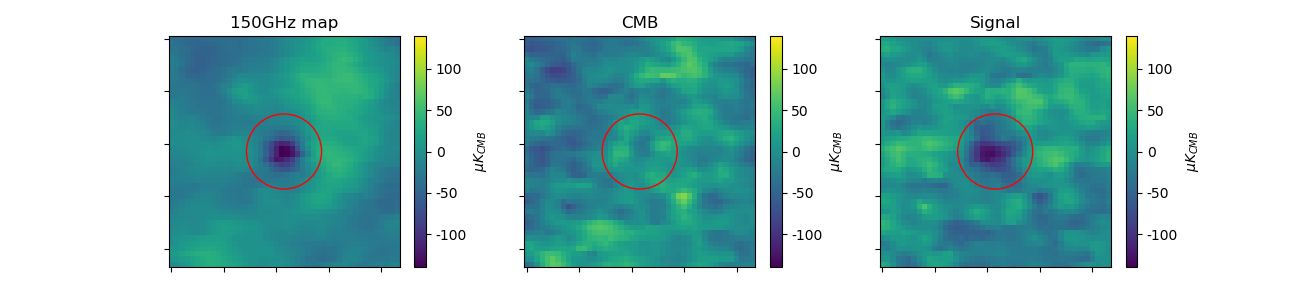}
        \caption{PSZ2 G262.73-40.92}
    \end{subfigure}
    \begin{subfigure}{\textwidth}
        \includegraphics[width=\textwidth]{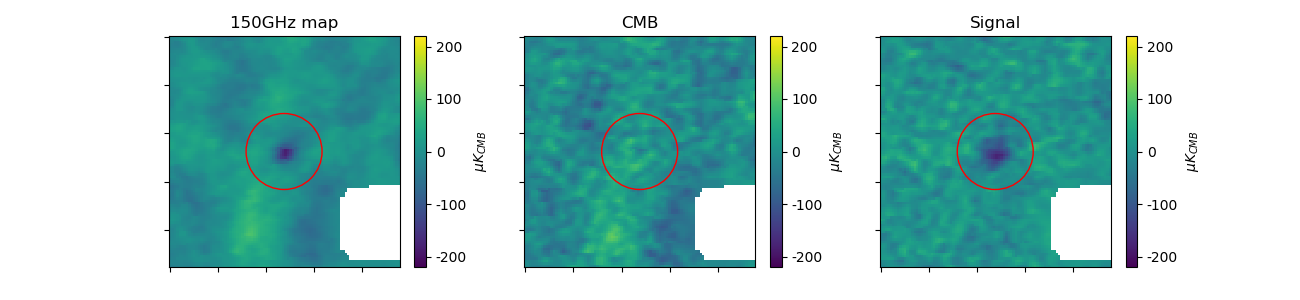}
        \caption{PSZ2 G263.68-22.55}
    \end{subfigure}
    \caption{SPT 150GHz frequency map (left panels), the estimated CMB signal (central panels), and the recovered SZ signal (right panels) for three clusters (see Fig. \ref{fig:back2} for the rest of the sample). The maps size is $6r_{500}\times6r_{500}$, the red circles mark $r_{500}$, the pixel size is 0.5 arcmin. Point sources are masked and visible in white. The top panels represent PSZ2 G259.98-63.43, the middle panels PSZ2 G262.73-40.92 and the bottom panels PSZ2 G263.68-22.55.}
    \label{fig:back1}
\end{figure*}
\subsubsection{Data processing of SPT observations}
\label{sec:sptback}

\begin{figure*}
\centering
    \begin{subfigure}{\textwidth}
        \includegraphics[width=\textwidth]{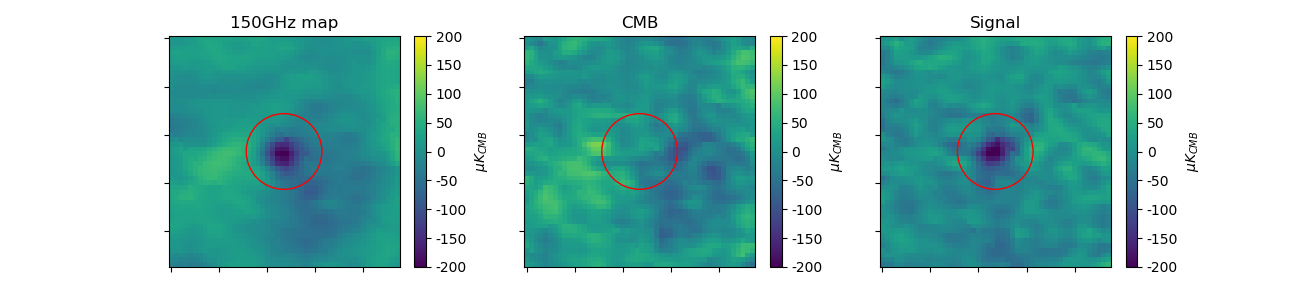}
        \caption{PSZ2 G271.18-30.95}
    \end{subfigure}
    \begin{subfigure}{\textwidth}
        \includegraphics[width=\textwidth]{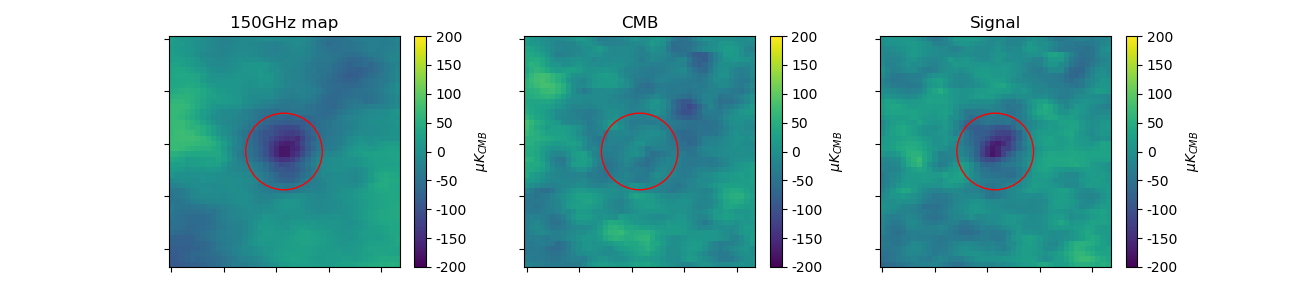}
        \caption{PSZ2 G277.76-51.74}
    \end{subfigure}
    \begin{subfigure}{\textwidth}
        \includegraphics[width=\textwidth]{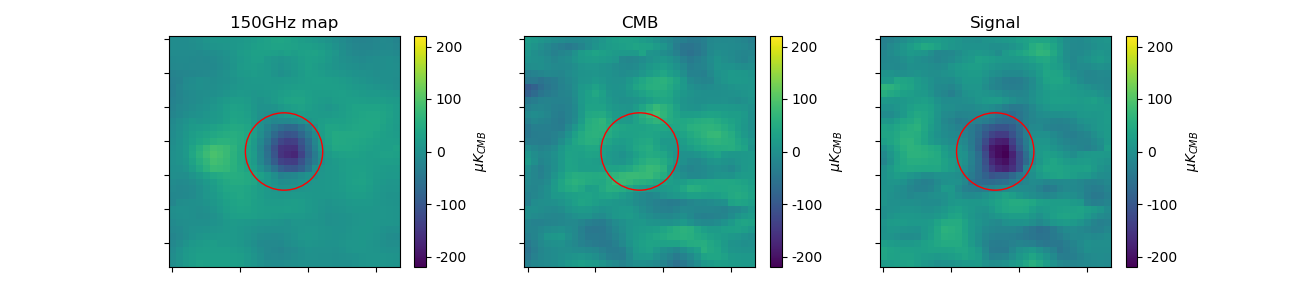}
        \caption{PSZ2 G339.63-69.34}
    \end{subfigure}
    \caption{As Figure \ref{fig:back1} for, from top to bottom: PSZ2 G271.18-30.95, PSZ2 G277.76-51.74, and PSZ2 G339.63-69.34.}
\label{fig:back2}
\end{figure*}

We preliminarly select patches of $1024\times1024$ pixels around the XMM-\textit{Newton} cluster centre.
We set the resolution of these maps to 0.5 arcmin/pixel.
Each tile covers a region of $\sim8.6^\circ\times8.6^\circ$.

These data differ from the {\em Planck} one in frequencies and spatial scales covered.
For this reason, we cannot process them with the component separation technique described in the previous section. 
It exploits the {\it Planck} data structure and thus is not adapted to SPT observations. 
In particular, SPT lacks the high-frequency channels from which we could derive the dust templates required for the multi-component fit described in the previous section; furthermore, the 220 GHz channel of SPT is too noisy to recover the CMB templates at the required precision. 
Without the templates to fit, we resort to a different solution.
In light of these limitations, we develop a method tailored to the SPT data structure, including information from {\em Planck} to improve the reconstruction on the larger spatial scales.

We recover the CMB signal from a linear combination (LC) of the three SPT channels and the 217GHz channel of {\em Planck}. 
Similar methodologies are widely used in CMB analysis, and various implementations have been applied to many millimetre data-sets released in the last decades \citep{2003ApJS..148...97B,2003PhRvD..68l3523T,2004ApJ...612..633E,2009A&A...493..835D,2011MNRAS.418..467R,2013A&A...558A.118H,2020JCAP...03..054O}.
To include the {\em Planck} 217 GHz channel in the analysis, we re-project it on a smaller region centred on the clusters with the same characteristic of the SPT ones (resolution: 0.5  arcmin/pixel, area: $8.6^\circ\times8.6^\circ$) .

The LC weights are computed to minimise the variance with respect to a signal constant in frequency ({\em i.e.} the CMB) and simultaneously to null the non-relativistic SZ component.
 Assuming that the data can be represented as the linear combination of different components (the so-called linear mixture model) the aforementioned conditions lead to the solution:
\begin{equation}
    \textbf{w}=\frac{eC^{-1}}{A^TC^{-1}A},
\end{equation}
in this notation, the elements of the vector $\textbf{w}$ correspond to the weights assigned to each frequency, $C$ is the data covariance matrix between the $N_{chan}$ channels, $e$ is a vector of ones of length $N_{chan}$ and A is a $N_{chan}\times 2$ matrix accounting for the emission the CMB and the SZ signal at the various frequency. 
Since our data are normalised to the CMB emission, the first column of A is constant and equal to one. while the second column contains the frequency scaling of the SZ signal:
\begin{equation}
    A=\begin{pmatrix}
        1  & s_{SZ,0} \\
       ... & ...      \\    
        1 & s_{SZ,N_{chan}}
\end{pmatrix},
\end{equation}
where the $s_{SZ,c}$ coefficients are derived as in Eq. \ref{eq:SZcoeff} with the channel spectral response of SPT $R_c(\nu)$ given in \cite{2018ApJS..239...10C}.

The SPT window function changes with the channels. 
To combine them, we first equalise the frequency maps, including the {\em Planck} one, to the 150 GHz spatial response.
Since the window function also depends on the sky coordinates, we compute it independently for each cluster.
The two instruments have different resolution, to combine them we follow a multi-scale approach.
We split each SPT map in a low-pass filtered map matching the {\em Planck} 217 GHz channel resolution, and in a high pass filtered map containing the residual small scale features. This operation corresponds to a convolution with a Gaussian beam of 5 arcmin to obtain the low-pass filtered map, while the residual of this map with the original one corresponds to the high-pass filtered map.

We then perform the LC separately on the low-passed maps and high-passed maps, including the {\em Planck map} in the combination of the large scales.
We compute the LC weights on a sub-region of $1.5\times1.5$ degree around the cluster centre.
This operation gives us two separate estimates of the large scales and small scales CMB fluctuations, that we re-combine into a single map.
Finally, we rescale the template to the three SPT channels window functions to obtain the final templates.
We finally subtract these templates to the SPT maps to obtain the background cleaned maps. 
Schematically, our pipeline can be summarised as follows:
\begin{enumerate}
    \item Equalize the maps at the 150GHz window function.
    \item Split the maps in low-pass and high-pass filtered maps.
    \item Compute the LC of the low and high spatial frequencies maps separately.
    \item Recombine the two scales into a single template.
    \item Reconvolve the map with the corresponding PSF to obtain the CMB template for each channel.
    \item Subtract the templates from the SPT maps to remove the CMB background.
\end{enumerate}

As stated before, we do not consider dust contamination in these data.
Different considerations justified this choice.
First, the dust emission is expected to be very low at the SPT frequencies where we will perform the fit of the SZ profile, namely 95 and 150 GHz.
In addition, these clusters are located quite far from the galactic plane, where the GTD contamination is especially significant.
Furthermore, due to the filter window function, SPT is blind to multipoles under $\ell=300$, corresponding to an angular separation $\gtrsim1^\circ$; as a consequence, any source of diffuse emission results highly suppressed in SPT data.
This feature has noticeable implications on the contamination from Galactic thermal dust: since it comes in large part as diffuse emission, after the SPT spatial filtering its residual contribution in the data is negligible.

We show the results of the background removal for the 150 GHz channel of SPT in the right panels of Fig. \ref{fig:back1} and Fig. \ref{fig:back2}, alongside the SPT raw maps in the left panels and the estimated CMB emission in the central panels, on a region spanning a radius of $3\times r_{500}$ around the cluster centres.
The CMB background, shown in the central panels, appears consistently separated from the SZ signal. 
We notice a significant noise residual due to the 220 GHz noise and perform several tests to address the issue.
We try alternative techniques to exclude this channel from the background reconstruction.
We also try to de-noise the CMB templates by applying a wavelet thresholding algorithm.
Unfortunately, we find out that, without the information from the 220 GHz channel, the risk of partially mixing a fraction of the cluster signal into the background template is very high.
On the other hand, while thresholding the templates efficiently suppress the noise, it also removes small scale fluctuations from the CMB reconstruction. 
The de-noised templates are too smoothed, and significant contamination from the CMB remains in the cleaned maps.
In conclusion, we address this issue by considering this additional noise contribution in the covariances, as described in the next section.

\subsubsection{Joint fit of {\it Planck} and SPT data}
\label{sec:fit}

Once we estimate the cluster signal from our component separation pipelines, we combine the SPT and {\em Planck} data in a joint fit of the cluster thermal SZ template.
The fitting functions are radially averaged SZ profiles $\overline{I}_{SZ,i}$ of the two-dimensional templates derived from Eq. \ref{eq:2dtemp}, we fit them to the profiles computed on the $y_c$ maps, that we denote by $\overline{y}_c$, estimated with the methods described beforehand.
We compute the profile value in a given radial bin defined as the average over the pixels whose centre falls inside an annulus delimited by the bin edges. 
We choose a linear binning scheme with circular bins of width $0.2r_{500}$ for SPT, and of 2 arcmin for {\em Planck}. The fit is performed up to $3 r_{500}$.
We include in the fit the 95 and 150 GHz channels of SPT, and the
100, 143, 353 GHz channels of Planck. The other channels, {\it i.e.} SPT 220 GHz and {\it Planck} 217 GHz, 545 GHz and 857 GHz, are used in the component separation but not included in the fit.
We expect the cluster signal to noise to be low at these frequencies. 
By excluding them, we improve the computational time with a minimal loss of information. 
At the same time, we also minimize the risk of residual contamination from the dust emission that dominates these channels.

\paragraph{Likelihood}
We start from a Gaussian likelihood for the profile parameters $\theta=\left[P_{0},c_{500},\alpha,\beta,\gamma\right]$:
\begin{equation}
    \ln{\mathcal{L}(\theta)}\propto\sum_i \left(\overline{y}_i-\overline{I}_{SZ,i}(\theta)\right)^TC_i^{-1}\left(\overline{y}_i- \overline{I}_{SZ,i} \right),
    \label{eq:gaulik}
\end{equation}
where, for each instrument $i$, $\overline{I}_{SZ,i}$ and $\overline{y}_i$ are column vectors whose entries run over the radial bins and the instrument frequency channels ({\it i.e.} for each instrument they contains the profiles computed on all the frequency channels concatenated).
Notice that we consider the SPT and Planck channels uncorrelated with one another.

\paragraph{Covariance}
We compute the covariances in order to take into account the residuals of the component separation step.
The use of common templates to clean different frequency channels raise the risk of introducing correlated residuals into the cleaned maps due to the propagation of errors on the reconstruction of the sky components.
We address this issue by estimating the covariances directly from the cleaned maps, taking into account the possible correlation between channels.
To do that, we first compute one hundred profiles centred outside the cluster, in regions treated with the same foreground cleaning technique but where we do not expect to find any signal of interest.
We masks the point sources and we exclude the gaps in the map in the computation of the profiles.
We then concatenate the profiles of the various frequency in a vector defined as in Eq. \ref{eq:gaulik} and we then compute the sample covariance between these profiles.
We repeat the process for both instruments, that we consider uncorrelated since we use different component separation techniques. 
The covariances obtained with this method account for both the instrumental noise and the correlation between channels.

\paragraph{Monte-Carlo sampling}
With the y maps and the Likelihood in hand, we perform the fit with the Cobaya MCMC sampler \citep{2021JCAP...05..057T}. 
We try different combinations of parameters to optimise the convergence of our chains.
To efficiently constrain all five parameters of the gNFW profile we need information on a wide range of spatial scales.
Our data-sets are very sensitive to the faint outskirts of the cluster but lack the resolution to fully characterise the inner core.
So that, while it is straightforward to leave free to vary the amplitude $P_0$ and the outer slope $\beta$, the choice among the remaining parameters is not trivial.
In case of limited resolution, strong degeneracies arise, being the case of $c_{500}$ and $\alpha$ particularly critical.
The concentration $c_{500}$ fixes the position of the transition region between the $r^{-\gamma}$ and the $r^{-\beta}$ regimes. 
Since the range of physically significant values is of the same order of magnitude as our bins width and way smaller than the SPT beam,  we decide to keep it fixed.
The other critical parameter is $\alpha$, which indeed governs the slope around $r_{500}/c_{500}$, but it is better understood as the inverse of the logarithmic width of the transition region between the $r^{-\gamma}$ and the $r^{-\beta}$ regimes, that is proportional to $\alpha^{-1}$.
Furthermore, it is strongly and non-linearly correlated with $P_{0}$ for values $\alpha<1$, and with both the other slopes.
The correlation with $\gamma$ is particularly problematic due to the limited resolution since enlarging the transition width can restrict the effects of varying the inner slope to unresolved regions.
These issues can be mitigated by changing the parametrization of Eq. \ref{eq:pressure} substituting $\alpha$ with $\alpha'=1/\alpha$ and fitting for the logarithm of the amplitude $\log{P_{0}}$ to make the correlation linear.
However, even with these precautions, the chains remained stuck into meaningless regions of the parameter space unless we put strict prior on $\alpha$.
Given these considerations and the results of our tests, we conclude to leave free to vary the amplitude $P_{0}$ and the inner and outer slopes $\gamma$ and $\beta$ and to keep the concentration parameter and the intermediate slope fixed to their universal values identified by \cite{2010A&A...517A..92A} $c_{500}=1.177$ and $\alpha=1.0510$.

\begin{figure*}
    \centering
    \begin{subfigure}{0.32\textwidth}
        \includegraphics[width=\textwidth]{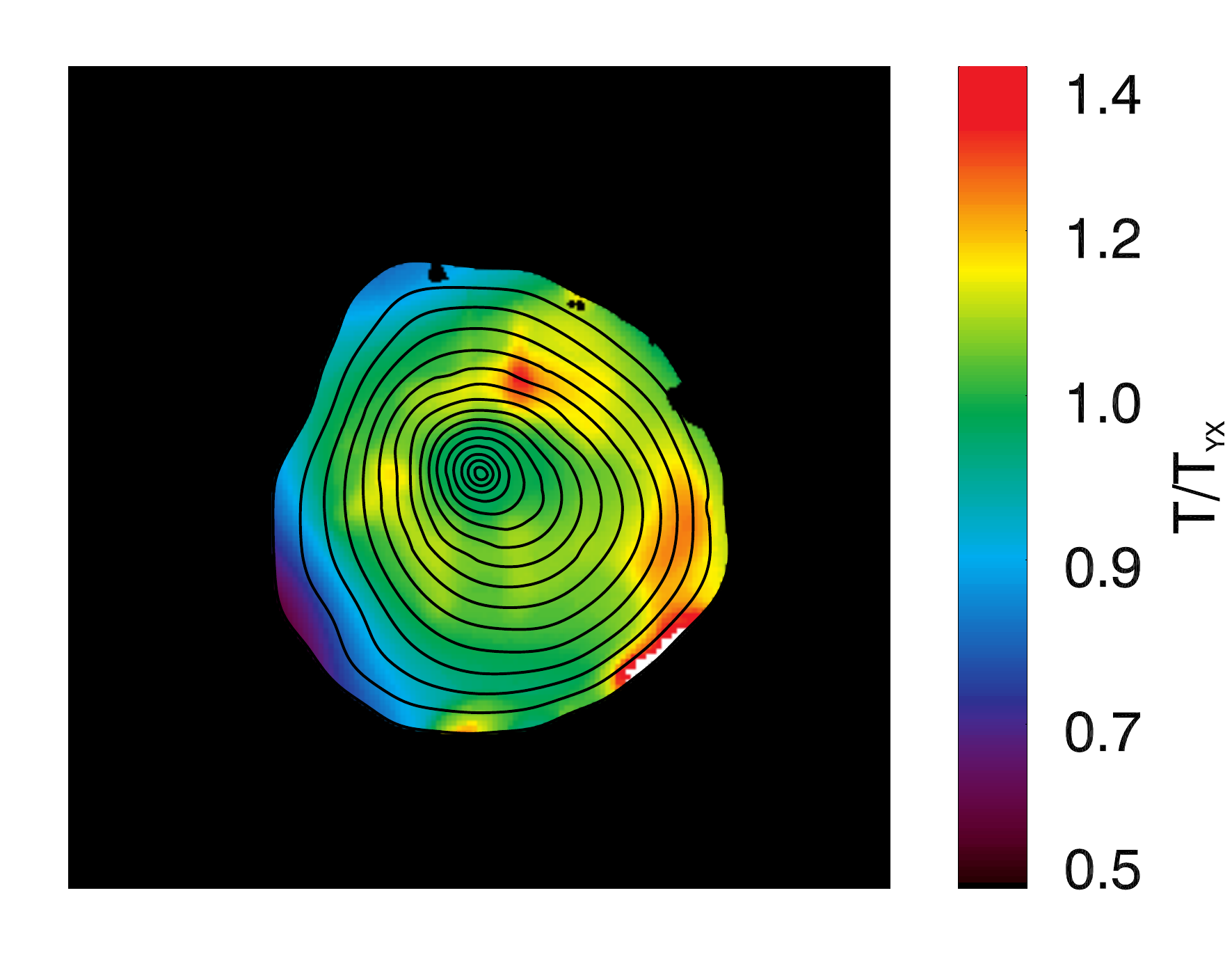}
        \caption{PSZ2G259.98-63.43}
    \end{subfigure}
    \begin{subfigure}{0.32\textwidth}
        \includegraphics[width=\textwidth]{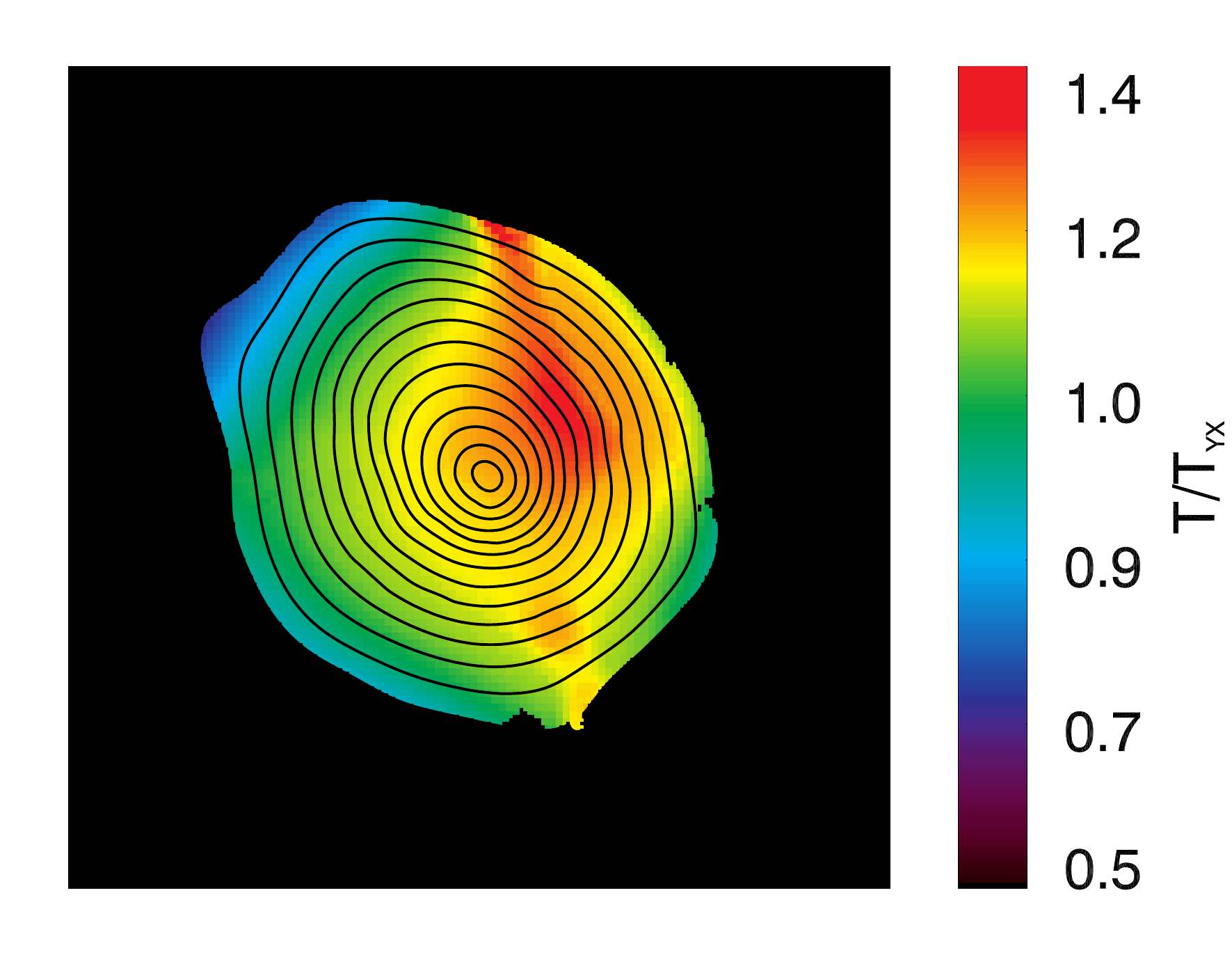}
        \caption{PSZ2G262.73-40.92}
    \end{subfigure}
    \begin{subfigure}{0.32\textwidth}
        \includegraphics[width=\textwidth]{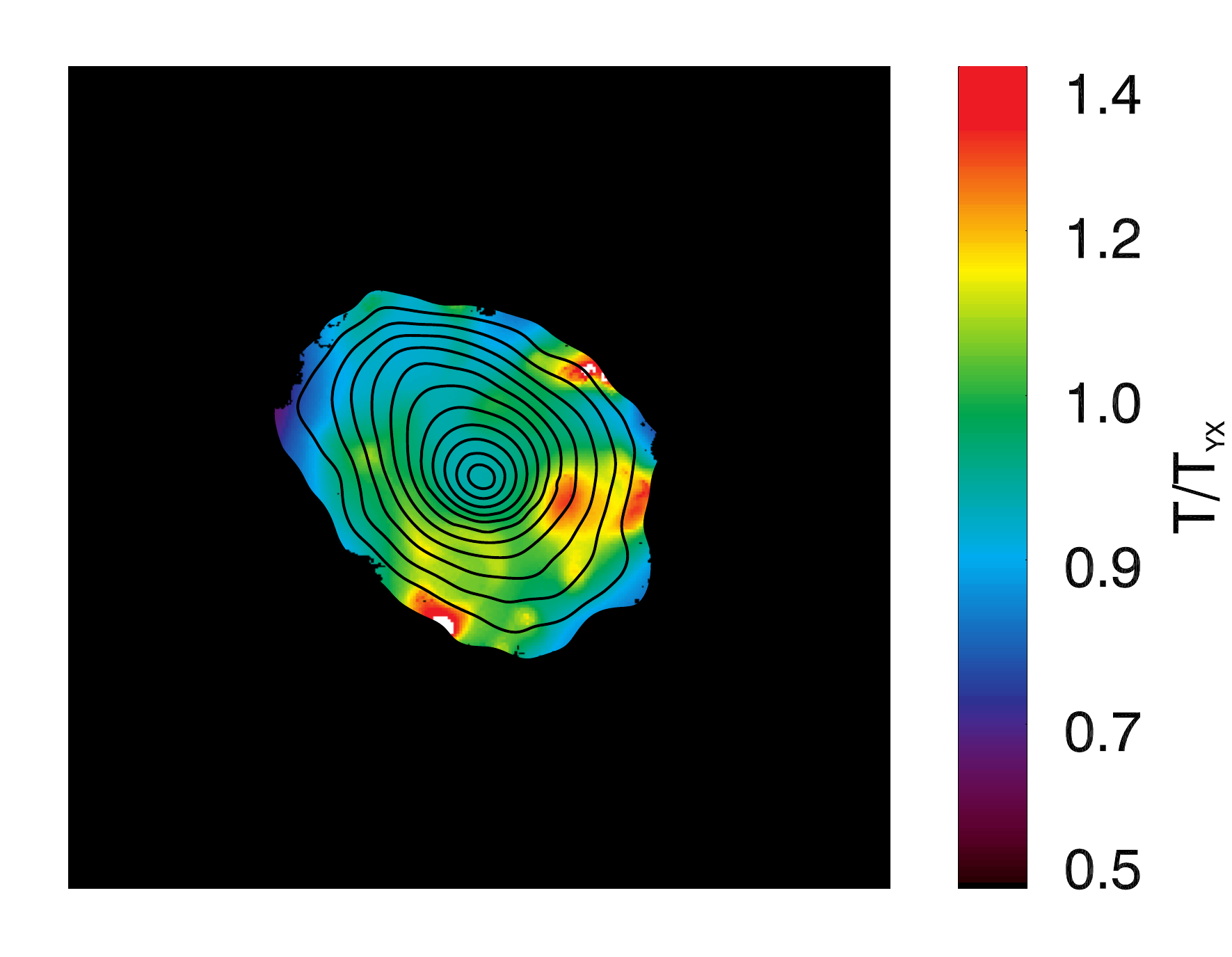}
        \caption{PSZ2G263.68-22.55}
    \end{subfigure}
    \begin{subfigure}{0.32\textwidth}
        \includegraphics[width=\textwidth]{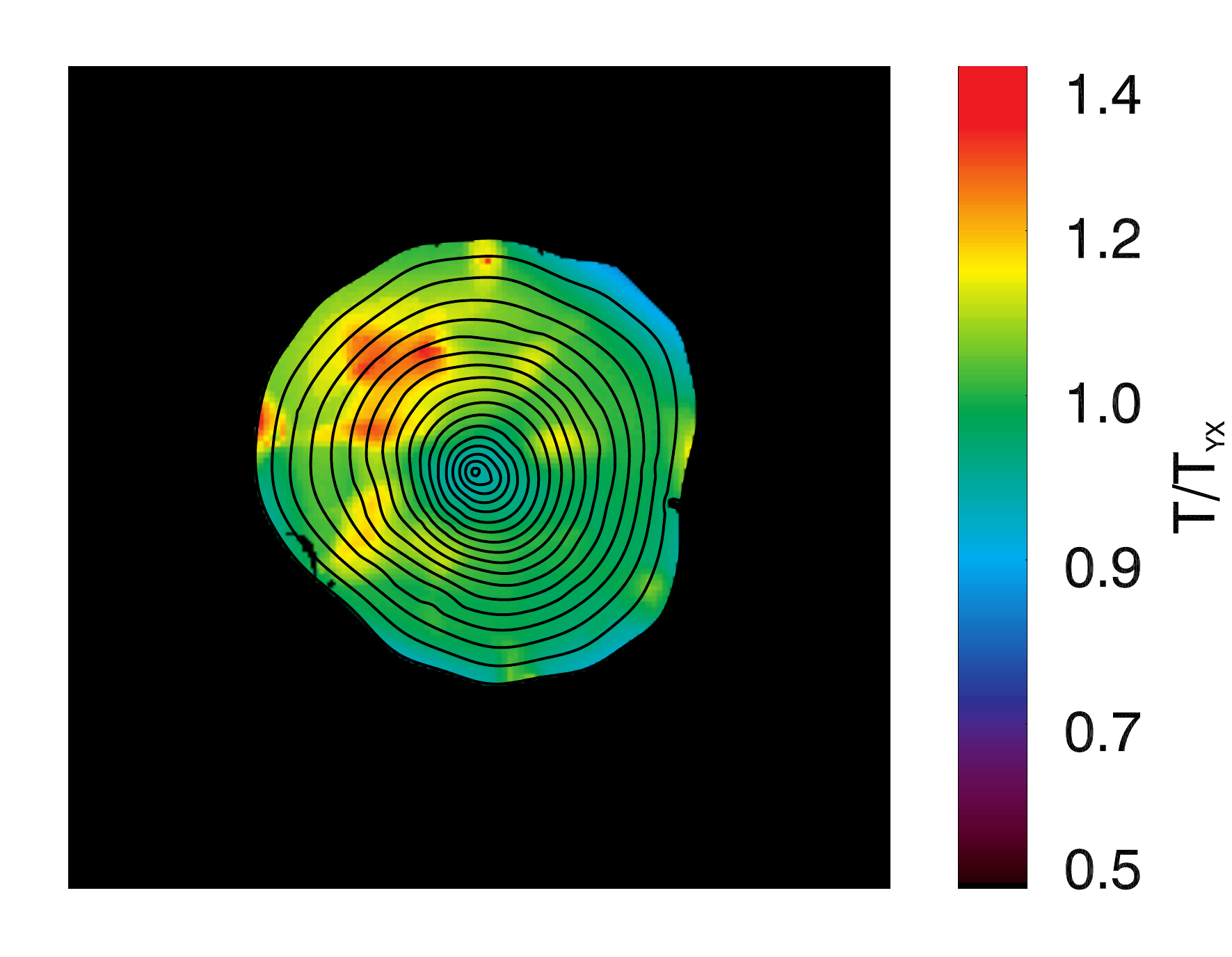}
        \caption{PSZ2G271.18-30.95}
    \end{subfigure}
    \begin{subfigure}{0.32\textwidth}
        \includegraphics[width=\textwidth]{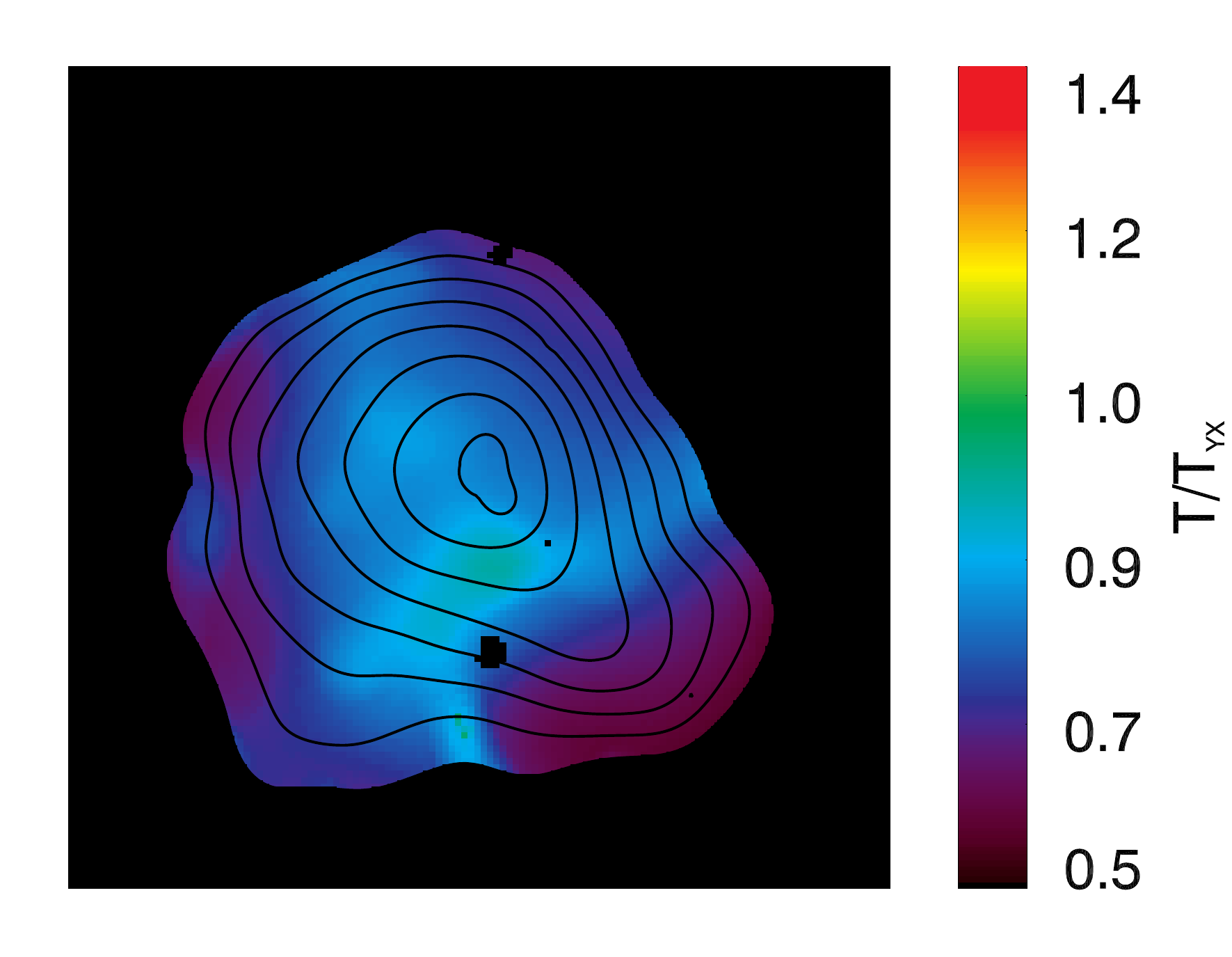}
        \caption{PSZ2G277.76-51.74}
    \end{subfigure}
    \begin{subfigure}{0.32\textwidth}
        \includegraphics[width=\textwidth]{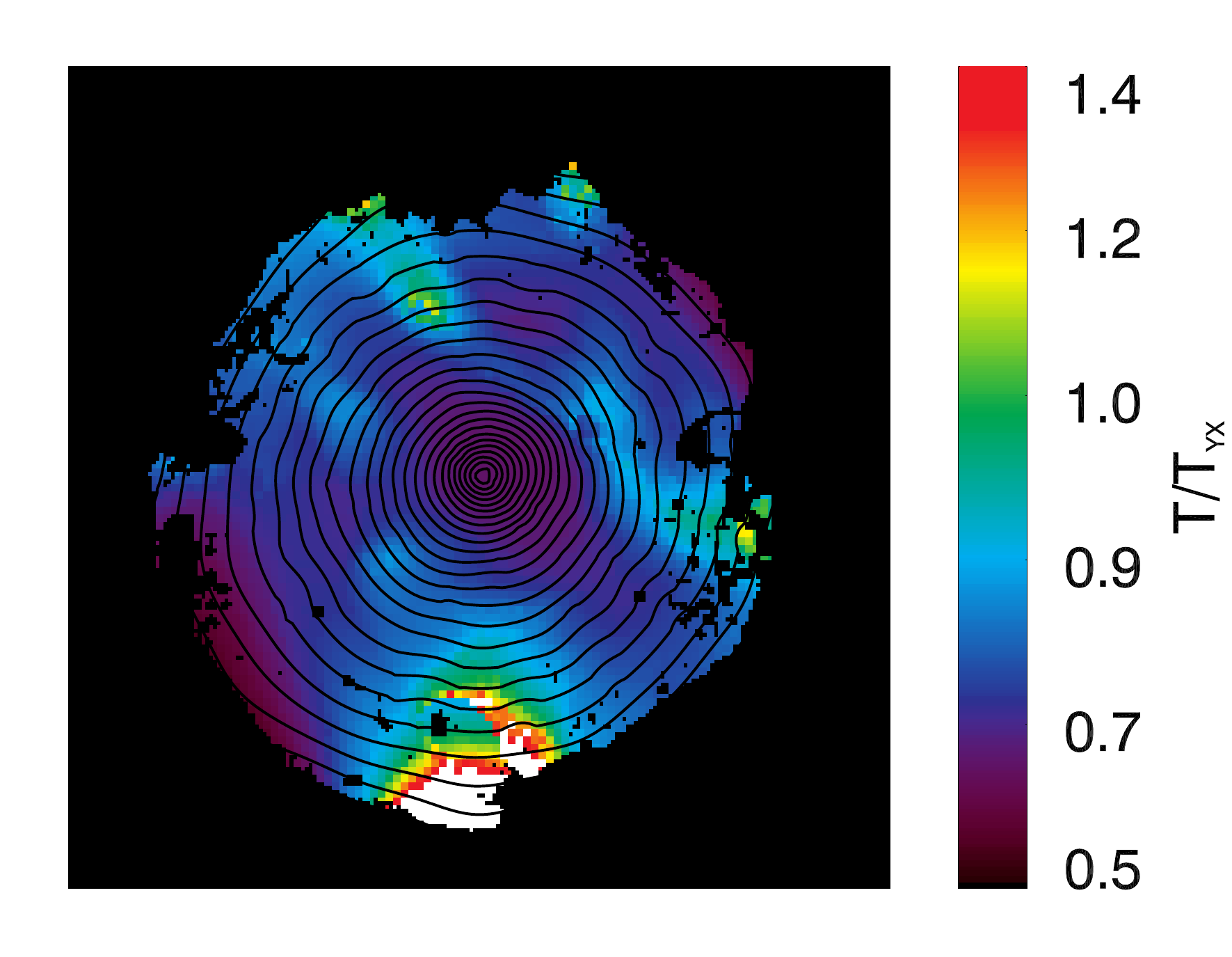}
        \caption{PSZ2G339.63-69}
    \end{subfigure}
    \caption{Normalised cluster temperature maps overlaid to the point sources removed X-ray surface brightness isocontours, as obtained from XMM-\textit{Newton} data. The maps size is $2r_{500}\times2r_{500}$ and are centred on the X-ray peak. Contour levels are logarithmic equispaced by a factor $\log\sqrt2$. The temperature maps are expressed in terms of $T_{Y_X}$, the mean spectroscopic temperature for the $Y_X$ scaling relation estimated in the intra-cluster radii range [0.15, 0.75] $r_{500}$. For all clusters but PSZ2G339.63-69, surface brightness contours and temperature maps have been extracted in the energy bands $[0.5-2.5]$ keV and $[0.3,12]$ keV, respectively. The surface brightness contours and temperature map of PSZ2G339.63-69 have been extracted in the energy band $[0.5-2]$ keV to reduce the contamination from the AGN, as detailed in section ~\ref{sec:Phoenix}.}
    \label{fig:Txmap}
\end{figure*}

\subsection{X-ray analysis}
\label{sec:Xana}

We analyse the X-ray data following the scheme used in \citet{2017ApJ...843...72B} and De Luca et al. (in prep.) that consists of three steps, detailed hereafter in this Section. We note that this pipeline differs from the standard analysis used by the CHEX-MATE collaboration to study the statistical properties of the cluster population (Bartalucci et al. in prep., Rossetti et al. in prep.). The products of the two X-ray pipelines are compared in a companion paper (De Luca et al. in prep.) to assess any differences in the results and any systematic uncertainties for a larger (and representative) sample of CHEX-MATE clusters. We found out that the two pipelines return compatible spectroscopic temperature profile.

\subsubsection{Data preparation}
The Observation Data Files (ODF) from the XMM-\textit{Newton} Science Archive are firstly pre-processed to generate calibrated event files for the data reduction with the \textsc{emchain} and \textsc{epchain} tools of \textsc{sas}, version 18.0.0. To threat flares or high-background periods we follow \citet{Bourdin2008, Bourdin2013}, removing all the events that deviate more than $3\sigma$ from the light curve profile. 
Point sources in the field of view are identified and masked using \textsc{SExtractor} \citep{SExtractor}.

\subsubsection{Diffuse background and foreground emissions} \label{sec:Xray bck}
The remaining X-ray foreground and background \footnote{For a summary of all the main background components that can afflict XMM-\textit{Newton} observations, see: \url{https://www.cosmos.esa.int/web/xmm-newton/epic-background-components}} are dominated by the quiescent particle background (QPB), the cosmic X-ray background (CXB), and the thermal emission associated with our Galaxy. We fit the normalisations of these components in a region where the cluster emission is negligible and with the spectral and spatial features described in \citet{Bourdin2013}. In particular, we exclude all the data within $1.5R_{500}$ from the X-ray cluster peak. We identify the X-ray peak as the coordinates of the maximum of a sparse wavelet-denoised \citep[][]{Starck2002, 2009A&A...504..641S} surface brightness map in the soft ($[0.5,2.5]$ keV) X-ray band. We set this position as the cluster centre to compute the radial profiles for the X-ray and SZ analysis. The normalisation of the QPB spectrum \citep{Kuntz2008} is estimated considering the energy band where this emission is dominant: $[10-12]$ keV for MOS and $[12-14]$ keV for PN cameras of the XMM-\textit{Newton} telescope.
As pointed out by many authors in the literature \citep[see e. g.][]{DeLuca2004, Kuntz2008, leccardi2008, Bourdin2013, Lovisari2019}, 
a residual focused component, originally attributed to soft protons (SP) but whose origin is still debated \citep[\textit{ e.g.}][]{2017ExA....44..309S}, can affect the X-ray observations even after the light curve filtering. To account for this additional component, we model it as a power-law with a fixed index ($-0.6$) and ratio ($0.3$) between MOS and PN. Despite its simplicity, this model is consistent with the preliminary results of a more detailed physical and predictive modelling of QPB and SP for CHEX-MATE observations (D. Eckert, private communication).
We estimate the normalisation of this component by performing a joint fit with the cluster spectrum (with the cluster metal abundance fixed to $Z=0.3$ and with temperature and normalisation as free parameters), considering an annulus centred on the X-ray peak with radii $[0.8-1]r_{500}$.

\subsubsection{Galaxy cluster spectral model} \label{sec:Xcl spec}
The cluster emission from the hot plasma present in the ICM is modelled combining the bremmsstrahlung continuum and the metal emission lines with the \textsc{apec}\footnote{\url{http://www.atomdb.org/index.php}} spectral library and the absorption due to the Galactic medium. For the latter, we consider the photoelectric cross-sections from the work of \citet{Verner1996}, the abundance table of \citet{Asplund2009}, and the total hydrogen column density defined as: $\rm N_{H, tot}= N_{H_I}+2N_{H_2}$. For the atomic hydrogen $\rm H_I$, we use the public data from the full-sky HI survey by the \citet{HI4PI2016}. For the molecular ($\rm N_{H_2}$) contribution, we use the results of Bourdin et al. (in prep.) that estimate the fraction of this component around the CHEX-MATE from the thermal dust emission excesses observed with the (sub)-mm sky survey of the \textit{Planck}-HFI in the combination of the HI4PI $\rm N_{H_I}$ maps.

The only exception for this procedure is the cluster PSZ2 G263.68-22.55. As we will detail more extensively in Appendix~\ref{app:NH263}, the molecular and the $\rm H_I$ hydrogen column density towards this specific cluster do not accurately describe the X-ray absorption in the soft band. In particular, if we assume the $\rm N_{H, tot}$ value, we observe an overestimate of the Galactic absorption for the cluster spectrum. Differently, using the atomic value in our modelling seems to overestimate the cluster spectrum in the soft band. 
Furthermore, if we consider the relation between $\rm N_{H_I}$ and $\rm N_{H, tot}$ for our sample, this cluster has a molecular fraction that is an outlier in the relation of Bourdin et al. (in prep.), with an increment in the hydrogen column density of $\rm \Delta N_{H}/N_{H_I}=70\%$. Thus, we decide to estimate $\rm N_{H, tot}$ directly from the X-ray data for a more accurate treatment of the X-ray spectrum of this cluster. In particular, we consider the cluster spectrum excluding the central region (where the higher metallicity could bias the fit) and the outskirt of the galaxy cluster, considering a circular annulus with radii $[0.15, 0.6]R_{500}$. We leave free to vary in the fit $\rm N_H$, the cluster metallicity, temperature, and the spectrum normalisation. The $\chi^2$ minimisation of our background (see Sec.~\ref{sec:Xray bck}) plus cluster emission models in the energy range $[0.3, 12.1]$ keV returns a value of $\rm N_H=(6.88^{+0.12}_{-0.13})\, 10^{20} \rm cm^{-2}$, with a reduced $\chi^2_\nu$ equal to $1.09$, significantly better than the molecular ($\chi^2_\nu=1.48)$ and the only atomic ($\chi^2_\nu=1.18)$ cases. We refer the reader to Appendix~\ref{app:NH263} for further details regarding the density content of the hydrogen column density towards this cluster of galaxies.

\subsubsection{Surface brightness and temperature maps}
The reconstruction of the clusters signal is shown in Fig. \ref{fig:Txmap}, where the contour plots of the soft X-ray surface brightness are superimposed on normalised temperature maps.
The surface brightness contours derive from wavelet analyses of photon images that we corrected for spatial variations of the effective area and background model. \\
The photon images are denoised via the 4-$\sigma$ soft-thresholding of variance stabilised wavelet transforms \citep{2008ITIP...17.1093Z,2009A&A...504..641S}, which are especially suited to process low photon counts. 
Image analyses include the inpainting of detected point sources and the spatial adaptation of wavelet coefficient thresholds to the spatial variations of the effective area.\\
The Temperature maps are computed using a spectral-imaging algorithm that combines spatially weighted likelihood estimates of the projected intra-cluster medium temperature with a curvelet analysis. 
This algorithm can be seen as an adaptation of the spectral-imaging algorithm presented in \citet{bourdin2015} to the X-ray data. 
Briefly, temperature log-likelihoods are first computed from spectral analysis in each pixel of the maps, then spatially weighted with kernels that correspond to the negative and positive parts of B3-spline wavelets functions. 
We can derive, with this method, wavelet coefficients of the temperature features and their expected fluctuation from spatially weighted Fisher information. 
We use such coefficients to derive wavelet and curvelet transforms that typically estimate features of apparent size in the range of [3.5, 60] arcsec. 
We finally reconstruct temperature maps from de-noised curvelet transforms that we inferred from a 4-$\sigma$ soft thresholding of the curvelet coefficients.
The temperature maps in Fig. \ref{fig:Txmap} are expressed in terms of the mean spectral temperatures estimated inside $[0.15, 0.75]r_{500}$ and, generally, used for $Y_X=M_{gas, 500}T_{Y_X}$, the X-ray counterpart of the SZ $Y$ parameter \citep{2006ApJ...650..128K}. 
In particular, $Y_X$ is estimated iteratively together with $M_{gas, 500}$, $T_{Y_X}$, and an only X-ray estimation of $r_{500}$ using the \citet{2010A&A...517A..92A} scaling relation. An exception to this procedure for the brightness and temperature maps (but also for the thermodynamical profiles) is adopted for PSZ2G339.63-69, also known as the Phoenix cluster, due to the presence of a strong AGN emission in the X-ray data. In section~\ref{sec:Phoenix} we discuss in more detail the different strategy adopted for this cluster. As can be seen from Fig.~\ref{fig:Txmap}, this overlap of the SPT and CHEX-MATE samples collect clusters with a variety of cluster morphology and temperature features. 

Regarding the cluster morphology, only two clusters (PSZ2G271.18-30.95, PSZ2G339.63-69) present regular and concentric X-ray isophotes and a cool core structure. PSZ2G259.98-63.43 and PSZ2G263.68-22.55 also show a temperature decrement towards the cluster centre, but exhibit local compressions of the X-ray isophotes that likely indicate cold fronts.
For the last two clusters, PSZ2G262.73-40.92 and PSZ2G277.76-51.74, we do not observe any relevant temperature drop towards the centre, with the latter cluster also presenting a more complex and disturbed morphology. This qualitative visual inspection of the cluster appearance is corroborated by the morphological analysis of the overall CHEX-MATE sample made by \citet{Campitiello2022}, where each cluster is analysed with several morphological parameters. According to their results, we consider in our sample two relaxed (PSZ2G271.18-30.95, PSZ2G339.63-69) galaxy clusters, one disturbed (PSZ2G277.76-51.74), and the remaining three with an intermediate, “mixed”, morphology.
\begin{figure*}
    \centering
    \includegraphics[width=\textwidth]{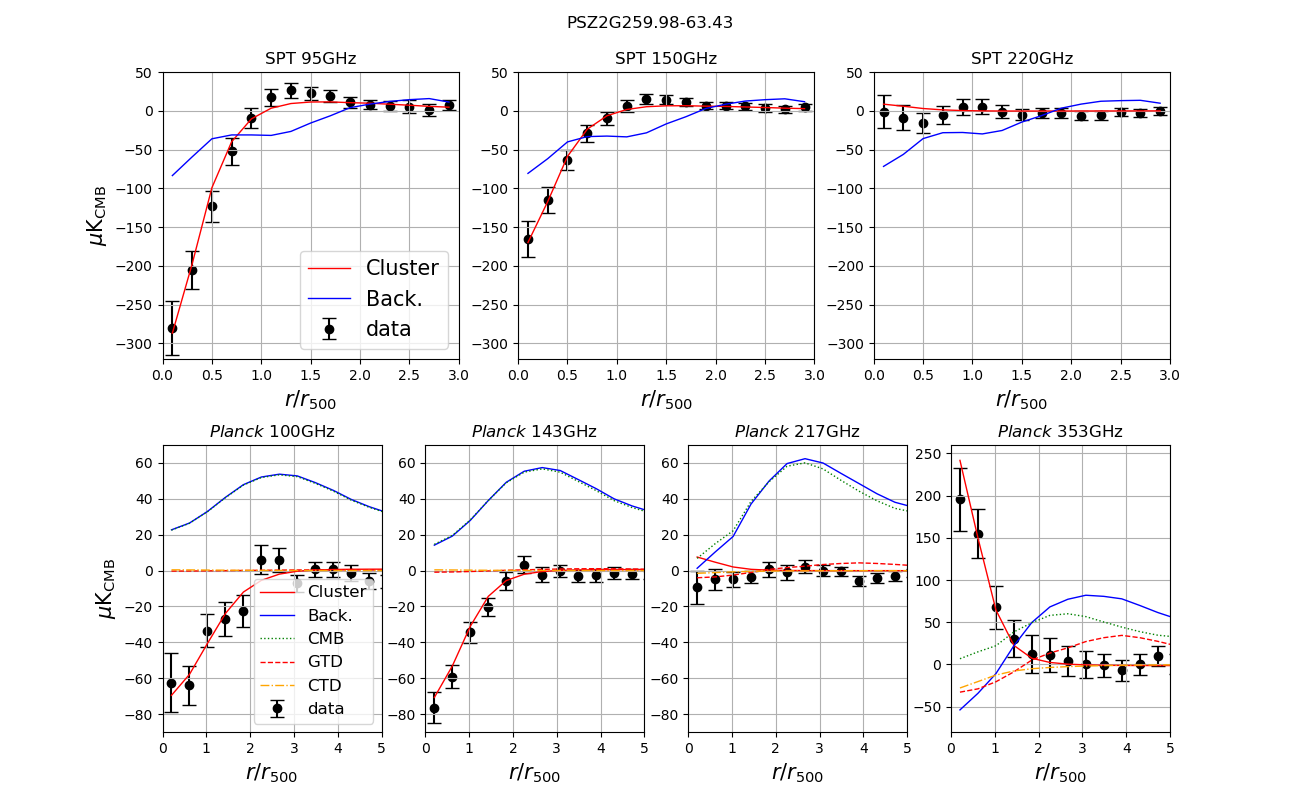}
    \caption{Radial profile around PSZ2G259.98-63.43 for the three SPT channels (95-150-220, upper panels) and four {\em Planck} channels (100-143-217-353, lower panels). The black dots represent the radial average on the SZ maps, the red line is the best fit of the cluster profile, the blue line is the radial mean of the total background signal. For {\em Planck} data we show also the individual components of the background model: the dotted cyan line is the CMB, the dashed red line is the GTD and the dash-dotted orange line is the cluster dust component.   }
    \label{fig:freqprof259}
\end{figure*}
\label{app:freqprof}
\begin{figure*}
    \centering
    \includegraphics[width=\textwidth]{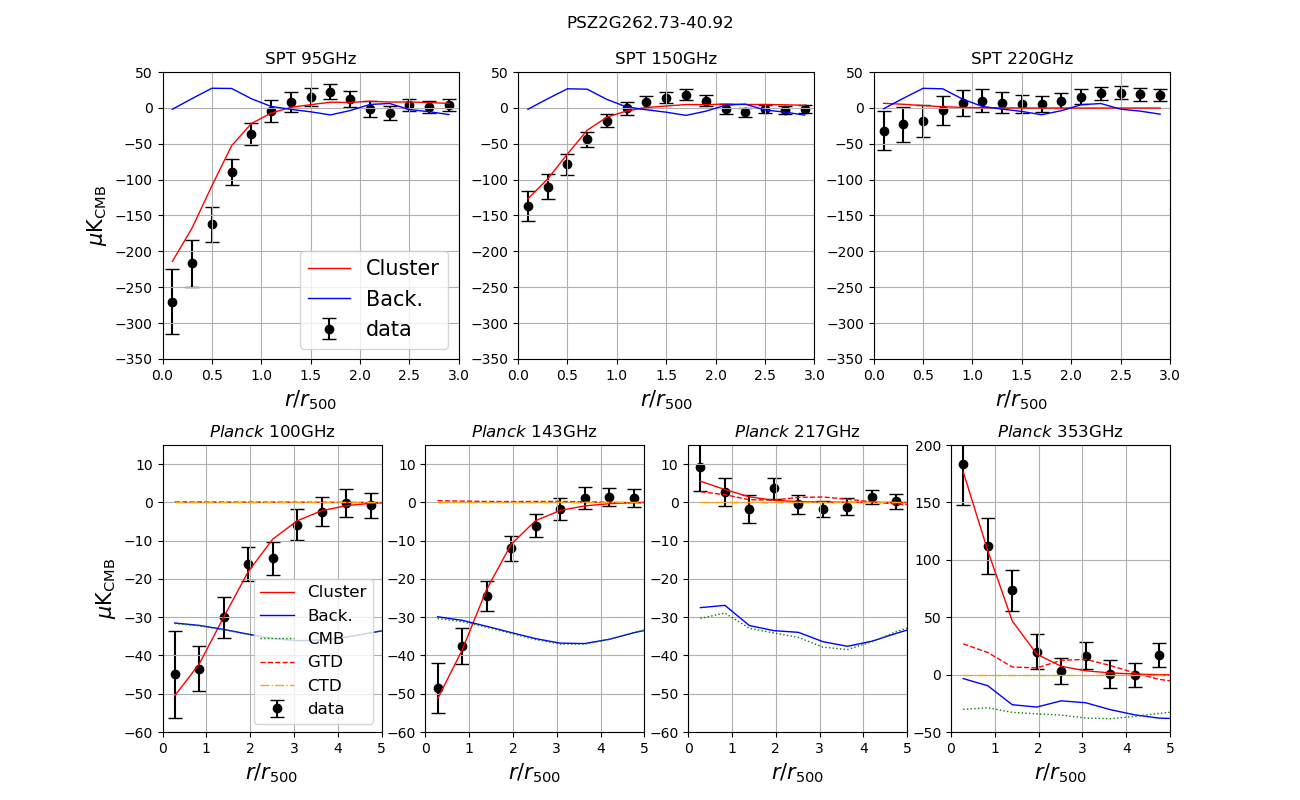}
    \caption{As Figure \ref{fig:freqprof259} but for PSZ2 G262.73-40.92}
    \label{fig:freqprof262}
\end{figure*}
\begin{figure*}
    \centering
    \includegraphics[width=\textwidth]{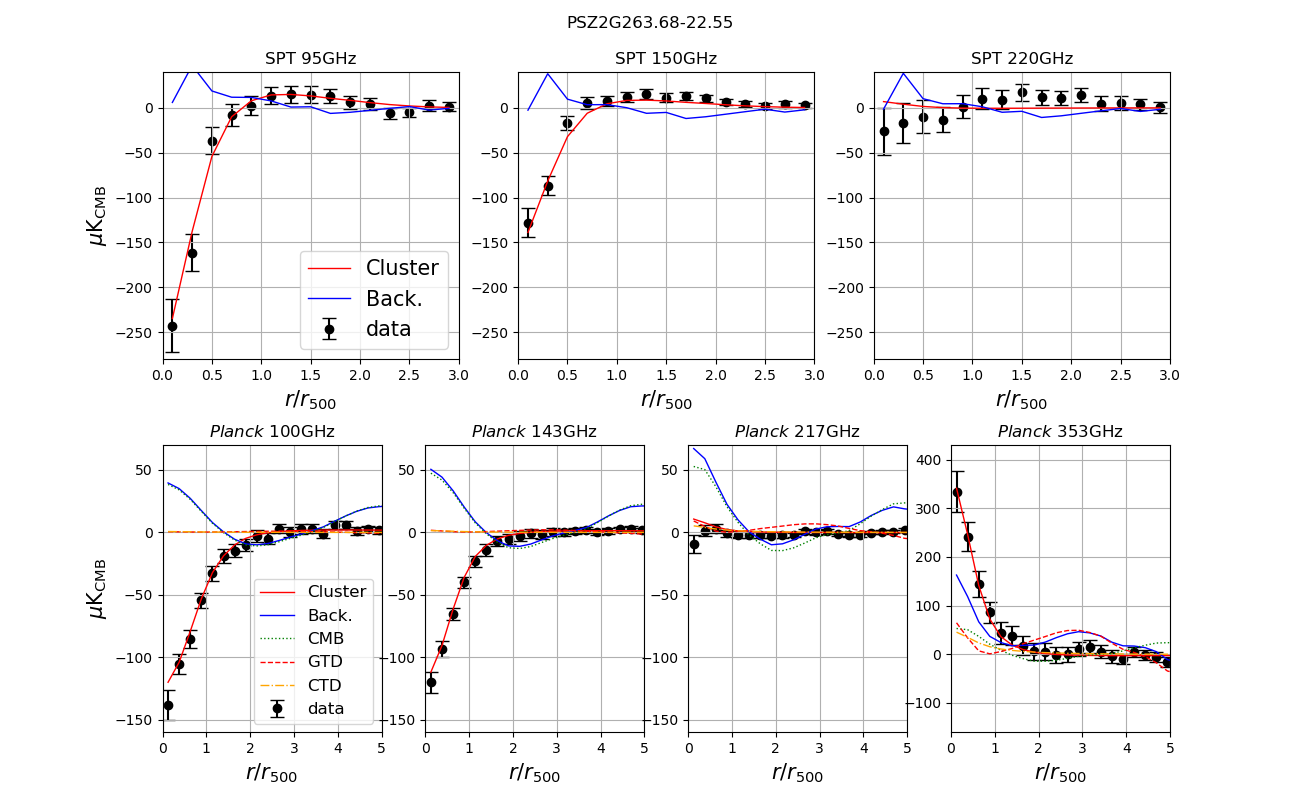}
    \caption{As Figure \ref{fig:freqprof259} but for PSZ2 G263.68-22.55}
    \label{fig:freqprof263}
\end{figure*}
\begin{figure*}
    \centering
    \includegraphics[width=\textwidth]{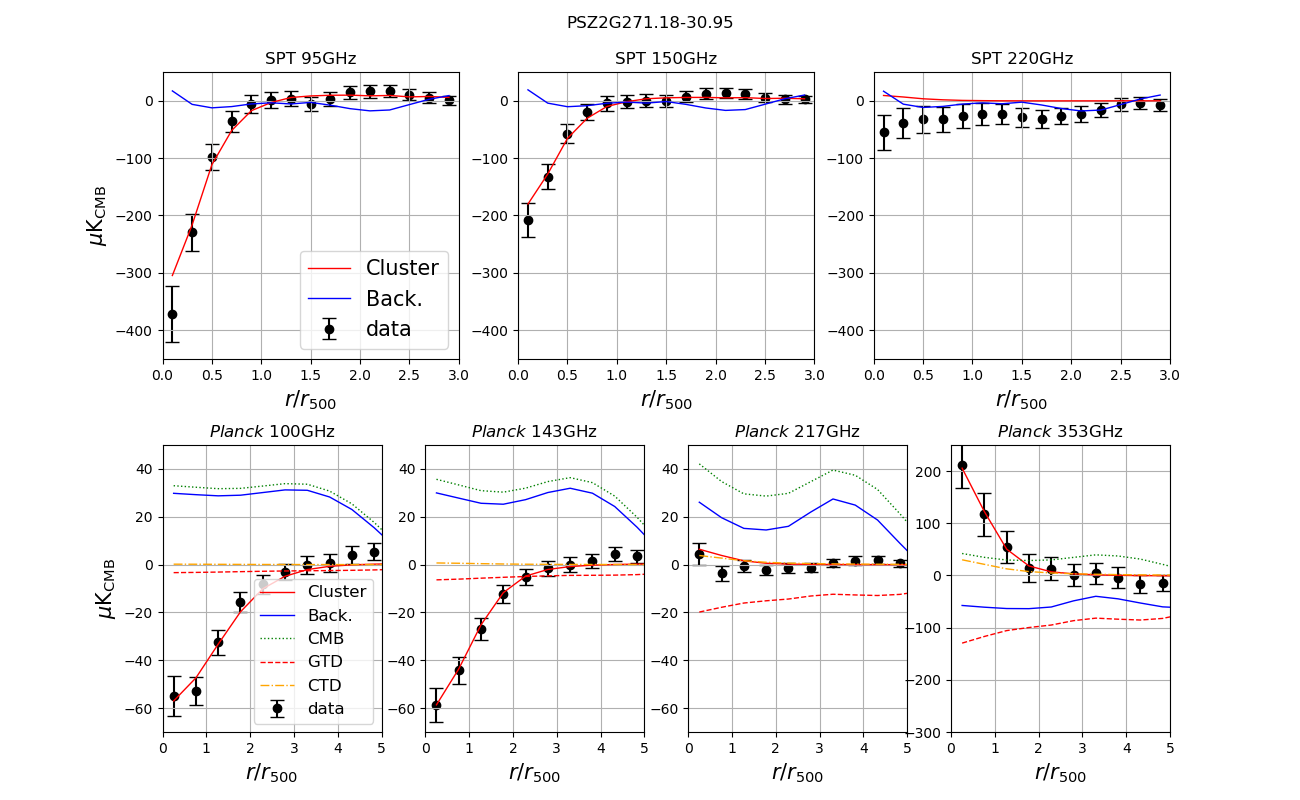}
    \caption{As Figure \ref{fig:freqprof259} but for PSZ2 G271.18-30.95}
    \label{fig:freqprof271}
\end{figure*}
\begin{figure*}
    \centering
    \includegraphics[width=\textwidth]{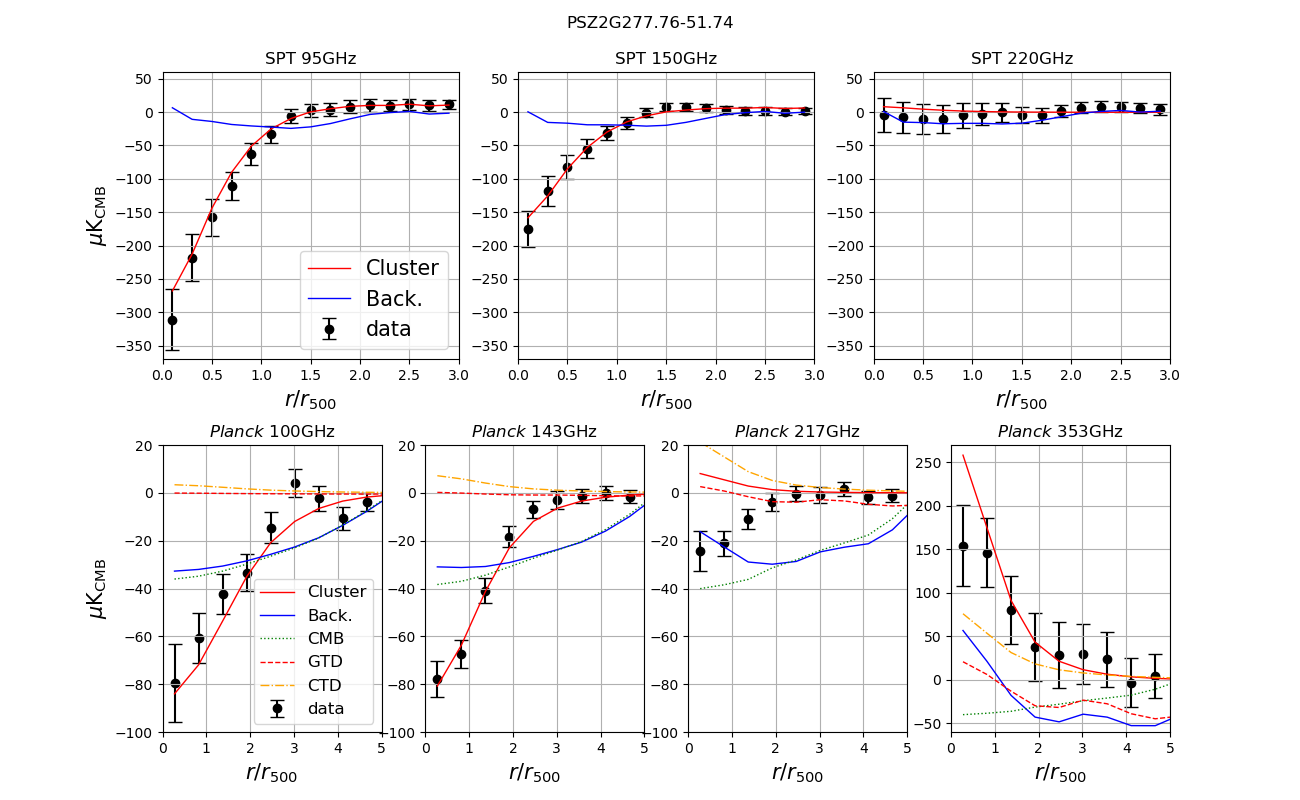}
    \caption{As Figure \ref{fig:freqprof259} but for PSZ2 G277.76-51.74}
    \label{fig:freqprof277}
\end{figure*}
\begin{figure*}
    \centering
    \includegraphics[width=\textwidth]{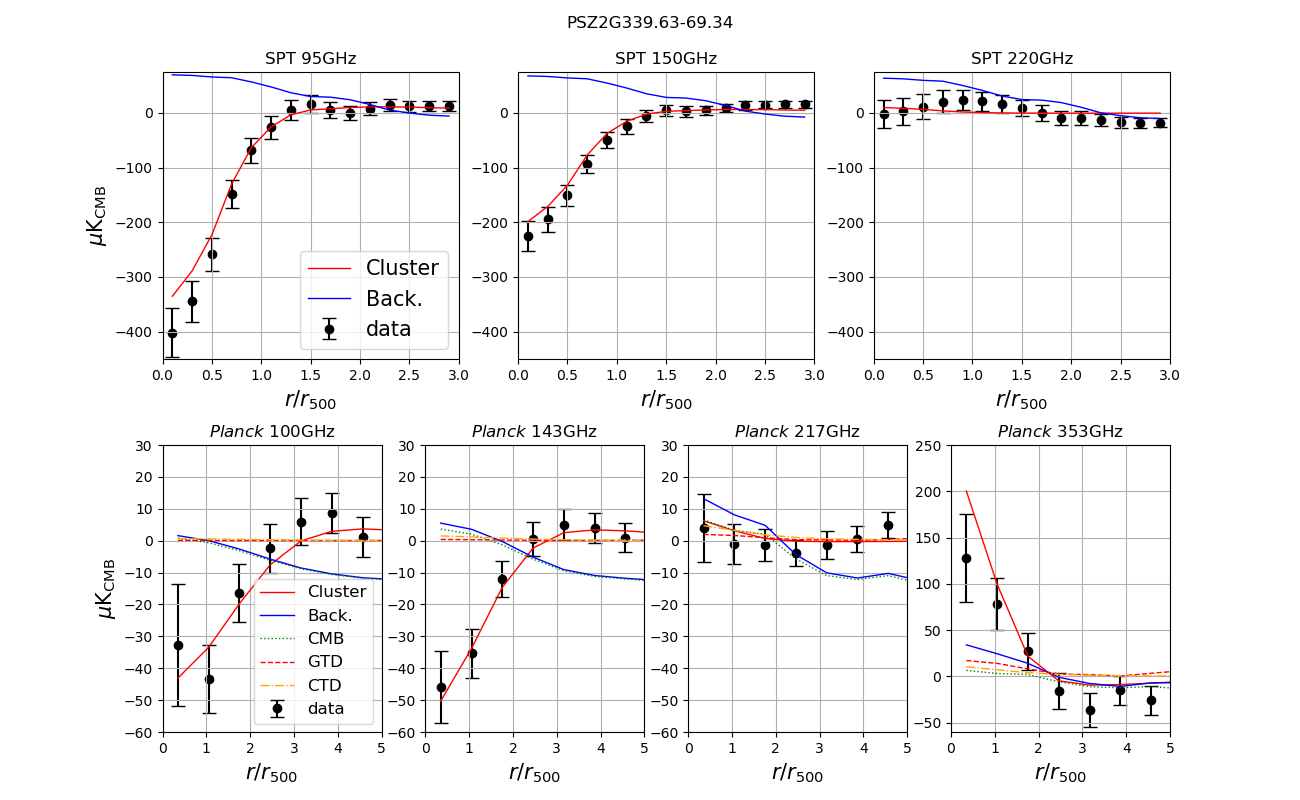}
    \caption{As Figure \ref{fig:freqprof259} but for PSZ2 G339.63-69.34}
    \label{fig:freqprof339}
\end{figure*}
\subsubsection{Radial profiles of thermodynamical quantities}
Considering the background model described in section~\ref{sec:Xray bck}, we estimate the thermodynamical X-ray profiles as done in \citet{2017ApJ...843...72B}. We parametrize the emission measure $[n_p n_e](r)$ and temperature $T(r)$ profiles using the analytical expressions of \citet{Vikhlinin2006}. These templates are integrated along the line of sight and fitted to the background-subtracted X-ray observable profiles: the projected temperature (estimated considering the energy band $[0.3,12.1]$ keV) and the soft ($[0.5, 2.5]$ keV) X-ray surface brightness $\Sigma_X$:
\begin{equation}
    \Sigma_X(r)= \frac{1}{4\pi(1+z)^3}\int[n_p n_e](r)\Lambda(T,Z)dl,
    \label{eq:X_SB}
\end{equation}
with $z$ the redshift of the cluster, $\Lambda(T,Z)$ the cooling function of the cluster emission, $T$ the temperature and $Z$ the element abundances.

The functional form suggested by \citet{Vikhlinin2006} for the emission measure can be written as:
\begin{align}
    n_p n_e (r)= & \frac{n_0^2(r/r_c)^{-\alpha}}{\left[1+(r/r_s)^\gamma\right]^{\epsilon/\gamma} \left[1+(r/r_{c})^2\right]^{3\beta_1-\alpha/2}} + \nonumber
    \\
    & + \frac{n_{0,2}^2}{\left[1+(r/r_{c,2})^2\right]^{3\beta_2}},
    \label{eq:3D_Xprof}
\end{align}
where we fix $\gamma$ to $\gamma=3$ and we leave the other parameters free to vary in the fit. We parametrise the behaviour of the temperature profiles as in \citet{Vikhlinin2006}, leaving all the eight parameters free to vary:
\begin{equation}
    T(r)= T_0 \frac{x+T_{min}/T_0}{x+1} \frac{(r/r_t)^{-a}}{\left[1+(r/r_t)^b\right]^{c/b}},
   \label{eq:3D_XTprof}
\end{equation}
with $x=(r/r_{cool})^a_{cool}$. For the temperature projection, we consider the spectroscopic-like temperature $T_s$ weighting scheme of \citet{Mazzotta2004}:
\begin{equation}
    kT_s= \frac{\int WkT(r)dl}{\int Wdl},
    \label{eq:TSl}
\end{equation}
where $k$ is the Boltzmann constant and the X-ray emissivity is used as a weight to better reproduce the effect of a single temperature fit in the spectroscopic analysis: $W=n_e^2/T^{3/4}$. 
This weighting scheme has been proven to be accurate in the temperature regime of the clusters in our sample ($>3$ keV).

All the fits are conducted with a least-squares minimization following the Levenberg–Marquardt algorithm. For uncertainties envelopes of the best-fit profiles, we perform $500$ parametric bootstrap realisations of the observed profiles, estimated considering twelve radial bins inside $R_{500}$ (and logarithmically spaced between $[0.15,0.8]R_{500}$) for the spectroscopic temperature, and $50$ logarithmic radial bins for the surface brightness inside $R_{500}$. We build the templates considering the XMM mirror PSF and the effect of the Galactic absorption described in Sec.~\ref{sec:Xcl spec}. The projection also considers a metal abundance constant normalisation ($Z=0.3$) for the cluster and the {\it Planck} \citep[eq.~79b][]{Planck2018VI} primordial helium abundance $Y^{BBN}_P=0.243$. As a consequence of these choices, the electron density can be estimated from Eq.~\ref{eq:3D_Xprof} with a particle mean weight of $\mu=0.592$ and a proton to electron ratio: $n_p/n_e=0.859$.
\subsubsection{The Phoenix cluster PSZ2G339.63-69} \label{sec:Phoenix}
The Phoenix cluster is a known astronomical object in X-ray, optical and radio astronomy. \citet{McDonald2019} show with their work based on the Karl Jansky Very Large Array, {\it Hubble} and {\it Chandra} space telescopes that the central temperature and entropy profiles are consistent with a pure cooling model. \citet{Kitayama2020} also measure a similar decrement towards the cluster centre by combining the SZ signal from ALMA and {\it Chandra} X-ray information. This efficient cooling in the centre of the cluster is highly related to the activity of the central galaxy, which hosts a powerful and obscured AGN \citep{McDonald2012, McDonald2013, Tozzi2015, McDonald2019} that with its feedback supports the formation of a multiphase condensation in the ICM. 

For our X-ray analysis, the strong AGN signal contaminates the XMM-\textit{Newton} observations up to the cluster periphery due to the larger point spread function of XMM (half-energy width at aimpoint: $15\arcsec$) compared to the {\it Chandra} telescope and the cluster apparent size ($z=0.596$, $R_{500}=2.85$ arcmin). As pointed out by \citet{Tozzi2015, McDonald2019}, the AGN is moderately obscured, with its emission that dominates the spectrum in the hard X-ray band above $2\, {\rm keV}$. Thus, the AGN signal makes it more difficult to reconstruct the cluster spectrum for XMM-\textit{Newton} EPIC cameras without proper modelling of the AGN emission over the entire spectral band in interest ($[0.3-12.1]$ keV). To reduce the impact of the AGN in the X-ray cluster modelling for the brightness and temperature maps or the thermodynamical profiles, we restrict the analysis only where the cluster dominates over the AGN signal, thus removing all the photons with $E>2$ keV.

\section{Results}
\label{sec:res}
In this section, we discuss the pressure profiles obtained with our SZ pipeline and compare them to the profiles derived from X-ray analysis.

\subsection{Millimetric fit}
We fit the profile on the 95 and 150 GHz channels of SPT, and 100, 143, 353 channels of {\em Planck}. 
We exclude the 220 GHz and the 217 GHz channels of SPT and {\em Planck}, respectively. 
These channels do not carry relevant information about the cluster, since the SZ signal is expected to be null around their frequencies.
We also exclude the high frequency channel of {\it Planck}, namely the 545 GHz and 857 GHz, since they are dominated by dust emission and the contribution to the fit of the cluster signal is marginal.
We show the comparison between the best joint fit clusters profiles and the SPT and {\em Planck} data, cleaned from the background and foreground signals in Fig. \ref{fig:freqprof259}-\ref{fig:freqprof339}.
We also show the profiles of the background and foreground components. 
Notice that none of the contaminants components is intrinsically positive or negative. 
Due to the high pass filter, we are looking at the fluctuations with respect to the underlying large scale signal.
So that even if the total GTD emission is always positive, its small scales fluctuation are not.
Furthermore, we remember that the CTD component is not the cluster dust emission but a correction term to the total dust template accounting for the different SED of the two dust components.
Thus, its sign varies depending on the specific features on each map.

Our sky model, including the cluster contribution and the cosmological and galactic contamination, provides an excellent reconstruction of both SPT and Planck data.
Furthermore, we verify that the two data sets are well consistent with one another. 
Notice that we do not show the \say{raw} data in the plots, but we include instead the foreground cleaned data. 
The raw signal corresponds to the sum between the black dots and the contaminant components.

These plots highlight the peculiarities of the two data-set that led us to implement specific component separation pipelines for each instrument. 
As it is evident, due to difference in resolution and spectral response, {\em Planck} raw data are more contaminated by large scale backgrounds and foregrounds signals than SPT. 
As stated before, SPT cannot detect these signals due to its intrinsic high-pass filter.

We recall that the 220 GHz SPT channel and the 217 GHz {\em Planck } channel are not included in the fit of the cluster profile since we expect the SZ amplitude to be lower than the noise at those frequencies.
However, they are more sensitive than the lower frequency channels to possible dust foreground residuals.
They thus provide a good benchmark for the component separation algorithms.
It is especially relevant for SPT since the 220GHz is the highest frequency channel, where the dust signal should be brighter.
In this regard, we do not found any evidence of dust contamination in SPT data above the noise level, as it is evident from the up-right panels of Fig. \ref{fig:freqprof259}-\ref{fig:freqprof339}.
The inspection of {\em Planck} data shows, instead, some level of dust contamination at high frequency in almost all the clusters.
However, the amplitude at the lower frequencies, common to SPT, is not relevant compared to both the cluster signal and the noise. 
The explanation of the relatively low level of dust contamination also in {\em Planck} data comes from the position on the sky of SPT clusters, far from the galactic plane, as well as from the high-pass filter that we applied to the data (described in section \ref{sec:planckback}).
Only in the case of PSZ2G271.18-30.95 (see Figure \ref{fig:freqprof271}) the dust emission results evident also at the lower frequencies common to SPT.
On the other hand, we do not find evidence of this emission in SPT data, but a possible hint comes from a systematic negative excess at 220 GHz. 
However, this deviation seems too faint with respect to the noise level to be relevant.
Since this signal comes in the form of diffuse emission, the SPT window function high-pass filters a large part of it.
These results confirm our choice, discussed in section \ref{sec:sptback}, to neglect the dust contribution in the SPT data reduction pipeline.

We conclude this section with a comment on the CMB reconstruction in SPT, which appears systematically different and lower than in the {\em Planck} data.
The explanation comes again from the different spatial filtering of the two instruments combined with the shape of the CMB power spectrum.
Since the amplitude of CMB fluctuations decreases with the scale, the CMB signal is lower in SPT data where the larger scales are suppressed. 
Furthermore, as commented in section \ref{sec:sptback}, the small scales of the CMB estimate derived for SPT are dominated by the 220 GHz noise. 

\begin{figure*}
    \centering
    \begin{subfigure}[a]{0.32\textwidth}
        \includegraphics[width=\textwidth]{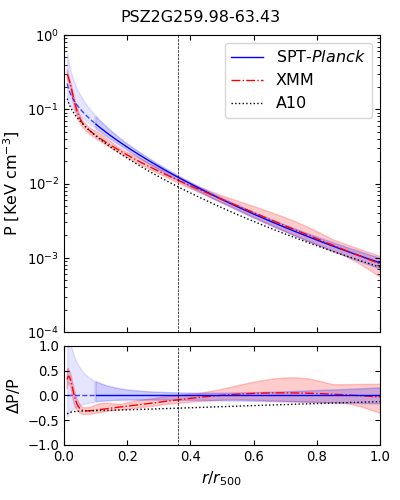}
    \end{subfigure}
    \begin{subfigure}[a]{0.32\textwidth}
        \includegraphics[width=\textwidth]{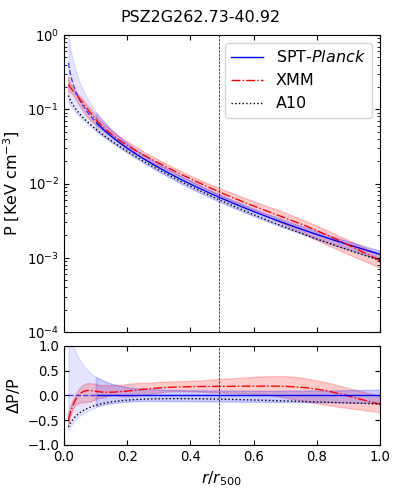}
    \end{subfigure}
        \begin{subfigure}[a]{0.32\textwidth}
    \includegraphics[width=\textwidth]{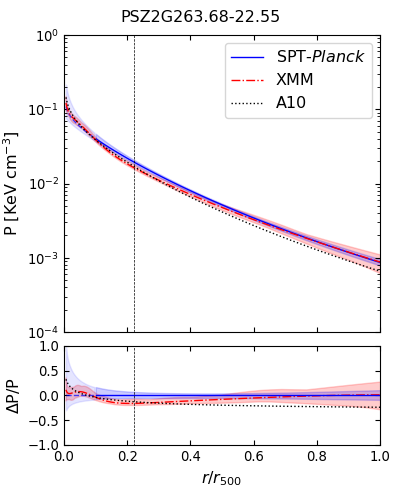}
    \end{subfigure}
    \begin{subfigure}[a]{0.32\textwidth}
        \includegraphics[width=\textwidth]{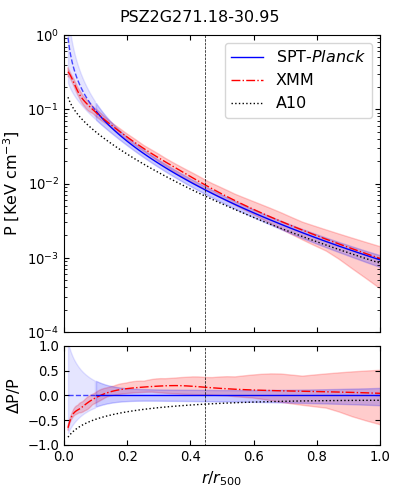}
    \end{subfigure}
    \begin{subfigure}[a]{0.32\textwidth}
        \includegraphics[width=\textwidth]{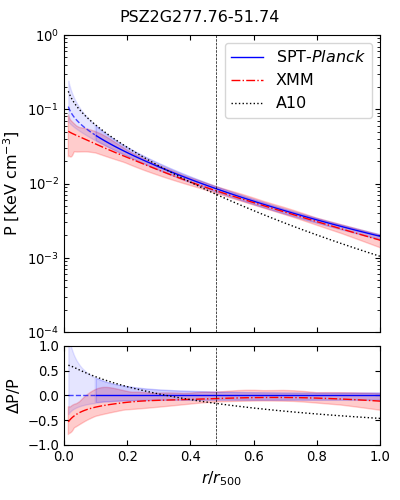}
    \end{subfigure}
    \begin{subfigure}[a]{0.32\textwidth}
        \includegraphics[width=\textwidth]{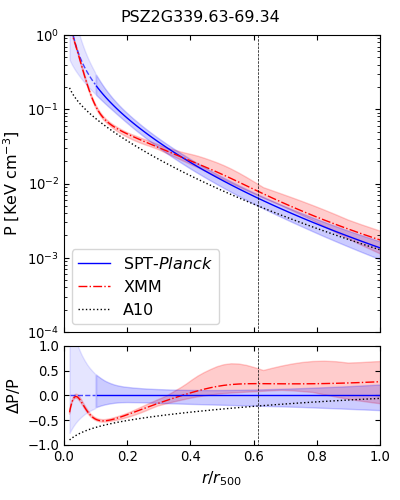}
    \end{subfigure}
    \caption{Comparison between pressure profiles from XMM-\textit{Newton} data (dash dotted red lines) and  from SPT-\textit{Planck} joint fit (blue lines). The shaded regions correspond to the $68\%$ credible intervals. The light blue dashed section of the SPT-\textit{Planck} line mark the region inside the innermost SPT data point. The dotted black lines show the universal profile from \cite{2010A&A...517A..92A} as comparison. The vertical dashed lines indicate the FWHM of the SPT beam. The lower panels represent the relative deviation with respect the SZ results. The ticks of the horizontal axis correspond to the edges of the bins on SPT data.}
    \label{fig:presprof}
\end{figure*}
\subsection{Comparison with X-ray profiles}

X-ray data provide an independent benchmark for our pipeline.
We compare the pressure profiles derived from the SZ maps with the expectations from X-ray observations.
X-ray and SZ data are significantly different in sensitivity and resolution and they probe the cluster structure at different regimes. 
Thus, we limit our comparison to the overlapping region between the two, {\em i.e.} inside $r_{500}$.
The XMM-\textit{Newton} sensitivity sets the upper limit: the X-ray cluster emission rapidly decline with the radius, given its quadratic dependence on the density, and it falls below the XMM-\textit{Newton} background for $r>r_{500}$.

We show the comparison between the two fits in Figure \ref{fig:presprof}. 
We find a perfect match between the two profiles outside the SPT FWHM (marked with a vertical dashed line in Fig. \ref{fig:presprof}).
Remarkably, we find good consistency also for smaller radii, for five out of six clusters.
The credibility intervals overlap along the whole radius range tested.
The agreement between SZ and X-ray profiles is good regardless of the actual cluster's angular size.
As the X-ray analysis highlights, our sample contains clusters with a various morphological features.
This comparison shows that our method provides consistent results regardless of the morphology.
In only one case, the Phoenix cluster PSZ2 G339.63-69.34, the results do not match perfectly, we will discuss this case in details later on.
These results confirm that our method correctly recover the profile up to the scale of the FWHM of the SPT beam. 
Some degree of divergence, not statistically relevant, can be observed in the very inner regions.
These small deviations are not source of concern since, due to the limited resolution, SPT and {\em Planck} are not expected to be sensitive to these small scales that are instead captured by XMM-\textit{Newton}.
Furthermore, the gNFW model fitted to the SZ data cannot reproduce small scale features as the changes of slope inside $r_{500}$ present in some clusters (in particular in PSZ2 G339.63-69.34, but also noticeable {\it e.g.} in PSZ2 G263.68-22.55 and PSZ2 G259.98-63.43).
We recall that the SZ fit is performed over radial bins of size $0.2r_{500}$, marked by the ticks on the horizontal axis in Fig. \ref{fig:presprof}.

The Phoenix cluster, PSZ2G339.63-69.34, represents the only case where the two data-sets appear slightly but systematically shifted.
It is the cluster with the smaller angular diameter in the sample (see Table \ref{table:sample}), and the one with the lowest signal to noise.
As we comment in section \ref{sec:Xana}, the strong emission of an AGN contaminate the X-ray signal and make the XMM-\textit{Newton} analysis challenging.
Furthermore, the profile obtained from X ray data is quite irregular, showing a number sharp changes of slope at small radii.
As commented before, the resolution of the millimetric data is to low to detect this behaviour, as well as the model cannot account for it.
In light of these limitations, we conclude that, despite the small deviations, the agreement between the two profile is very good and probe that our method perform well also in poor conditions.

For comparison, in Figure \ref{fig:presprof} we show also in the \say{universal} profile of \cite{2010A&A...517A..92A}. 
We observe a scatter around the profile but the sample is too small to draw any conclusion in this respect.
We will investigate further this topic in future works on larger samples. 

We report the best fit value for the parameters in Table \ref{tab:bestfit}, with the universal profiles parameters as reference. 
The credibility intervals for each parameter are obtained excluding the tails of the marginalised posterior symmetrically so that they contain the 68\% of the parameter space. 

\subsection{Joint-fit vs SPT}
\begin{figure*}
    \centering
    \begin{subfigure}[a]{0.32\textwidth}
        \includegraphics[width=\textwidth]{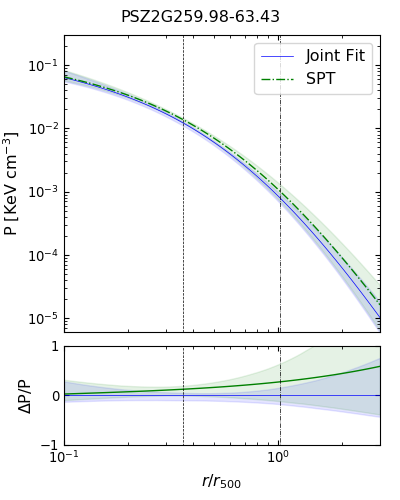}
    \end{subfigure}
    \begin{subfigure}[a]{0.32\textwidth}
        \includegraphics[width=\textwidth]{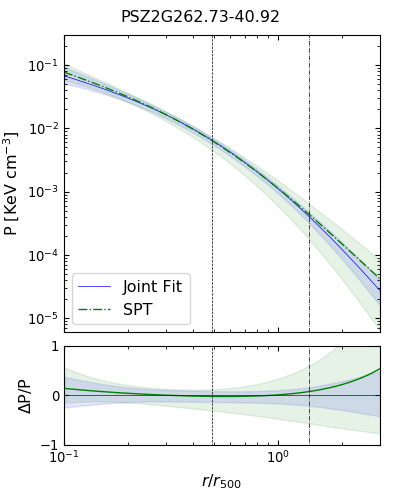}
    \end{subfigure}
        \begin{subfigure}[a]{0.32\textwidth}
    \includegraphics[width=\textwidth]{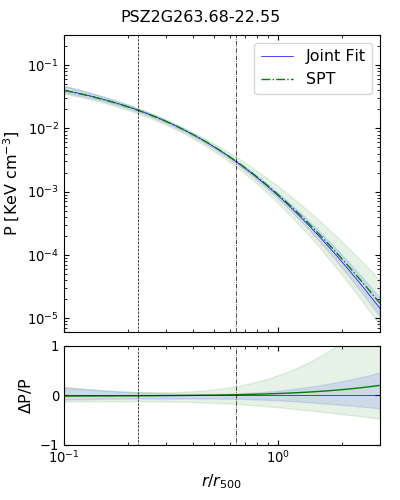}
    \end{subfigure}
    \begin{subfigure}[a]{0.32\textwidth}
        \includegraphics[width=\textwidth]{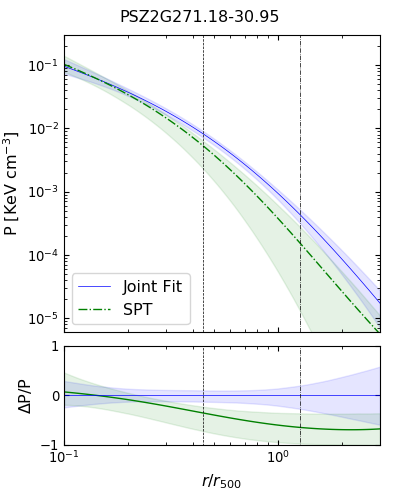}
    \end{subfigure}
    \begin{subfigure}[a]{0.32\textwidth}
        \includegraphics[width=\textwidth]{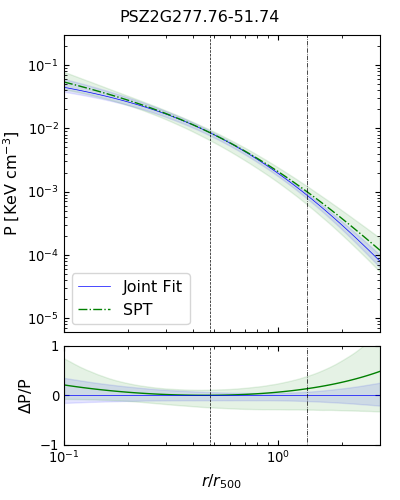}
    \end{subfigure}
    \begin{subfigure}[a]{0.32\textwidth}
        \includegraphics[width=\textwidth]{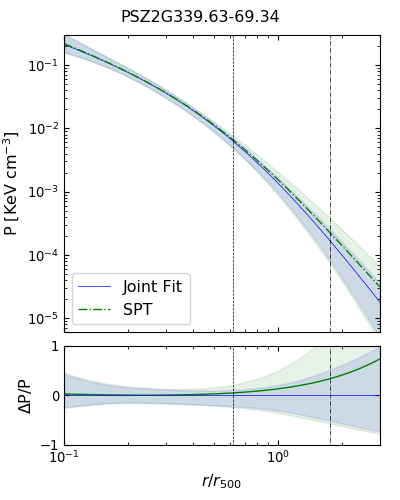}
    \end{subfigure}
    \caption{Comparison between pressure profiles from SPT data (dash dotted green lines) and from the joint fit on the two datasets (blue line). The vertical dashed lines indicate the FWHM of the SPT beam and the dash-dotted line the lower {\em Planck} beam FWHM (5 arcmin). The lower panels represent the relative deviation with respect the joint fit.}
    \label{fig:sptplanck}
\end{figure*}

In this section, we highlight the advantages of combining different probes. 
In Fig. \ref{fig:sptplanck} we compare the results of the SPT-{\em Planck} joint fit with the profiles obtained with SPT alone. 
We do not compare the fit on {\it Planck} data alone since it does not converge with our choice of free parameters due to the limited resolution.
A different combination of free parameters ({\it e.g.} to fix the central slope $\gamma$) would make the comparison unfair.
Notice that, unlike Fig \ref{fig:presprof}, here we show all the scales considered in the fit so that the plot extends up to $3r_{500}$.
We mark the SPT  and {\em Planck} beam FWHM with vertical dashed and dash-dotted lines, respectively.
For {\em Planck} we report the FWHM of the beam of the channels with the best resolution, corresponding to 5 arcmin.
From these plots it is evident how {\em Planck} contribution significantly improves the fit on the outer regions. 
The uncertainties on the profiles become significantly more narrow when we include the {\em Planck} channels. 
The inner parts of the profiles instead coincide for the two fits. 
This show that the small scales are dominated by the information from SPT, as expected.
\begin{table*}
\caption{Best fit parameters for the gNFW fit}     
\centering          
\begin{tabular}{|c| c| c| c|}     
\hline       
    Cluster    & $P_0$            & $\beta$             & $\gamma$  \\  \hline

 PSZ2 G259.98-63.43 & $12.3 \ \left[+11.7,-3.4\right]$ & $5.87\ \left[+0.71,-0.59\right]$ & $<0.42$ \\ 
 PSZ2 G262.73-40.92 & $4.8 \ \left[+7.5,-2.2\right]$ & $4.78\pm0.73$          & $<0.82$ \\ 
 PSZ2 G263.68-22.55 & $11.2 \ \left[+5.8,-1.9\right]$ & $5.46\ \left[+0.46,-0.34\right]$ & $<0.26$  \\ 
 PSZ2 G271.18-30.95 & $5.7 \ \left[10.1,-3.2\right]$ & $5.24\ \left[+0.88,-1.0 \right]$  & $0.75 \ \left[+0.36,-0.57\right]$         \\ 
 PSZ2 G277.76-51.74 & $5.9 \ \left[+3.7,-1.3\right]$ & $4.23\ \left[+0.39,-0.32\right]$ & $<0.36$ \\ 
 PSZ2 G339.63-69.34 & $11.7 \ \left[+28.1,-5.6\right]$ &  $6.4\pm1.1$           & $<0.85$ \\ \hline
 A10                & $8.4$ &  $5.49$           & $0.31$ \\
\hline       

\end{tabular}
\label{tab:bestfit}
\tablefoot{Best fit values of the pressure profiles free parameters obtained from the joint fit of SPT and {\it Planck} data for our cluster sample and the parameters of the universal profile from \cite{2010A&A...517A..92A} (A10). We show the uncertainties that includes the $68\%$ credibility interval. }
\end{table*} 
\section{Conclusions}
\label{sec:con}
In this paper, we presented a technique to measure the pressure profile of galaxy clusters in millimetric and submillimetric observation from SPT and {\em Planck}.

To fully exploit the potential of the two data-sets, we developed two independent component separation pipelines specifically tailored to the characteristic of each instrument.
For {\em Planck} we took advantage of the high number of frequency channels to build a complete analytical model of the galactic foreground and the cosmological background.
For SPT, we adopt a linear combination to reconstruct the CMB signal and we verified that the intrinsic spatial filtering minimises the dust foreground.
We use a parametric model of the profiles to efficiently deconvolve the instrumental response to the cluster signal taking into account the angular resolution of each channel.

We applied our algorithm to a sample of six galaxy cluster representing the intersection between the SPT and CHEX-MATE catalogues.
We compared the profiles derived from the SPT-{\em Planck} joint fit to the ones obtained from XMM-\textit{Newton} data.
We found an excellent agreement between the SZ and X-ray pressure profiles for all the cluster in the sample. 
The results are stable over a variety of angular size and for clusters of different morphology.

This consistency provides us with evidence of the robustness of our method.
We probe the potential of large sky millimetric surveys to constrain the profiles of galaxy clusters. 
We show that combining the data from {\it Planck} and SPT greatly improves the constraints compared to the fit on a single instrument.
Thanks to its high resolution, SPT can track the pressure profile up to the inner region of the cluster ($\sim 0.1r_{500}$.
{\it Planck} provides a robust anchor at large scale and a reliable absolute measurement.
We obtain results consistent and comparable with X-ray analysis tracking the pressure of the cluster from the outskirt up to intermediate radii ($\sim 10^{-1}r_{500}$). 
The results of this comparison motivate us to extend our analysis to derive the SZ pressure profiles of all the clusters detected both by {\it Planck} and SPT to investigates the structure and evolution of the profiles through cosmic time.

\begin{acknowledgements}
Based on observations obtained with XMM-\textit{Newton}, an ESA science mission with instruments and contributions directly funded by ESA Member States and NASA.
F.O., F.D., H.B. and P.M. acknowledge financial contribution from the contracts ASI-INAF Athena 2019- 27-HH.0, “Attività di Studio per la comunità scientifica di Astrofisica delle Alte Energie e Fisica Astroparticellare” (Accordo Attuativo ASI-INAF n. 2017-14- H.0), from the European Union’s Horizon 2020 Programme under the AHEAD2020 project (grant agreement n. 871158) and support from INFN through the InDark initiative. J.S. acknowledge the NASA/ADAP Grant Number 80NSSC21K1571.
\end{acknowledgements}

%
%
\bibliographystyle{aa}
\bibliography{SPT-Planck_pressure_profile}

\begin{thebibliography}{70}
\expandafter\ifx\csname natexlab\endcsname\relax\def\natexlab#1{#1}\fi

\bibitem[{{Aghanim} {et~al.}(2019){Aghanim}, {Douspis}, {Hurier}, {Crichton},
  {Diego}, {Hasselfield}, {Macias-Perez}, {Marriage}, {Pointecouteau},
  {Remazeilles}, \& {Soubri{\'e}}}]{2019A&A...632A..47A}
{Aghanim}, N., {Douspis}, M., {Hurier}, G., {et~al.} 2019, \aap, 632, A47

\bibitem[{{Arnaud} {et~al.}(2010){Arnaud}, {Pratt}, {Piffaretti},
  {B{\"o}hringer}, {Croston}, \& {Pointecouteau}}]{2010A&A...517A..92A}
{Arnaud}, M., {Pratt}, G.~W., {Piffaretti}, R., {et~al.} 2010, \aap, 517, A92

\bibitem[{{Asplund} {et~al.}(2009){Asplund}, {Grevesse}, {Sauval}, \&
  {Scott}}]{Asplund2009}
{Asplund}, M., {Grevesse}, N., {Sauval}, A.~J., \& {Scott}, P. 2009, \araa, 47,
  481

\bibitem[{{Battaglia} {et~al.}(2012){Battaglia}, {Bond}, {Pfrommer}, \&
  {Sievers}}]{2012ApJ...758...75B}
{Battaglia}, N., {Bond}, J.~R., {Pfrommer}, C., \& {Sievers}, J.~L. 2012, \apj,
  758, 75

\bibitem[{{Bennett} {et~al.}(2003){Bennett}, {Hill}, {Hinshaw}, {Nolta},
  {Odegard}, {Page}, {Spergel}, {Weiland}, {Wright}, {Halpern}, {Jarosik},
  {Kogut}, {Limon}, {Meyer}, {Tucker}, \& {Wollack}}]{2003ApJS..148...97B}
{Bennett}, C.~L., {Hill}, R.~S., {Hinshaw}, G., {et~al.} 2003, \apjs, 148, 97

\bibitem[{{Bertin} \& {Arnouts}(1996)}]{SExtractor}
{Bertin}, E. \& {Arnouts}, S. 1996, \aaps, 117, 393

\bibitem[{{Bertschinger}(1998)}]{1998ARA&A..36..599B}
{Bertschinger}, E. 1998, \araa, 36, 599

\bibitem[{{Birkinshaw}(1999)}]{1999PhR...310...97B}
{Birkinshaw}, M. 1999, \physrep, 310, 97

\bibitem[{{Bleem} {et~al.}(2015){Bleem}, {Stalder}, {de Haan}, {Aird}, {Allen},
  {Applegate}, {Ashby}, {Bautz}, {Bayliss}, {Benson}, {Bocquet}, {Brodwin},
  {Carlstrom}, {Chang}, {Chiu}, {Cho}, {Clocchiatti}, {Crawford}, {Crites},
  {Desai}, {Dietrich}, {Dobbs}, {Foley}, {Forman}, {George}, {Gladders},
  {Gonzalez}, {Halverson}, {Hennig}, {Hoekstra}, {Holder}, {Holzapfel},
  {Hrubes}, {Jones}, {Keisler}, {Knox}, {Lee}, {Leitch}, {Liu}, {Lueker},
  {Luong-Van}, {Mantz}, {Marrone}, {McDonald}, {McMahon}, {Meyer}, {Mocanu},
  {Mohr}, {Murray}, {Padin}, {Pryke}, {Reichardt}, {Rest}, {Ruel}, {Ruhl},
  {Saliwanchik}, {Saro}, {Sayre}, {Schaffer}, {Schrabback}, {Shirokoff},
  {Song}, {Spieler}, {Stanford}, {Staniszewski}, {Stark}, {Story}, {Stubbs},
  {Vanderlinde}, {Vieira}, {Vikhlinin}, {Williamson}, {Zahn}, \&
  {Zenteno}}]{2015ApJS..216...27B}
{Bleem}, L.~E., {Stalder}, B., {de Haan}, T., {et~al.} 2015, \apjs, 216, 27

\bibitem[{{Bourdin} {et~al.}(2017){Bourdin}, {Mazzotta}, {Kozmanyan}, {Jones},
  \& {Vikhlinin}}]{2017ApJ...843...72B}
{Bourdin}, H., {Mazzotta}, P., {Kozmanyan}, A., {Jones}, C., \& {Vikhlinin}, A.
  2017, \apj, 843, 72

\bibitem[{Bourdin {et~al.}(2013)Bourdin, Mazzotta, Markevitch, Giacintucci, \&
  Brunetti}]{Bourdin2013}
Bourdin, H., Mazzotta, P., Markevitch, M., Giacintucci, S., \& Brunetti, G.
  2013, The Astrophysical Journal, 764, 82

\bibitem[{Bourdin {et~al.}(2015)Bourdin, Mazzotta, \& Rasia}]{bourdin2015}
Bourdin, H., Mazzotta, P., \& Rasia, E. 2015, The Astrophysical Journal, 815,
  92

\bibitem[{{Bourdin, H.} \& {Mazzotta, P.}(2008)}]{Bourdin2008}
{Bourdin, H.} \& {Mazzotta, P.} 2008, A\&A, 479, 307

\bibitem[{Campitiello {et~al.}(2022)Campitiello, Ettori, Lovisari, Bartalucci,
  Eckert, Rasia, Rossetti, Gastaldello, Pratt, Maughan, Pointecouteau, Sereno,
  Biffi, Borgani, De~Luca, De~Petris, Gaspari, Ghizzardi, Mazzotta, \&
  Molendi}]{Campitiello2022}
Campitiello, M.~G., Ettori, S., Lovisari, L., {et~al.} 2022, CHEX-MATE:
  Morphological analysis of the sample

\bibitem[{{Carlstrom} {et~al.}(2011){Carlstrom}, {Ade}, {Aird}, {Benson},
  {Bleem}, {Busetti}, {Chang}, {Chauvin}, {Cho}, {Crawford}, {Crites}, {Dobbs},
  {Halverson}, {Heimsath}, {Holzapfel}, {Hrubes}, {Joy}, {Keisler}, {Lanting},
  {Lee}, {Leitch}, {Leong}, {Lu}, {Lueker}, {Luong-Van}, {McMahon}, {Mehl},
  {Meyer}, {Mohr}, {Montroy}, {Padin}, {Plagge}, {Pryke}, {Ruhl}, {Schaffer},
  {Schwan}, {Shirokoff}, {Spieler}, {Staniszewski}, {Stark}, {Tucker},
  {Vanderlinde}, {Vieira}, \& {Williamson}}]{2011PASP..123..568C}
{Carlstrom}, J.~E., {Ade}, P.~A.~R., {Aird}, K.~A., {et~al.} 2011, \pasp, 123,
  568

\bibitem[{{Chown} {et~al.}(2018){Chown}, {Omori}, {Aylor}, {Benson}, {Bleem},
  {Carlstrom}, {Chang}, {Cho}, {Crawford}, {Crites}, {de Haan}, {Dobbs},
  {Everett}, {George}, {Henning}, {Halverson}, {Harrington}, {Holder},
  {Holzapfel}, {Hou}, {Hrubes}, {Knox}, {Lee}, {Luong-Van}, {Marrone},
  {McMahon}, {Meyer}, {Millea}, {Mocanu}, {Mohr}, {Natoli}, {Padin}, {Pryke},
  {Reichardt}, {Ruhl}, {Sayre}, {Schaffer}, {Shirokoff}, {Simard},
  {Staniszewski}, {Stark}, {Story}, {Vanderlinde}, {Vieira}, {Williamson},
  {Wu}, \& {South Pole Telescope Collaboration}}]{2018ApJS..239...10C}
{Chown}, R., {Omori}, Y., {Aylor}, K., {et~al.} 2018, \apjs, 239, 10

\bibitem[{Curry \& Schoenberg(1966)}]{Curry1966OnPF}
Curry, H.~B. \& Schoenberg, I.~J. 1966, Journal D Analyse Mathematique, 17, 347

\bibitem[{{De Luca, A.} \& {Molendi, S.}(2004)}]{DeLuca2004}
{De Luca, A.} \& {Molendi, S.} 2004, A\&A, 419, 837

\bibitem[{{Delabrouille} {et~al.}(2009){Delabrouille}, {Cardoso}, {Le Jeune},
  {Betoule}, {Fay}, \& {Guilloux}}]{2009A&A...493..835D}
{Delabrouille}, J., {Cardoso}, J.~F., {Le Jeune}, M., {et~al.} 2009, \aap, 493,
  835

\bibitem[{{Eriksen} {et~al.}(2004){Eriksen}, {Banday}, {G{\'o}rski}, \&
  {Lilje}}]{2004ApJ...612..633E}
{Eriksen}, H.~K., {Banday}, A.~J., {G{\'o}rski}, K.~M., \& {Lilje}, P.~B. 2004,
  \apj, 612, 633

\bibitem[{{Finkbeiner} {et~al.}(1999){Finkbeiner}, {Davis}, \&
  {Schlegel}}]{1999ApJ...524..867F}
{Finkbeiner}, D.~P., {Davis}, M., \& {Schlegel}, D.~J. 1999, \apj, 524, 867

\bibitem[{{HI4PI Collaboration} {et~al.}(2016){HI4PI Collaboration}, {Ben
  Bekhti, N.}, {Fl\"oer, L.}, {Keller, R.}, {Kerp, J.}, {Lenz, D.}, {Winkel,
  B.}, {Bailin, J.}, {Calabretta, M. R.}, {Dedes, L.}, {Ford, H. A.}, {Gibson,
  B. K.}, {Haud, U.}, {Janowiecki, S.}, {Kalberla, P. M. W.}, {Lockman, F. J.},
  {McClure-Griffiths, N. M.}, {Murphy, T.}, {Nakanishi, H.}, {Pisano, D. J.},
  \& {Staveley-Smith, L.}}]{HI4PI2016}
{HI4PI Collaboration}, {Ben Bekhti, N.}, {Fl\"oer, L.}, {et~al.} 2016, A\&A,
  594, A116

\bibitem[{{Hurier} {et~al.}(2013){Hurier}, {Mac{\'\i}as-P{\'e}rez}, \&
  {Hildebrandt}}]{2013A&A...558A.118H}
{Hurier}, G., {Mac{\'\i}as-P{\'e}rez}, J.~F., \& {Hildebrandt}, S. 2013, \aap,
  558, A118

\bibitem[{{Kaiser} {et~al.}(1995){Kaiser}, {Squires}, \&
  {Broadhurst}}]{1995ApJ...449..460K}
{Kaiser}, N., {Squires}, G., \& {Broadhurst}, T. 1995, \apj, 449, 460

\bibitem[{Kitayama {et~al.}(2020)Kitayama, Ueda, Akahori, Komatsu, Kawabe,
  Kohno, Takakuwa, Takizawa, Tsutsumi, \& Yoshikawa}]{Kitayama2020}
Kitayama, T., Ueda, S., Akahori, T., {et~al.} 2020, Publications of the
  Astronomical Society of Japan, 72
  [\eprint{https://academic.oup.com/pasj/article-pdf/72/2/33/33122640/psaa009.pdf}],
  33

\bibitem[{{Kravtsov} \& {Borgani}(2012)}]{2012ARA&A..50..353K}
{Kravtsov}, A.~V. \& {Borgani}, S. 2012, \araa, 50, 353

\bibitem[{{Kravtsov} {et~al.}(2006){Kravtsov}, {Vikhlinin}, \&
  {Nagai}}]{2006ApJ...650..128K}
{Kravtsov}, A.~V., {Vikhlinin}, A., \& {Nagai}, D. 2006, \apj, 650, 128

\bibitem[{{Kuntz, K. D.} \& {Snowden, S. L.}(2008)}]{Kuntz2008}
{Kuntz, K. D.} \& {Snowden, S. L.} 2008, A\&A, 478, 575

\bibitem[{{Lau} {et~al.}(2009){Lau}, {Kravtsov}, \&
  {Nagai}}]{2009ApJ...705.1129L}
{Lau}, E.~T., {Kravtsov}, A.~V., \& {Nagai}, D. 2009, \apj, 705, 1129

\bibitem[{{Lau} {et~al.}(2015){Lau}, {Nagai}, {Avestruz}, {Nelson}, \&
  {Vikhlinin}}]{2015ApJ...806...68L}
{Lau}, E.~T., {Nagai}, D., {Avestruz}, C., {Nelson}, K., \& {Vikhlinin}, A.
  2015, \apj, 806, 68

\bibitem[{{Leccardi, A.} \& {Molendi, S.}(2008)}]{leccardi2008}
{Leccardi, A.} \& {Molendi, S.} 2008, A\&A, 486, 359

\bibitem[{Lovisari \& Reiprich(2019)}]{Lovisari2019}
Lovisari, L. \& Reiprich, T.~H. 2019, Monthly Notices of the Royal Astronomical
  Society, 483, 540

\bibitem[{Mazzotta {et~al.}(2004)Mazzotta, Rasia, Moscardini, \&
  Tormen}]{Mazzotta2004}
Mazzotta, P., Rasia, E., Moscardini, L., \& Tormen, G. 2004, Monthly Notices of
  the Royal Astronomical Society, 354, 10

\bibitem[{{McCarthy} {et~al.}(2014){McCarthy}, {Le Brun}, {Schaye}, \&
  {Holder}}]{2014MNRAS.440.3645M}
{McCarthy}, I.~G., {Le Brun}, A.~M.~C., {Schaye}, J., \& {Holder}, G.~P. 2014,
  \mnras, 440, 3645

\bibitem[{{McDonald} {et~al.}(2012){McDonald}, {Bayliss}, {Benson}, {Foley},
  {Ruel}, {Sullivan}, {Veilleux}, {Aird}, {Ashby}, {Bautz}, {Bazin}, {Bleem},
  {Brodwin}, {Carlstrom}, {Chang}, {Cho}, {Clocchiatti}, {Crawford}, {Crites},
  {de Haan}, {Desai}, {Dobbs}, {Dudley}, {Egami}, {Forman}, {Garmire},
  {George}, {Gladders}, {Gonzalez}, {Halverson}, {Harrington}, {High},
  {Holder}, {Holzapfel}, {Hoover}, {Hrubes}, {Jones}, {Joy}, {Keisler}, {Knox},
  {Lee}, {Leitch}, {Liu}, {Lueker}, {Luong-van}, {Mantz}, {Marrone}, {McMahon},
  {Mehl}, {Meyer}, {Miller}, {Mocanu}, {Mohr}, {Montroy}, {Murray}, {Natoli},
  {Padin}, {Plagge}, {Pryke}, {Rawle}, {Reichardt}, {Rest}, {Rex}, {Ruhl},
  {Saliwanchik}, {Saro}, {Sayre}, {Schaffer}, {Shaw}, {Shirokoff}, {Simcoe},
  {Song}, {Spieler}, {Stalder}, {Staniszewski}, {Stark}, {Story}, {Stubbs},
  {{\v{S}}uhada}, {van Engelen}, {Vanderlinde}, {Vieira}, {Vikhlinin},
  {Williamson}, {Zahn}, \& {Zenteno}}]{McDonald2012}
{McDonald}, M., {Bayliss}, M., {Benson}, B.~A., {et~al.} 2012, \nat, 488, 349

\bibitem[{McDonald {et~al.}(2013)McDonald, Benson, Veilleux, Bautz, \&
  Reichardt}]{McDonald2013}
McDonald, M., Benson, B., Veilleux, S., Bautz, M.~W., \& Reichardt, C.~L. 2013,
  The Astrophysical Journal, 765, L37

\bibitem[{{McDonald} {et~al.}(2014){McDonald}, {Benson}, {Vikhlinin}, {Aird},
  {Allen}, {Bautz}, {Bayliss}, {Bleem}, {Bocquet}, {Brodwin}, {Carlstrom},
  {Chang}, {Cho}, {Clocchiatti}, {Crawford}, {Crites}, {de Haan}, {Dobbs},
  {Foley}, {Forman}, {George}, {Gladders}, {Gonzalez}, {Halverson},
  {Hlavacek-Larrondo}, {Holder}, {Holzapfel}, {Hrubes}, {Jones}, {Keisler},
  {Knox}, {Lee}, {Leitch}, {Liu}, {Lueker}, {Luong-Van}, {Mantz}, {Marrone},
  {McMahon}, {Meyer}, {Miller}, {Mocanu}, {Mohr}, {Murray}, {Padin}, {Pryke},
  {Reichardt}, {Rest}, {Ruhl}, {Saliwanchik}, {Saro}, {Sayre}, {Schaffer},
  {Shirokoff}, {Spieler}, {Stalder}, {Stanford}, {Staniszewski}, {Stark},
  {Story}, {Stubbs}, {Vanderlinde}, {Vieira}, {Williamson}, {Zahn}, \&
  {Zenteno}}]{2014ApJ...794...67M}
{McDonald}, M., {Benson}, B.~A., {Vikhlinin}, A., {et~al.} 2014, \apj, 794, 67

\bibitem[{McDonald {et~al.}(2019)McDonald, McNamara, Voit, Bayliss, Benson,
  Brodwin, Canning, Florian, Garmire, Gaspari, Gladders, Hlavacek-Larrondo,
  Kara, Reichardt, Russell, Saro, Sharon, Somboonpanyakul, Tremblay, \& van
  Weeren}]{McDonald2019}
McDonald, M., McNamara, B.~R., Voit, G.~M., {et~al.} 2019, The Astrophysical
  Journal, 885, 63

\bibitem[{{Meisner} \& {Finkbeiner}(2015)}]{2015ApJ...798...88M}
{Meisner}, A.~M. \& {Finkbeiner}, D.~P. 2015, \apj, 798, 88

\bibitem[{{Melin} {et~al.}(2021){Melin}, {Bartlett}, {Tarr{\'\i}o}, \&
  {Pratt}}]{2021A&A...647A.106M}
{Melin}, J.~B., {Bartlett}, J.~G., {Tarr{\'\i}o}, P., \& {Pratt}, G.~W. 2021,
  \aap, 647, A106

\bibitem[{{Nagai} {et~al.}(2007){Nagai}, {Kravtsov}, \&
  {Vikhlinin}}]{2007ApJ...668....1N}
{Nagai}, D., {Kravtsov}, A.~V., \& {Vikhlinin}, A. 2007, \apj, 668, 1

\bibitem[{{Navarro} {et~al.}(1997){Navarro}, {Frenk}, \&
  {White}}]{1997ApJ...490..493N}
{Navarro}, J.~F., {Frenk}, C.~S., \& {White}, S. D.~M. 1997, \apj, 490, 493

\bibitem[{{Nelson} {et~al.}(2014){Nelson}, {Lau}, \&
  {Nagai}}]{2014ApJ...792...25N}
{Nelson}, K., {Lau}, E.~T., \& {Nagai}, D. 2014, \apj, 792, 25

\bibitem[{{Oppizzi} {et~al.}(2020){Oppizzi}, {Renzi}, {Liguori}, {Hansen},
  {Marinucci}, {Baccigalupi}, {Bertacca}, \& {Poletti}}]{2020JCAP...03..054O}
{Oppizzi}, F., {Renzi}, A., {Liguori}, M., {et~al.} 2020, \jcap, 2020, 054

\bibitem[{{Planck Collaboration} {et~al.}(2014{\natexlab{a}}){Planck
  Collaboration}, {Abergel}, {Ade}, {Aghanim}, {Alves}, {Aniano},
  {Armitage-Caplan}, {Arnaud}, {Ashdown}, {Atrio-Barandela}, {Aumont},
  {Baccigalupi}, {Banday}, {Barreiro}, {Bartlett}, {Battaner}, {Benabed},
  {Beno{\^\i}t}, {Benoit-L{\'e}vy}, {Bernard}, {Bersanelli}, {Bielewicz},
  {Bobin}, {Bock}, {Bonaldi}, {Bond}, {Borrill}, {Bouchet}, {Boulanger},
  {Bridges}, {Bucher}, {Burigana}, {Butler}, {Cardoso}, {Catalano},
  {Chamballu}, {Chary}, {Chiang}, {Chiang}, {Christensen}, {Church}, {Clemens},
  {Clements}, {Colombi}, {Colombo}, {Combet}, {Couchot}, {Coulais}, {Crill},
  {Curto}, {Cuttaia}, {Danese}, {Davies}, {Davis}, {de Bernardis}, {de Rosa},
  {de Zotti}, {Delabrouille}, {Delouis}, {D{\'e}sert}, {Dickinson}, {Diego},
  {Dole}, {Donzelli}, {Dor{\'e}}, {Douspis}, {Draine}, {Dupac}, {Efstathiou},
  {En{\ss}lin}, {Eriksen}, {Falgarone}, {Finelli}, {Forni}, {Frailis},
  {Fraisse}, {Franceschi}, {Galeotta}, {Ganga}, {Ghosh}, {Giard}, {Giardino},
  {Giraud-H{\'e}raud}, {Gonz{\'a}lez-Nuevo}, {G{\'o}rski}, {Gratton},
  {Gregorio}, {Grenier}, {Gruppuso}, {Guillet}, {Hansen}, {Hanson}, {Harrison},
  {Helou}, {Henrot-Versill{\'e}}, {Hern{\'a}ndez-Monteagudo}, {Herranz},
  {Hildebrandt}, {Hivon}, {Hobson}, {Holmes}, {Hornstrup}, {Hovest},
  {Huffenberger}, {Jaffe}, {Jaffe}, {Jewell}, {Joncas}, {Jones}, {Juvela},
  {Keih{\"a}nen}, {Keskitalo}, {Kisner}, {Knoche}, {Knox}, {Kunz},
  {Kurki-Suonio}, {Lagache}, {L{\"a}hteenm{\"a}ki}, {Lamarre}, {Lasenby},
  {Laureijs}, {Lawrence}, {Leonardi}, {Le{\'o}n-Tavares}, {Lesgourgues},
  {Levrier}, {Liguori}, {Lilje}, {Linden-V{\o}rnle}, {L{\'o}pez-Caniego},
  {Lubin}, {Mac{\'\i}as-P{\'e}rez}, {Maffei}, {Maino}, {Mandolesi}, {Maris},
  {Marshall}, {Martin}, {Mart{\'\i}nez-Gonz{\'a}lez}, {Masi}, {Massardi},
  {Matarrese}, {Matthai}, {Mazzotta}, {McGehee}, {Melchiorri}, {Mendes},
  {Mennella}, {Migliaccio}, {Mitra}, {Miville-Desch{\^e}nes}, {Moneti},
  {Montier}, {Morgante}, {Mortlock}, {Munshi}, {Murphy}, {Naselsky}, {Nati},
  {Natoli}, {Netterfield}, {N{\o}rgaard-Nielsen}, {Noviello}, {Novikov},
  {Novikov}, {Osborne}, {Oxborrow}, {Paci}, {Pagano}, {Pajot}, {Paladini},
  {Paoletti}, {Pasian}, {Patanchon}, {Perdereau}, {Perotto}, {Perrotta},
  {Piacentini}, {Piat}, {Pierpaoli}, {Pietrobon}, {Plaszczynski},
  {Pointecouteau}, {Polenta}, {Ponthieu}, {Popa}, {Poutanen}, {Pratt},
  {Pr{\'e}zeau}, {Prunet}, {Puget}, {Rachen}, {Reach}, {Rebolo}, {Reinecke},
  {Remazeilles}, {Renault}, {Ricciardi}, {Riller}, {Ristorcelli}, {Rocha},
  {Rosset}, {Roudier}, {Rowan-Robinson}, {Rubi{\~n}o-Mart{\'\i}n}, {Rusholme},
  {Sandri}, {Santos}, {Savini}, {Scott}, {Seiffert}, {Shellard}, {Spencer},
  {Starck}, {Stolyarov}, {Stompor}, {Sudiwala}, {Sunyaev}, {Sureau}, {Sutton},
  {Suur-Uski}, {Sygnet}, {Tauber}, {Tavagnacco}, {Terenzi}, {Toffolatti},
  {Tomasi}, {Tristram}, {Tucci}, {Tuovinen}, {T{\"u}rler}, {Umana},
  {Valenziano}, {Valiviita}, {Van Tent}, {Verstraete}, {Vielva}, {Villa},
  {Vittorio}, {Wade}, {Wandelt}, {Welikala}, {Ysard}, {Yvon}, {Zacchei}, \&
  {Zonca}}]{2014A&A...571A..11P}
{Planck Collaboration}, {Abergel}, A., {Ade}, P.~A.~R., {et~al.}
  2014{\natexlab{a}}, \aap, 571, A11

\bibitem[{{Planck Collaboration} {et~al.}(2016{\natexlab{a}}){Planck
  Collaboration}, {Adam}, {Ade}, {Aghanim}, {Arnaud}, {Ashdown}, {Aumont},
  {Baccigalupi}, {Banday}, {Barreiro}, {Bartolo}, {Battaner}, {Benabed},
  {Beno{\^\i}t}, {Benoit-L{\'e}vy}, {Bernard}, {Bersanelli}, {Bertincourt},
  {Bielewicz}, {Bock}, {Bonavera}, {Bond}, {Borrill}, {Bouchet}, {Boulanger},
  {Bucher}, {Burigana}, {Calabrese}, {Cardoso}, {Catalano}, {Challinor},
  {Chamballu}, {Chary}, {Chiang}, {Christensen}, {Clements}, {Colombi},
  {Colombo}, {Combet}, {Couchot}, {Coulais}, {Crill}, {Curto}, {Cuttaia},
  {Danese}, {Davies}, {Davis}, {de Bernardis}, {de Rosa}, {de Zotti},
  {Delabrouille}, {Delouis}, {D{\'e}sert}, {Diego}, {Dole}, {Donzelli},
  {Dor{\'e}}, {Douspis}, {Ducout}, {Dupac}, {Efstathiou}, {Elsner},
  {En{\ss}lin}, {Eriksen}, {Falgarone}, {Fergusson}, {Finelli}, {Forni},
  {Frailis}, {Fraisse}, {Franceschi}, {Frejsel}, {Galeotta}, {Galli}, {Ganga},
  {Ghosh}, {Giard}, {Giraud-H{\'e}raud}, {Gjerl{\o}w}, {Gonz{\'a}lez-Nuevo},
  {G{\'o}rski}, {Gratton}, {Gruppuso}, {Gudmundsson}, {Hansen}, {Hanson},
  {Harrison}, {Henrot-Versill{\'e}}, {Herranz}, {Hildebrandt}, {Hivon},
  {Hobson}, {Holmes}, {Hornstrup}, {Hovest}, {Huffenberger}, {Hurier}, {Jaffe},
  {Jaffe}, {Jones}, {Juvela}, {Keih{\"a}nen}, {Keskitalo}, {Kisner}, {Kneissl},
  {Knoche}, {Kunz}, {Kurki-Suonio}, {Lagache}, {Lamarre}, {Lasenby},
  {Lattanzi}, {Lawrence}, {Le Jeune}, {Leahy}, {Lellouch}, {Leonardi},
  {Lesgourgues}, {Levrier}, {Liguori}, {Lilje}, {Linden-V{\o}rnle},
  {L{\'o}pez-Caniego}, {Lubin}, {Mac{\'\i}as-P{\'e}rez}, {Maggio}, {Maino},
  {Mandolesi}, {Mangilli}, {Maris}, {Martin}, {Mart{\'\i}nez-Gonz{\'a}lez},
  {Masi}, {Matarrese}, {McGehee}, {Melchiorri}, {Mendes}, {Mennella},
  {Migliaccio}, {Mitra}, {Miville-Desch{\^e}nes}, {Moneti}, {Montier},
  {Moreno}, {Morgante}, {Mortlock}, {Moss}, {Mottet}, {Munshi}, {Murphy},
  {Naselsky}, {Nati}, {Natoli}, {Netterfield}, {N{\o}rgaard-Nielsen},
  {Noviello}, {Novikov}, {Novikov}, {Oxborrow}, {Paci}, {Pagano}, {Pajot},
  {Paoletti}, {Pasian}, {Patanchon}, {Pearson}, {Perdereau}, {Perotto},
  {Perrotta}, {Pettorino}, {Piacentini}, {Piat}, {Pierpaoli}, {Pietrobon},
  {Plaszczynski}, {Pointecouteau}, {Polenta}, {Pratt}, {Pr{\'e}zeau}, {Prunet},
  {Puget}, {Rachen}, {Reinecke}, {Remazeilles}, {Renault}, {Renzi},
  {Ristorcelli}, {Rocha}, {Rosset}, {Rossetti}, {Roudier}, {Rowan-Robinson},
  {Rusholme}, {Sandri}, {Santos}, {Sauv{\'e}}, {Savelainen}, {Savini}, {Scott},
  {Seiffert}, {Shellard}, {Spencer}, {Stolyarov}, {Stompor}, {Sudiwala},
  {Sutton}, {Suur-Uski}, {Sygnet}, {Tauber}, {Terenzi}, {Toffolatti}, {Tomasi},
  {Tristram}, {Tucci}, {Tuovinen}, {Valenziano}, {Valiviita}, {Van Tent},
  {Vibert}, {Vielva}, {Villa}, {Wade}, {Wandelt}, {Watson}, {Wehus}, {Yvon},
  {Zacchei}, \& {Zonca}}]{2016A&A...594A...7P}
{Planck Collaboration}, {Adam}, R., {Ade}, P.~A.~R., {et~al.}
  2016{\natexlab{a}}, \aap, 594, A7

\bibitem[{{Planck Collaboration} {et~al.}(2016{\natexlab{b}}){Planck
  Collaboration}, {Adam}, {Ade}, {Aghanim}, {Arnaud}, {Ashdown}, {Aumont},
  {Baccigalupi}, {Banday}, {Barreiro}, {Bartolo}, {Battaner}, {Benabed},
  {Beno{\^\i}t}, {Benoit-L{\'e}vy}, {Bernard}, {Bersanelli}, {Bertincourt},
  {Bielewicz}, {Bock}, {Bonavera}, {Bond}, {Borrill}, {Bouchet}, {Boulanger},
  {Bucher}, {Burigana}, {Calabrese}, {Cardoso}, {Catalano}, {Challinor},
  {Chamballu}, {Chiang}, {Christensen}, {Clements}, {Colombi}, {Colombo},
  {Combet}, {Couchot}, {Coulais}, {Crill}, {Curto}, {Cuttaia}, {Danese},
  {Davies}, {Davis}, {de Bernardis}, {de Rosa}, {de Zotti}, {Delabrouille},
  {Delouis}, {D{\'e}sert}, {Diego}, {Dole}, {Donzelli}, {Dor{\'e}}, {Douspis},
  {Ducout}, {Dupac}, {Efstathiou}, {Elsner}, {En{\ss}lin}, {Eriksen},
  {Falgarone}, {Fergusson}, {Finelli}, {Forni}, {Frailis}, {Fraisse},
  {Franceschi}, {Frejsel}, {Galeotta}, {Galli}, {Ganga}, {Ghosh}, {Giard},
  {Giraud-H{\'e}raud}, {Gjerl{\o}w}, {Gonz{\'a}lez-Nuevo}, {G{\'o}rski},
  {Gratton}, {Gruppuso}, {Gudmundsson}, {Hansen}, {Hanson}, {Harrison},
  {Henrot-Versill{\'e}}, {Herranz}, {Hildebrandt}, {Hivon}, {Hobson}, {Holmes},
  {Hornstrup}, {Hovest}, {Huffenberger}, {Hurier}, {Jaffe}, {Jaffe}, {Jones},
  {Juvela}, {Keih{\"a}nen}, {Keskitalo}, {Kisner}, {Kneissl}, {Knoche}, {Kunz},
  {Kurki-Suonio}, {Lagache}, {Lamarre}, {Lasenby}, {Lattanzi}, {Lawrence}, {Le
  Jeune}, {Leahy}, {Lellouch}, {Leonardi}, {Lesgourgues}, {Levrier}, {Liguori},
  {Lilje}, {Linden-V{\o}rnle}, {L{\'o}pez-Caniego}, {Lubin},
  {Mac{\'\i}as-P{\'e}rez}, {Maggio}, {Maino}, {Mandolesi}, {Mangilli}, {Maris},
  {Martin}, {Mart{\'\i}nez-Gonz{\'a}lez}, {Masi}, {Matarrese}, {McGehee},
  {Melchiorri}, {Mendes}, {Mennella}, {Migliaccio}, {Mitra},
  {Miville-Desch{\^e}nes}, {Moneti}, {Montier}, {Moreno}, {Morgante},
  {Mortlock}, {Moss}, {Mottet}, {Munshi}, {Murphy}, {Naselsky}, {Nati},
  {Natoli}, {Netterfield}, {N{\o}rgaard-Nielsen}, {Noviello}, {Novikov},
  {Novikov}, {Oxborrow}, {Paci}, {Pagano}, {Pajot}, {Paoletti}, {Pasian},
  {Patanchon}, {Pearson}, {Perdereau}, {Perotto}, {Perrotta}, {Pettorino},
  {Piacentini}, {Piat}, {Pierpaoli}, {Pietrobon}, {Plaszczynski},
  {Pointecouteau}, {Polenta}, {Pratt}, {Pr{\'e}zeau}, {Prunet}, {Puget},
  {Rachen}, {Reinecke}, {Remazeilles}, {Renault}, {Renzi}, {Ristorcelli},
  {Rocha}, {Rosset}, {Rossetti}, {Roudier}, {Rusholme}, {Sandri}, {Santos},
  {Sauv{\'e}}, {Savelainen}, {Savini}, {Scott}, {Seiffert}, {Shellard},
  {Spencer}, {Stolyarov}, {Stompor}, {Sudiwala}, {Sutton}, {Suur-Uski},
  {Sygnet}, {Tauber}, {Terenzi}, {Toffolatti}, {Tomasi}, {Tristram}, {Tucci},
  {Tuovinen}, {Valenziano}, {Valiviita}, {Van Tent}, {Vibert}, {Vielva},
  {Villa}, {Wade}, {Wandelt}, {Watson}, {Wehus}, {Yvon}, {Zacchei}, \&
  {Zonca}}]{2016A&A...594A...8P}
{Planck Collaboration}, {Adam}, R., {Ade}, P.~A.~R., {et~al.}
  2016{\natexlab{b}}, \aap, 594, A8

\bibitem[{{Planck Collaboration} {et~al.}(2016{\natexlab{c}}){Planck
  Collaboration}, {Adam}, {Ade}, {Aghanim}, {Ashdown}, {Aumont}, {Baccigalupi},
  {Banday}, {Barreiro}, {Bartolo}, {Battaner}, {Benabed}, {Benoit-L{\'e}vy},
  {Bersanelli}, {Bielewicz}, {Bikmaev}, {Bonaldi}, {Bond}, {Borrill},
  {Bouchet}, {Burenin}, {Burigana}, {Calabrese}, {Cardoso}, {Catalano},
  {Chiang}, {Christensen}, {Churazov}, {Colombo}, {Combet}, {Comis}, {Couchot},
  {Crill}, {Curto}, {Cuttaia}, {Danese}, {Davis}, {de Bernardis}, {de Rosa},
  {de Zotti}, {Delabrouille}, {D{\'e}sert}, {Diego}, {Dole}, {Dor{\'e}},
  {Douspis}, {Ducout}, {Dupac}, {Elsner}, {En{\ss}lin}, {Finelli}, {Forni},
  {Frailis}, {Fraisse}, {Franceschi}, {Galeotta}, {Ganga}, {G{\'e}nova-Santos},
  {Giard}, {Giraud-H{\'e}raud}, {Gjerl{\o}w}, {Gonz{\'a}lez-Nuevo},
  {G{\'o}rski}, {Gregorio}, {Gruppuso}, {Gudmundsson}, {Hansen}, {Harrison},
  {Hern{\'a}ndez-Monteagudo}, {Herranz}, {Hildebrandt}, {Hivon}, {Hobson},
  {Hornstrup}, {Hovest}, {Hurier}, {Jaffe}, {Jaffe}, {Jones}, {Keih{\"a}nen},
  {Keskitalo}, {Khamitov}, {Kisner}, {Kneissl}, {Knoche}, {Kunz},
  {Kurki-Suonio}, {Lagache}, {L{\"a}hteenm{\"a}ki}, {Lamarre}, {Lasenby},
  {Lattanzi}, {Lawrence}, {Leonardi}, {Levrier}, {Liguori}, {Lilje},
  {Linden-V{\o}rnle}, {L{\'o}pez-Caniego}, {Mac{\'\i}as-P{\'e}rez}, {Maffei},
  {Maggio}, {Mandolesi}, {Mangilli}, {Maris}, {Martin},
  {Mart{\'\i}nez-Gonz{\'a}lez}, {Masi}, {Matarrese}, {Melchiorri}, {Mennella},
  {Migliaccio}, {Miville-Desch{\^e}nes}, {Moneti}, {Montier}, {Morgante},
  {Mortlock}, {Munshi}, {Murphy}, {Naselsky}, {Nati}, {Natoli},
  {N{\o}rgaard-Nielsen}, {Novikov}, {Novikov}, {Oxborrow}, {Pagano}, {Pajot},
  {Paoletti}, {Pasian}, {Perdereau}, {Perotto}, {Pettorino}, {Piacentini},
  {Piat}, {Plaszczynski}, {Pointecouteau}, {Polenta}, {Ponthieu}, {Pratt},
  {Prunet}, {Puget}, {Rachen}, {Rebolo}, {Reinecke}, {Remazeilles}, {Renault},
  {Renzi}, {Ristorcelli}, {Rocha}, {Rosset}, {Rossetti}, {Roudier},
  {Rubi{\~n}o-Mart{\'\i}n}, {Rusholme}, {Santos}, {Savelainen}, {Savini},
  {Scott}, {Stolyarov}, {Stompor}, {Sudiwala}, {Sunyaev}, {Sutton},
  {Suur-Uski}, {Sygnet}, {Tauber}, {Terenzi}, {Toffolatti}, {Tomasi},
  {Tristram}, {Tucci}, {Valenziano}, {Valiviita}, {Van Tent}, {Vielva},
  {Villa}, {Wade}, {Wehus}, {Yvon}, {Zacchei}, \&
  {Zonca}}]{2016A&A...596A.104P}
{Planck Collaboration}, {Adam}, R., {Ade}, P.~A.~R., {et~al.}
  2016{\natexlab{c}}, \aap, 596, A104

\bibitem[{{Planck Collaboration} {et~al.}(2016{\natexlab{d}}){Planck
  Collaboration}, {Ade}, {Aghanim}, {Alves}, {Aniano}, {Arnaud}, {Ashdown},
  {Aumont}, {Baccigalupi}, {Banday}, {Barreiro}, {Bartolo}, {Battaner},
  {Benabed}, {Benoit-L{\'e}vy}, {Bernard}, {Bersanelli}, {Bielewicz},
  {Bonaldi}, {Bonavera}, {Bond}, {Borrill}, {Bouchet}, {Boulanger}, {Burigana},
  {Butler}, {Calabrese}, {Cardoso}, {Catalano}, {Chamballu}, {Chiang},
  {Christensen}, {Clements}, {Colombi}, {Colombo}, {Couchot}, {Crill}, {Curto},
  {Cuttaia}, {Danese}, {Davies}, {Davis}, {de Bernardis}, {de Rosa}, {de
  Zotti}, {Delabrouille}, {Dickinson}, {Diego}, {Dole}, {Donzelli}, {Dor{\'e}},
  {Douspis}, {Draine}, {Ducout}, {Dupac}, {Efstathiou}, {Elsner}, {En{\ss}lin},
  {Eriksen}, {Falgarone}, {Finelli}, {Forni}, {Frailis}, {Fraisse},
  {Franceschi}, {Frejsel}, {Galeotta}, {Galli}, {Ganga}, {Ghosh}, {Giard},
  {Gjerl{\o}w}, {Gonz{\'a}lez-Nuevo}, {G{\'o}rski}, {Gregorio}, {Gruppuso},
  {Guillet}, {Hansen}, {Hanson}, {Harrison}, {Henrot-Versill{\'e}},
  {Hern{\'a}ndez-Monteagudo}, {Herranz}, {Hildebrandt}, {Hivon}, {Holmes},
  {Hovest}, {Huffenberger}, {Hurier}, {Jaffe}, {Jaffe}, {Jones},
  {Keih{\"a}nen}, {Keskitalo}, {Kisner}, {Kneissl}, {Knoche}, {Kunz},
  {Kurki-Suonio}, {Lagache}, {Lamarre}, {Lasenby}, {Lattanzi}, {Lawrence},
  {Leonardi}, {Levrier}, {Liguori}, {Lilje}, {Linden-V{\o}rnle},
  {L{\'o}pez-Caniego}, {Lubin}, {Mac{\'\i}as-P{\'e}rez}, {Maffei}, {Maino},
  {Mandolesi}, {Maris}, {Marshall}, {Martin}, {Mart{\'\i}nez-Gonz{\'a}lez},
  {Masi}, {Matarrese}, {Mazzotta}, {Melchiorri}, {Mendes}, {Mennella},
  {Migliaccio}, {Miville-Desch{\^e}nes}, {Moneti}, {Montier}, {Morgante},
  {Mortlock}, {Munshi}, {Murphy}, {Naselsky}, {Natoli}, {N{\o}rgaard-Nielsen},
  {Novikov}, {Novikov}, {Oxborrow}, {Pagano}, {Pajot}, {Paladini}, {Paoletti},
  {Pasian}, {Perdereau}, {Perotto}, {Perrotta}, {Pettorino}, {Piacentini},
  {Piat}, {Plaszczynski}, {Pointecouteau}, {Polenta}, {Ponthieu}, {Popa},
  {Pratt}, {Prunet}, {Puget}, {Rachen}, {Reach}, {Rebolo}, {Reinecke},
  {Remazeilles}, {Renault}, {Ristorcelli}, {Rocha}, {Roudier},
  {Rubi{\~n}o-Mart{\'\i}n}, {Rusholme}, {Sandri}, {Santos}, {Scott}, {Spencer},
  {Stolyarov}, {Sudiwala}, {Sunyaev}, {Sutton}, {Suur-Uski}, {Sygnet},
  {Tauber}, {Terenzi}, {Toffolatti}, {Tomasi}, {Tristram}, {Tucci}, {Umana},
  {Valenziano}, {Valiviita}, {Van Tent}, {Vielva}, {Villa}, {Wade}, {Wandelt},
  {Wehus}, {Ysard}, {Yvon}, {Zacchei}, \& {Zonca}}]{2016A&A...586A.132P}
{Planck Collaboration}, {Ade}, P.~A.~R., {Aghanim}, N., {et~al.}
  2016{\natexlab{d}}, \aap, 586, A132

\bibitem[{{Planck Collaboration} {et~al.}(2014{\natexlab{b}}){Planck
  Collaboration}, {Ade}, {Aghanim}, {Alves}, {Armitage-Caplan}, {Arnaud},
  {Ashdown}, {Atrio-Barandela}, {Aumont}, {Aussel}, {Baccigalupi}, {Banday},
  {Barreiro}, {Barrena}, {Bartelmann}, {Bartlett}, {Bartolo}, {Basak},
  {Battaner}, {Battye}, {Benabed}, {Beno{\^\i}t}, {Benoit-L{\'e}vy}, {Bernard},
  {Bersanelli}, {Bertincourt}, {Bethermin}, {Bielewicz}, {Bikmaev},
  {Blanchard}, {Bobin}, {Bock}, {B{\"o}hringer}, {Bonaldi}, {Bonavera}, {Bond},
  {Borrill}, {Bouchet}, {Boulanger}, {Bourdin}, {Bowyer}, {Bridges}, {Brown},
  {Bucher}, {Burenin}, {Burigana}, {Butler}, {Calabrese}, {Cappellini},
  {Cardoso}, {Carr}, {Carvalho}, {Casale}, {Castex}, {Catalano}, {Challinor},
  {Chamballu}, {Chary}, {Chen}, {Chiang}, {Chiang}, {Chon}, {Christensen},
  {Churazov}, {Church}, {Clemens}, {Clements}, {Colombi}, {Colombo}, {Combet},
  {Comis}, {Couchot}, {Coulais}, {Crill}, {Cruz}, {Curto}, {Cuttaia}, {Da
  Silva}, {Dahle}, {Danese}, {Davies}, {Davis}, {de Bernardis}, {de Rosa}, {de
  Zotti}, {D{\'e}chelette}, {Delabrouille}, {Delouis}, {D{\'e}mocl{\`e}s},
  {D{\'e}sert}, {Dick}, {Dickinson}, {Diego}, {Dolag}, {Dole}, {Donzelli},
  {Dor{\'e}}, {Douspis}, {Ducout}, {Dunkley}, {Dupac}, {Efstathiou}, {Elsner},
  {En{\ss}lin}, {Eriksen}, {Fabre}, {Falgarone}, {Falvella}, {Fantaye},
  {Fergusson}, {Filliard}, {Finelli}, {Flores-Cacho}, {Foley}, {Forni},
  {Fosalba}, {Frailis}, {Fraisse}, {Franceschi}, {Freschi}, {Fromenteau},
  {Frommert}, {Gaier}, {Galeotta}, {Gallegos}, {Galli}, {Gandolfo}, {Ganga},
  {Gauthier}, {G{\'e}nova-Santos}, {Ghosh}, {Giard}, {Giardino}, {Gilfanov},
  {Girard}, {Giraud-H{\'e}raud}, {Gjerl{\o}w}, {Gonz{\'a}lez-Nuevo},
  {G{\'o}rski}, {Gratton}, {Gregorio}, {Gruppuso}, {Gudmundsson}, {Haissinski},
  {Hamann}, {Hansen}, {Hansen}, {Hanson}, {Harrison}, {Heavens}, {Helou},
  {Hempel}, {Henrot-Versill{\'e}}, {Hern{\'a}ndez-Monteagudo}, {Herranz},
  {Hildebrandt}, {Hivon}, {Ho}, {Hobson}, {Holmes}, {Hornstrup}, {Hou},
  {Hovest}, {Huey}, {Huffenberger}, {Hurier}, {Ili{\'c}}, {Jaffe}, {Jaffe},
  {Jasche}, {Jewell}, {Jones}, {Juvela}, {Kalberla}, {Kangaslahti},
  {Keih{\"a}nen}, {Kerp}, {Keskitalo}, {Khamitov}, {Kiiveri}, {Kim}, {Kisner},
  {Kneissl}, {Knoche}, {Knox}, {Kunz}, {Kurki-Suonio}, {Lacasa}, {Lagache},
  {L{\"a}hteenm{\"a}ki}, {Lamarre}, {Langer}, {Lasenby}, {Lattanzi},
  {Laureijs}, {Lavabre}, {Lawrence}, {Le Jeune}, {Leach}, {Leahy}, {Leonardi},
  {Le{\'o}n-Tavares}, {Leroy}, {Lesgourgues}, {Lewis}, {Li}, {Liddle},
  {Liguori}, {Lilje}, {Linden-V{\o}rnle}, {Lindholm}, {L{\'o}pez-Caniego},
  {Lowe}, {Lubin}, {Mac{\'\i}as-P{\'e}rez}, {MacTavish}, {Maffei}, {Maggio},
  {Maino}, {Mandolesi}, {Mangilli}, {Marcos-Caballero}, {Marinucci}, {Maris},
  {Marleau}, {Marshall}, {Martin}, {Mart{\'\i}nez-Gonz{\'a}lez}, {Masi},
  {Massardi}, {Matarrese}, {Matsumura}, {Matthai}, {Maurin}, {Mazzotta},
  {McDonald}, {McEwen}, {McGehee}, {Mei}, {Meinhold}, {Melchiorri}, {Melin},
  {Mendes}, {Menegoni}, {Mennella}, {Migliaccio}, {Mikkelsen}, {Millea},
  {Miniscalco}, {Mitra}, {Miville-Desch{\^e}nes}, {Molinari}, {Moneti},
  {Montier}, {Morgante}, {Morisset}, {Mortlock}, {Moss}, {Munshi}, {Murphy},
  {Naselsky}, {Nati}, {Natoli}, {Negrello}, {Nesvadba}, {Netterfield},
  {N{\o}rgaard-Nielsen}, {North}, {Noviello}, {Novikov}, {Novikov}, {O'Dwyer},
  {Orieux}, {Osborne}, {O'Sullivan}, {Oxborrow}, {Paci}, {Pagano}, {Pajot},
  {Paladini}, {Pandolfi}, {Paoletti}, {Partridge}, {Pasian}, {Patanchon},
  {Paykari}, {Pearson}, {Pearson}, {Peel}, {Peiris}, {Perdereau}, {Perotto},
  {Perrotta}, {Pettorino}, {Piacentini}, {Piat}, {Pierpaoli}, {Pietrobon},
  {Plaszczynski}, {Platania}, {Pogosyan}, {Pointecouteau}, {Polenta},
  {Ponthieu}, {Popa}, {Poutanen}, {Pratt}, {Pr{\'e}zeau}, {Prunet}, {Puget},
  {Pullen}, {Rachen}, {Racine}, {Rahlin}, {R{\"a}th}, {Reach}, {Rebolo},
  {Reinecke}, {Remazeilles}, {Renault}, {Renzi}, {Riazuelo}, {Ricciardi},
  {Riller}, {Ringeval}, {Ristorcelli}, {Robbers}, {Rocha}, {Roman}, {Rosset},
  {Rossetti}, {Roudier}, {Rowan-Robinson}, {Rubi{\~n}o-Mart{\'\i}n},
  {Ruiz-Granados}, {Rusholme}, {Salerno}, {Sandri}, {Sanselme}, {Santos},
  {Savelainen}, {Savini}, {Schaefer}, {Schiavon}, {Scott}, {Seiffert}, {Serra},
  {Shellard}, {Smith}, {Smoot}, {Souradeep}, {Spencer}, {Starck}, {Stolyarov},
  {Stompor}, {Sudiwala}, {Sunyaev}, {Sureau}, {Sutter}, {Sutton}, {Suur-Uski},
  {Sygnet}, {Tauber}, {Tavagnacco}, {Taylor}, {Terenzi}, {Texier},
  {Toffolatti}, {Tomasi}, {Torre}, {Tristram}, {Tucci}, {Tuovinen},
  {T{\"u}rler}, {Tuttlebee}, {Umana}, {Valenziano}, {Valiviita}, {Van Tent},
  {Varis}, {Vibert}, {Viel}, {Vielva}, {Villa}, {Vittorio}, {Wade}, {Wandelt},
  {Watson}, {Watson}, {Wehus}, {Welikala}, {Weller}, {White}, {White},
  {Wilkinson}, {Winkel}, {Xia}, {Yvon}, {Zacchei}, {Zibin}, \&
  {Zonca}}]{2014A&A...571A...1P}
{Planck Collaboration}, {Ade}, P.~A.~R., {Aghanim}, N., {et~al.}
  2014{\natexlab{b}}, \aap, 571, A1

\bibitem[{{Planck Collaboration} {et~al.}(2013){Planck Collaboration}, {Ade},
  {Aghanim}, {Arnaud}, {Ashdown}, {Atrio-Barandela}, {Aumont}, {Baccigalupi},
  {Balbi}, {Banday}, {Barreiro}, {Bartlett}, {Battaner}, {Benabed},
  {Beno{\^\i}t}, {Bernard}, {Bersanelli}, {Bhatia}, {Bikmaev}, {Bobin},
  {B{\"o}hringer}, {Bonaldi}, {Bond}, {Borgani}, {Borrill}, {Bouchet},
  {Bourdin}, {Brown}, {Burenin}, {Burigana}, {Cabella}, {Cardoso}, {Carvalho},
  {Castex}, {Catalano}, {Cay{\'o}n}, {Chamballu}, {Chiang}, {Chon},
  {Christensen}, {Churazov}, {Clements}, {Colafrancesco}, {Colombi}, {Colombo},
  {Comis}, {Coulais}, {Crill}, {Cuttaia}, {Da Silva}, {Dahle}, {Danese},
  {Davis}, {de Bernardis}, {de Gasperis}, {de Zotti}, {Delabrouille},
  {D{\'e}mocl{\`e}s}, {D{\'e}sert}, {Diego}, {Dolag}, {Dole}, {Donzelli},
  {Dor{\'e}}, {D{\"o}rl}, {Douspis}, {Dupac}, {Efstathiou}, {En{\ss}lin},
  {Eriksen}, {Finelli}, {Flores-Cacho}, {Forni}, {Fosalba}, {Frailis},
  {Franceschi}, {Frommert}, {Galeotta}, {Ganga}, {G{\'e}nova-Santos}, {Giard},
  {Giraud-H{\'e}raud}, {Gonz{\'a}lez-Nuevo}, {G{\'o}rski}, {Gregorio},
  {Gruppuso}, {Hansen}, {Harrison}, {Hempel}, {Henrot-Versill{\'e}},
  {Hern{\'a}ndez-Monteagudo}, {Herranz}, {Hildebrandt}, {Hivon}, {Hobson},
  {Holmes}, {Hurier}, {Jaffe}, {Jaffe}, {Jagemann}, {Jones}, {Juvela},
  {Keih{\"a}nen}, {Khamitov}, {Kisner}, {Kneissl}, {Knoche}, {Knox}, {Kunz},
  {Kurki-Suonio}, {Lagache}, {L{\"a}hteenm{\"a}ki}, {Lamarre}, {Lasenby},
  {Lawrence}, {Le Jeune}, {Leonardi}, {Liddle}, {Lilje}, {L{\'o}pez-Caniego},
  {Luzzi}, {Mac{\'\i}as-P{\'e}rez}, {Maino}, {Mandolesi}, {Maris}, {Marleau},
  {Marshall}, {Mart{\'\i}nez-Gonz{\'a}lez}, {Masi}, {Massardi}, {Matarrese},
  {Mazzotta}, {Mei}, {Melchiorri}, {Melin}, {Mendes}, {Mennella}, {Mitra},
  {Miville-Desch{\^e}nes}, {Moneti}, {Montier}, {Morgante}, {Mortlock},
  {Munshi}, {Murphy}, {Naselsky}, {Nati}, {Natoli}, {N{\o}rgaard-Nielsen},
  {Noviello}, {Novikov}, {Novikov}, {Osborne}, {Pajot}, {Paoletti}, {Pasian},
  {Patanchon}, {Perdereau}, {Perotto}, {Perrotta}, {Piacentini}, {Piat},
  {Pierpaoli}, {Piffaretti}, {Plaszczynski}, {Pointecouteau}, {Polenta},
  {Ponthieu}, {Popa}, {Poutanen}, {Pratt}, {Prunet}, {Puget}, {Rachen},
  {Reach}, {Rebolo}, {Reinecke}, {Remazeilles}, {Renault}, {Ricciardi},
  {Riller}, {Ristorcelli}, {Rocha}, {Roman}, {Rosset}, {Rossetti},
  {Rubi{\~n}o-Mart{\'\i}n}, {Rusholme}, {Sandri}, {Savini}, {Scott}, {Smoot},
  {Starck}, {Sudiwala}, {Sunyaev}, {Sutton}, {Suur-Uski}, {Sygnet}, {Tauber},
  {Terenzi}, {Toffolatti}, {Tomasi}, {Tristram}, {Tuovinen}, {Valenziano}, {Van
  Tent}, {Varis}, {Vielva}, {Villa}, {Vittorio}, {Wade}, {Wandelt}, {Welikala},
  {White}, {White}, {Yvon}, {Zacchei}, \& {Zonca}}]{2013A&A...550A.131P}
{Planck Collaboration}, {Ade}, P.~A.~R., {Aghanim}, N., {et~al.} 2013, \aap,
  550, A131

\bibitem[{{Planck Collaboration} {et~al.}(2016{\natexlab{e}}){Planck
  Collaboration}, {Ade}, {Aghanim}, {Arnaud}, {Ashdown}, {Aumont},
  {Baccigalupi}, {Banday}, {Barreiro}, {Barrena}, {Bartlett}, {Bartolo},
  {Battaner}, {Battye}, {Benabed}, {Beno{\^\i}t}, {Benoit-L{\'e}vy}, {Bernard},
  {Bersanelli}, {Bielewicz}, {Bikmaev}, {B{\"o}hringer}, {Bonaldi}, {Bonavera},
  {Bond}, {Borrill}, {Bouchet}, {Bucher}, {Burenin}, {Burigana}, {Butler},
  {Calabrese}, {Cardoso}, {Carvalho}, {Catalano}, {Challinor}, {Chamballu},
  {Chary}, {Chiang}, {Chon}, {Christensen}, {Clements}, {Colombi}, {Colombo},
  {Combet}, {Comis}, {Couchot}, {Coulais}, {Crill}, {Curto}, {Cuttaia},
  {Dahle}, {Danese}, {Davies}, {Davis}, {de Bernardis}, {de Rosa}, {de Zotti},
  {Delabrouille}, {D{\'e}sert}, {Dickinson}, {Diego}, {Dolag}, {Dole},
  {Donzelli}, {Dor{\'e}}, {Douspis}, {Ducout}, {Dupac}, {Efstathiou},
  {Eisenhardt}, {Elsner}, {En{\ss}lin}, {Eriksen}, {Falgarone}, {Fergusson},
  {Feroz}, {Ferragamo}, {Finelli}, {Forni}, {Frailis}, {Fraisse}, {Franceschi},
  {Frejsel}, {Galeotta}, {Galli}, {Ganga}, {G{\'e}nova-Santos}, {Giard},
  {Giraud-H{\'e}raud}, {Gjerl{\o}w}, {Gonz{\'a}lez-Nuevo}, {G{\'o}rski},
  {Grainge}, {Gratton}, {Gregorio}, {Gruppuso}, {Gudmundsson}, {Hansen},
  {Hanson}, {Harrison}, {Hempel}, {Henrot-Versill{\'e}},
  {Hern{\'a}ndez-Monteagudo}, {Herranz}, {Hildebrandt}, {Hivon}, {Hobson},
  {Holmes}, {Hornstrup}, {Hovest}, {Huffenberger}, {Hurier}, {Jaffe}, {Jaffe},
  {Jin}, {Jones}, {Juvela}, {Keih{\"a}nen}, {Keskitalo}, {Khamitov}, {Kisner},
  {Kneissl}, {Knoche}, {Kunz}, {Kurki-Suonio}, {Lagache}, {Lamarre}, {Lasenby},
  {Lattanzi}, {Lawrence}, {Leonardi}, {Lesgourgues}, {Levrier}, {Liguori},
  {Lilje}, {Linden-V{\o}rnle}, {L{\'o}pez-Caniego}, {Lubin},
  {Mac{\'\i}as-P{\'e}rez}, {Maggio}, {Maino}, {Mak}, {Mandolesi}, {Mangilli},
  {Martin}, {Mart{\'\i}nez-Gonz{\'a}lez}, {Masi}, {Matarrese}, {Mazzotta},
  {McGehee}, {Mei}, {Melchiorri}, {Melin}, {Mendes}, {Mennella}, {Migliaccio},
  {Mitra}, {Miville-Desch{\^e}nes}, {Moneti}, {Montier}, {Morgante},
  {Mortlock}, {Moss}, {Munshi}, {Murphy}, {Naselsky}, {Nastasi}, {Nati},
  {Natoli}, {Netterfield}, {N{\o}rgaard-Nielsen}, {Noviello}, {Novikov},
  {Novikov}, {Olamaie}, {Oxborrow}, {Paci}, {Pagano}, {Pajot}, {Paoletti},
  {Pasian}, {Patanchon}, {Pearson}, {Perdereau}, {Perotto}, {Perrott},
  {Perrotta}, {Pettorino}, {Piacentini}, {Piat}, {Pierpaoli}, {Pietrobon},
  {Plaszczynski}, {Pointecouteau}, {Polenta}, {Pratt}, {Pr{\'e}zeau}, {Prunet},
  {Puget}, {Rachen}, {Reach}, {Rebolo}, {Reinecke}, {Remazeilles}, {Renault},
  {Renzi}, {Ristorcelli}, {Rocha}, {Rosset}, {Rossetti}, {Roudier}, {Rozo},
  {Rubi{\~n}o-Mart{\'\i}n}, {Rumsey}, {Rusholme}, {Rykoff}, {Sandri}, {Santos},
  {Saunders}, {Savelainen}, {Savini}, {Schammel}, {Scott}, {Seiffert},
  {Shellard}, {Shimwell}, {Spencer}, {Stanford}, {Stern}, {Stolyarov},
  {Stompor}, {Streblyanska}, {Sudiwala}, {Sunyaev}, {Sutton}, {Suur-Uski},
  {Sygnet}, {Tauber}, {Terenzi}, {Toffolatti}, {Tomasi}, {Tramonte},
  {Tristram}, {Tucci}, {Tuovinen}, {Umana}, {Valenziano}, {Valiviita}, {Van
  Tent}, {Vielva}, {Villa}, {Wade}, {Wandelt}, {Wehus}, {White}, {Wright},
  {Yvon}, {Zacchei}, \& {Zonca}}]{2016A&A...594A..27P}
{Planck Collaboration}, {Ade}, P.~A.~R., {Aghanim}, N., {et~al.}
  2016{\natexlab{e}}, \aap, 594, A27

\bibitem[{{Planck Collaboration} {et~al.}(2016{\natexlab{f}}){Planck
  Collaboration}, {Ade}, {Aghanim}, {Arnaud}, {Aumont}, {Baccigalupi},
  {Banday}, {Barreiro}, {Bartlett}, {Bartolo}, {Battaner}, {Benabed},
  {Benoit-L{\'e}vy}, {Bernard}, {Bersanelli}, {Bielewicz}, {Bock}, {Bonaldi},
  {Bonavera}, {Bond}, {Borrill}, {Bouchet}, {Burigana}, {Butler}, {Calabrese},
  {Catalano}, {Chamballu}, {Chiang}, {Christensen}, {Churazov}, {Clements},
  {Colombo}, {Combet}, {Comis}, {Couchot}, {Coulais}, {Crill}, {Curto},
  {Cuttaia}, {Danese}, {Davies}, {Davis}, {de Bernardis}, {de Rosa}, {de
  Zotti}, {Delabrouille}, {Dickinson}, {Diego}, {Dole}, {Donzelli}, {Dor{\'e}},
  {Douspis}, {Ducout}, {Dupac}, {Efstathiou}, {Elsner}, {En{\ss}lin},
  {Eriksen}, {Finelli}, {Flores-Cacho}, {Forni}, {Frailis}, {Fraisse},
  {Franceschi}, {Galeotta}, {Galli}, {Ganga}, {G{\'e}nova-Santos}, {Giard},
  {Giraud-H{\'e}raud}, {Gjerl{\o}w}, {Gonz{\'a}lez-Nuevo}, {G{\'o}rski},
  {Gregorio}, {Gruppuso}, {Gudmundsson}, {Hansen}, {Harrison}, {Helou},
  {Hern{\'a}ndez-Monteagudo}, {Herranz}, {Hildebrandt}, {Hivon}, {Hobson},
  {Hornstrup}, {Hovest}, {Huffenberger}, {Hurier}, {Jaffe}, {Jaffe}, {Jones},
  {Keih{\"a}nen}, {Keskitalo}, {Kisner}, {Kneissl}, {Knoche}, {Kunz},
  {Kurki-Suonio}, {Lagache}, {Lamarre}, {Langer}, {Lasenby}, {Lattanzi},
  {Lawrence}, {Leonardi}, {Levrier}, {Lilje}, {Linden-V{\o}rnle},
  {L{\'o}pez-Caniego}, {Lubin}, {Mac{\'\i}as-P{\'e}rez}, {Maffei}, {Maggio},
  {Maino}, {Mak}, {Mandolesi}, {Mangilli}, {Maris}, {Martin},
  {Mart{\'\i}nez-Gonz{\'a}lez}, {Masi}, {Matarrese}, {Melchiorri}, {Mennella},
  {Migliaccio}, {Mitra}, {Miville-Desch{\^e}nes}, {Moneti}, {Montier},
  {Morgante}, {Mortlock}, {Munshi}, {Murphy}, {Nati}, {Natoli}, {Noviello},
  {Novikov}, {Novikov}, {Oxborrow}, {Paci}, {Pagano}, {Pajot}, {Paoletti},
  {Partridge}, {Pasian}, {Pearson}, {Perdereau}, {Perotto}, {Pettorino},
  {Piacentini}, {Piat}, {Pierpaoli}, {Plaszczynski}, {Pointecouteau},
  {Polenta}, {Ponthieu}, {Pratt}, {Prunet}, {Puget}, {Rachen}, {Reinecke},
  {Remazeilles}, {Renault}, {Renzi}, {Ristorcelli}, {Rocha}, {Rosset},
  {Rossetti}, {Roudier}, {Rubi{\~n}o-Mart{\'\i}n}, {Rusholme}, {Sandri},
  {Santos}, {Savelainen}, {Savini}, {Scott}, {Spencer}, {Stolyarov}, {Stompor},
  {Sunyaev}, {Sutton}, {Suur-Uski}, {Sygnet}, {Tauber}, {Terenzi},
  {Toffolatti}, {Tomasi}, {Tristram}, {Tucci}, {Umana}, {Valenziano},
  {Valiviita}, {Van Tent}, {Vielva}, {Villa}, {Wade}, {Wandelt}, {Wehus},
  {Welikala}, {Yvon}, {Zacchei}, \& {Zonca}}]{2016A&A...594A..23P}
{Planck Collaboration}, {Ade}, P.~A.~R., {Aghanim}, N., {et~al.}
  2016{\natexlab{f}}, \aap, 594, A23

\bibitem[{{Planck Collaboration} {et~al.}(2020{\natexlab{a}}){Planck
  Collaboration}, {Aghanim}, {Akrami}, {Arroja}, {Ashdown}, {Aumont},
  {Baccigalupi}, {Ballardini}, {Banday}, {Barreiro}, {Bartolo}, {Basak},
  {Battye}, {Benabed}, {Bernard}, {Bersanelli}, {Bielewicz}, {Bock}, {Bond},
  {Borrill}, {Bouchet}, {Boulanger}, {Bucher}, {Burigana}, {Butler},
  {Calabrese}, {Cardoso}, {Carron}, {Casaponsa}, {Challinor}, {Chiang},
  {Colombo}, {Combet}, {Contreras}, {Crill}, {Cuttaia}, {de Bernardis}, {de
  Zotti}, {Delabrouille}, {Delouis}, {D{\'e}sert}, {Di Valentino}, {Dickinson},
  {Diego}, {Donzelli}, {Dor{\'e}}, {Douspis}, {Ducout}, {Dupac}, {Efstathiou},
  {Elsner}, {En{\ss}lin}, {Eriksen}, {Falgarone}, {Fantaye}, {Fergusson},
  {Fernandez-Cobos}, {Finelli}, {Forastieri}, {Frailis}, {Franceschi},
  {Frolov}, {Galeotta}, {Galli}, {Ganga}, {G{\'e}nova-Santos}, {Gerbino},
  {Ghosh}, {Gonz{\'a}lez-Nuevo}, {G{\'o}rski}, {Gratton}, {Gruppuso},
  {Gudmundsson}, {Hamann}, {Handley}, {Hansen}, {Helou}, {Herranz},
  {Hildebrandt}, {Hivon}, {Huang}, {Jaffe}, {Jones}, {Karakci}, {Keih{\"a}nen},
  {Keskitalo}, {Kiiveri}, {Kim}, {Kisner}, {Knox}, {Krachmalnicoff}, {Kunz},
  {Kurki-Suonio}, {Lagache}, {Lamarre}, {Langer}, {Lasenby}, {Lattanzi},
  {Lawrence}, {Le Jeune}, {Leahy}, {Lesgourgues}, {Levrier}, {Lewis},
  {Liguori}, {Lilje}, {Lilley}, {Lindholm}, {L{\'o}pez-Caniego}, {Lubin}, {Ma},
  {Mac{\'\i}as-P{\'e}rez}, {Maggio}, {Maino}, {Mandolesi}, {Mangilli},
  {Marcos-Caballero}, {Maris}, {Martin}, {Martinelli},
  {Mart{\'\i}nez-Gonz{\'a}lez}, {Matarrese}, {Mauri}, {McEwen}, {Meerburg},
  {Meinhold}, {Melchiorri}, {Mennella}, {Migliaccio}, {Millea}, {Mitra},
  {Miville-Desch{\^e}nes}, {Molinari}, {Moneti}, {Montier}, {Morgante}, {Moss},
  {Mottet}, {M{\"u}nchmeyer}, {Natoli}, {N{\o}rgaard-Nielsen}, {Oxborrow},
  {Pagano}, {Paoletti}, {Partridge}, {Patanchon}, {Pearson}, {Peel}, {Peiris},
  {Perrotta}, {Pettorino}, {Piacentini}, {Polastri}, {Polenta}, {Puget},
  {Rachen}, {Reinecke}, {Remazeilles}, {Renault}, {Renzi}, {Rocha}, {Rosset},
  {Roudier}, {Rubi{\~n}o-Mart{\'\i}n}, {Ruiz-Granados}, {Salvati}, {Sandri},
  {Savelainen}, {Scott}, {Shellard}, {Shiraishi}, {Sirignano}, {Sirri},
  {Spencer}, {Sunyaev}, {Suur-Uski}, {Tauber}, {Tavagnacco}, {Tenti},
  {Terenzi}, {Toffolatti}, {Tomasi}, {Trombetti}, {Valiviita}, {Van Tent},
  {Vibert}, {Vielva}, {Villa}, {Vittorio}, {Wandelt}, {Wehus}, {White},
  {White}, {Zacchei}, \& {Zonca}}]{2020A&A...641A...1P}
{Planck Collaboration}, {Aghanim}, N., {Akrami}, Y., {et~al.}
  2020{\natexlab{a}}, \aap, 641, A1

\bibitem[{{Planck Collaboration} {et~al.}(2020{\natexlab{b}}){Planck
  Collaboration}, {Aghanim, N.}, {Akrami, Y.}, {Ashdown, M.}, {Aumont, J.},
  {Baccigalupi, C.}, {Ballardini, M.}, {Banday, A. J.}, {Barreiro, R. B.},
  {Bartolo, N.}, {Basak, S.}, {Battye, R.}, {Benabed, K.}, {Bernard, J.-P.},
  {Bersanelli, M.}, {Bielewicz, P.}, {Bock, J. J.}, {Bond, J. R.}, {Borrill,
  J.}, {Bouchet, F. R.}, {Boulanger, F.}, {Bucher, M.}, {Burigana, C.},
  {Butler, R. C.}, {Calabrese, E.}, {Cardoso, J.-F.}, {Carron, J.}, {Challinor,
  A.}, {Chiang, H. C.}, {Chluba, J.}, {Colombo, L. P. L.}, {Combet, C.},
  {Contreras, D.}, {Crill, B. P.}, {Cuttaia, F.}, {de Bernardis, P.}, {de
  Zotti, G.}, {Delabrouille, J.}, {Delouis, J.-M.}, {Di Valentino, E.}, {Diego,
  J. M.}, {Dor\'e, O.}, {Douspis, M.}, {Ducout, A.}, {Dupac, X.}, {Dusini, S.},
  {Efstathiou, G.}, {Elsner, F.}, {En\ss{}lin, T. A.}, {Eriksen, H. K.},
  {Fantaye, Y.}, {Farhang, M.}, {Fergusson, J.}, {Fernandez-Cobos, R.},
  {Finelli, F.}, {Forastieri, F.}, {Frailis, M.}, {Fraisse, A. A.},
  {Franceschi, E.}, {Frolov, A.}, {Galeotta, S.}, {Galli, S.}, {Ganga, K.},
  {G\'enova-Santos, R. T.}, {Gerbino, M.}, {Ghosh, T.}, {Gonz\'alez-Nuevo, J.},
  {G\'orski, K. M.}, {Gratton, S.}, {Gruppuso, A.}, {Gudmundsson, J. E.},
  {Hamann, J.}, {Handley, W.}, {Hansen, F. K.}, {Herranz, D.}, {Hildebrandt, S.
  R.}, {Hivon, E.}, {Huang, Z.}, {Jaffe, A. H.}, {Jones, W. C.}, {Karakci, A.},
  {Keih\"anen, E.}, {Keskitalo, R.}, {Kiiveri, K.}, {Kim, J.}, {Kisner, T. S.},
  {Knox, L.}, {Krachmalnicoff, N.}, {Kunz, M.}, {Kurki-Suonio, H.}, {Lagache,
  G.}, {Lamarre, J.-M.}, {Lasenby, A.}, {Lattanzi, M.}, {Lawrence, C. R.}, {Le
  Jeune, M.}, {Lemos, P.}, {Lesgourgues, J.}, {Levrier, F.}, {Lewis, A.},
  {Liguori, M.}, {Lilje, P. B.}, {Lilley, M.}, {Lindholm, V.},
  {L\'opez-Caniego, M.}, {Lubin, P. M.}, {Ma, Y.-Z.}, {Mac\'{\i}as-P\'erez, J.
  F.}, {Maggio, G.}, {Maino, D.}, {Mandolesi, N.}, {Mangilli, A.},
  {Marcos-Caballero, A.}, {Maris, M.}, {Martin, P. G.}, {Martinelli, M.},
  {Mart\'{\i}nez-Gonz\'alez, E.}, {Matarrese, S.}, {Mauri, N.}, {McEwen, J.
  D.}, {Meinhold, P. R.}, {Melchiorri, A.}, {Mennella, A.}, {Migliaccio, M.},
  {Millea, M.}, {Mitra, S.}, {Miville-Desch\^enes, M.-A.}, {Molinari, D.},
  {Montier, L.}, {Morgante, G.}, {Moss, A.}, {Natoli, P.},
  {N\o{}rgaard-Nielsen, H. U.}, {Pagano, L.}, {Paoletti, D.}, {Partridge, B.},
  {Patanchon, G.}, {Peiris, H. V.}, {Perrotta, F.}, {Pettorino, V.},
  {Piacentini, F.}, {Polastri, L.}, {Polenta, G.}, {Puget, J.-L.}, {Rachen, J.
  P.}, {Reinecke, M.}, {Remazeilles, M.}, {Renzi, A.}, {Rocha, G.}, {Rosset,
  C.}, {Roudier, G.}, {Rubi\~no-Mart\'{\i}n, J. A.}, {Ruiz-Granados, B.},
  {Salvati, L.}, {Sandri, M.}, {Savelainen, M.}, {Scott, D.}, {Shellard, E. P.
  S.}, {Sirignano, C.}, {Sirri, G.}, {Spencer, L. D.}, {Sunyaev, R.},
  {Suur-Uski, A.-S.}, {Tauber, J. A.}, {Tavagnacco, D.}, {Tenti, M.},
  {Toffolatti, L.}, {Tomasi, M.}, {Trombetti, T.}, {Valenziano, L.},
  {Valiviita, J.}, {Van Tent, B.}, {Vibert, L.}, {Vielva, P.}, {Villa, F.},
  {Vittorio, N.}, {Wandelt, B. D.}, {Wehus, I. K.}, {White, M.}, {White, S. D.
  M.}, {Zacchei, A.}, \& {Zonca, A.}}]{Planck2018VI}
{Planck Collaboration}, {Aghanim, N.}, {Akrami, Y.}, {et~al.}
  2020{\natexlab{b}}, A\&A, 641, A6

\bibitem[{{Pointecouteau} {et~al.}(2021){Pointecouteau}, {Santiago-Bautista},
  {Douspis}, {Aghanim}, {Crichton}, {Diego}, {Hurier}, {Macias-Perez},
  {Marriage}, {Remazeilles}, {Caretta}, \&
  {Bravo-Alfaro}}]{2021A&A...651A..73P}
{Pointecouteau}, E., {Santiago-Bautista}, I., {Douspis}, M., {et~al.} 2021,
  \aap, 651, A73

\bibitem[{{Remazeilles} {et~al.}(2011){Remazeilles}, {Delabrouille}, \&
  {Cardoso}}]{2011MNRAS.418..467R}
{Remazeilles}, M., {Delabrouille}, J., \& {Cardoso}, J.-F. 2011, \mnras, 418,
  467

\bibitem[{{Salvati} {et~al.}(2021){Salvati}, {Saro}, {Bocquet}, {Costanzi},
  {Ansarinejad}, {Benson}, {Bleem}, {Calzadilla}, {Carlstrom}, {Chang},
  {Chown}, {Crites}, {deHaan}, {Dobbs}, {Everett}, {Floyd}, {Grandis},
  {George}, {Halverson}, {Holder}, {Holzapfel}, {Hrubes}, {Lee}, {Luong-Van},
  {McDonald}, {McMahon}, {Meyer}, {Millea}, {Mocanu}, {Mohr}, {Natoli},
  {Omori}, {Padin}, {Pryke}, {Reichardt}, {Ruhl}, {Ruppin}, {Schaffer},
  {Schrabback}, {Shirokoff}, {Staniszewski}, {Stark}, {Vieira}, \&
  {Williamson}}]{2021arXiv211203606S}
{Salvati}, L., {Saro}, A., {Bocquet}, S., {et~al.} 2021, arXiv e-prints,
  arXiv:2112.03606

\bibitem[{{Salvetti} {et~al.}(2017){Salvetti}, {Marelli}, {Gastaldello},
  {Ghizzardi}, {Molendi}, {De Luca}, {Moretti}, {Rossetti}, \&
  {Tiengo}}]{2017ExA....44..309S}
{Salvetti}, D., {Marelli}, M., {Gastaldello}, F., {et~al.} 2017, Experimental
  Astronomy, 44, 309

\bibitem[{{Starck} {et~al.}(2009){Starck}, {Fadili}, {Digel}, {Zhang}, \&
  {Chiang}}]{2009A&A...504..641S}
{Starck}, J.~L., {Fadili}, J.~M., {Digel}, S., {Zhang}, B., \& {Chiang}, J.
  2009, \aap, 504, 641

\bibitem[{{Starck} {et~al.}(2002){Starck}, {Pantin}, \& {Murtagh}}]{Starck2002}
{Starck}, J.~L., {Pantin}, E., \& {Murtagh}, F. 2002, \pasp, 114, 1051

\bibitem[{{Sunyaev} \& {Zeldovich}(1972)}]{1972CoASP...4..173S}
{Sunyaev}, R.~A. \& {Zeldovich}, Y.~B. 1972, Comments on Astrophysics and Space
  Physics, 4, 173

\bibitem[{{Tegmark} {et~al.}(2003){Tegmark}, {de Oliveira-Costa}, \&
  {Hamilton}}]{2003PhRvD..68l3523T}
{Tegmark}, M., {de Oliveira-Costa}, A., \& {Hamilton}, A.~J. 2003, \prd, 68,
  123523

\bibitem[{{The CHEX-MATE Collaboration} {et~al.}(2021){The CHEX-MATE
  Collaboration}, {Arnaud, M.}, {Ettori, S.}, {Pratt, G. W.}, {Rossetti, M.},
  {Eckert, D.}, {Gastaldello, F.}, {Gavazzi, R.}, {Kay, S.T.}, {Lovisari, L.},
  {Maughan, B.J.}, {Pointecouteau, E.}, {Sereno, M.}, {Bartalucci, I.},
  {Bonafede, A.}, {Bourdin, H.}, {Cassano, R.}, {Duffy, R.T.}, {Iqbal, A.},
  {Maurogordato, S.}, {Rasia, E.}, {Sayers, J.}, {Andrade-Santos, F.}, {Aussel,
  H.}, {Barnes, D.J.}, {Barrena, R.}, {Borgani, S.}, {Burkutean, S.}, {Clerc,
  N.}, {Corasaniti, P.-S.}, {Cuillandre, J.-C.}, {De Grandi, S.}, {De Petris,
  M.}, {Dolag, K.}, {Donahue, M.}, {Ferragamo, A.}, {Gaspari, M.}, {Ghizzardi,
  S.}, {Gitti, M.}, {Haines, C.P.}, {Jauzac, M.}, {Johnston-Hollitt, M.},
  {Jones, C.}, {K\'eruzor\'e, F.}, {LeBrun, A.M.C.}, {Mayet, F.}, {Mazzotta,
  P.}, {Melin, J.-B.}, {Molendi, S.}, {Nonino, M.}, {Okabe, N.}, {Paltani, S.},
  {Perotto, L.}, {Pires, S.}, {Radovich, M.}, {Rubino-Martin, J.-A.}, {Salvati,
  L.}, {Saro, A.}, {Sartoris, B.}, {Schellenberger, G.}, {Streblyanska, A.},
  {Tarr\'{\i}o, P.}, {Tozzi, P.}, {Umetsu, K.}, {van der Burg, R.F.J.}, {Vazza,
  F.}, {Venturi, T.}, {Yepes, G.}, \& {Zarattini, S.}}]{CHEX-MATE}
{The CHEX-MATE Collaboration}, {Arnaud, M.}, {Ettori, S.}, {et~al.} 2021, A\&A,
  650, A104

\bibitem[{{Torrado} \& {Lewis}(2021)}]{2021JCAP...05..057T}
{Torrado}, J. \& {Lewis}, A. 2021, \jcap, 2021, 057

\bibitem[{{Tozzi, P.} {et~al.}(2015){Tozzi, P.}, {Gastaldello, F.}, {Molendi,
  S.}, {Ettori, S.}, {Santos, J. S.}, {De Grandi, S.}, {Balestra, I.}, {Rosati,
  P.}, {Altieri, B.}, {Cresci, G.}, {Menanteau, F.}, \& {Valtchanov,
  I.}}]{Tozzi2015}
{Tozzi, P.}, {Gastaldello, F.}, {Molendi, S.}, {et~al.} 2015, A\&A, 580, A6

\bibitem[{{Verner} \& {Ferland}(1996)}]{Verner1996}
{Verner}, D.~A. \& {Ferland}, G.~J. 1996, \apjs, 103, 467

\bibitem[{{Vikhlinin} {et~al.}(2006){Vikhlinin}, {Kravtsov}, {Forman}, {Jones},
  {Markevitch}, {Murray}, \& {Van Speybroeck}}]{Vikhlinin2006}
{Vikhlinin}, A., {Kravtsov}, A., {Forman}, W., {et~al.} 2006, \apj, 640, 691

\bibitem[{{Voit}(2005)}]{2005RvMP...77..207V}
{Voit}, G.~M. 2005, Reviews of Modern Physics, 77, 207

\bibitem[{{Zhang} {et~al.}(2008){Zhang}, {Fadili}, \&
  {Starck}}]{2008ITIP...17.1093Z}
{Zhang}, B., {Fadili}, J.~M., \& {Starck}, J.-L. 2008, IEEE Transactions on
  Image Processing, 17, 1093

\end{thebibliography}

\begin{appendix} 
\section{X-ray absorption toward the CHEX-MATE+SPT cluster sample} \label{app:NH263}

\begin{figure}
	\includegraphics[width=\columnwidth]{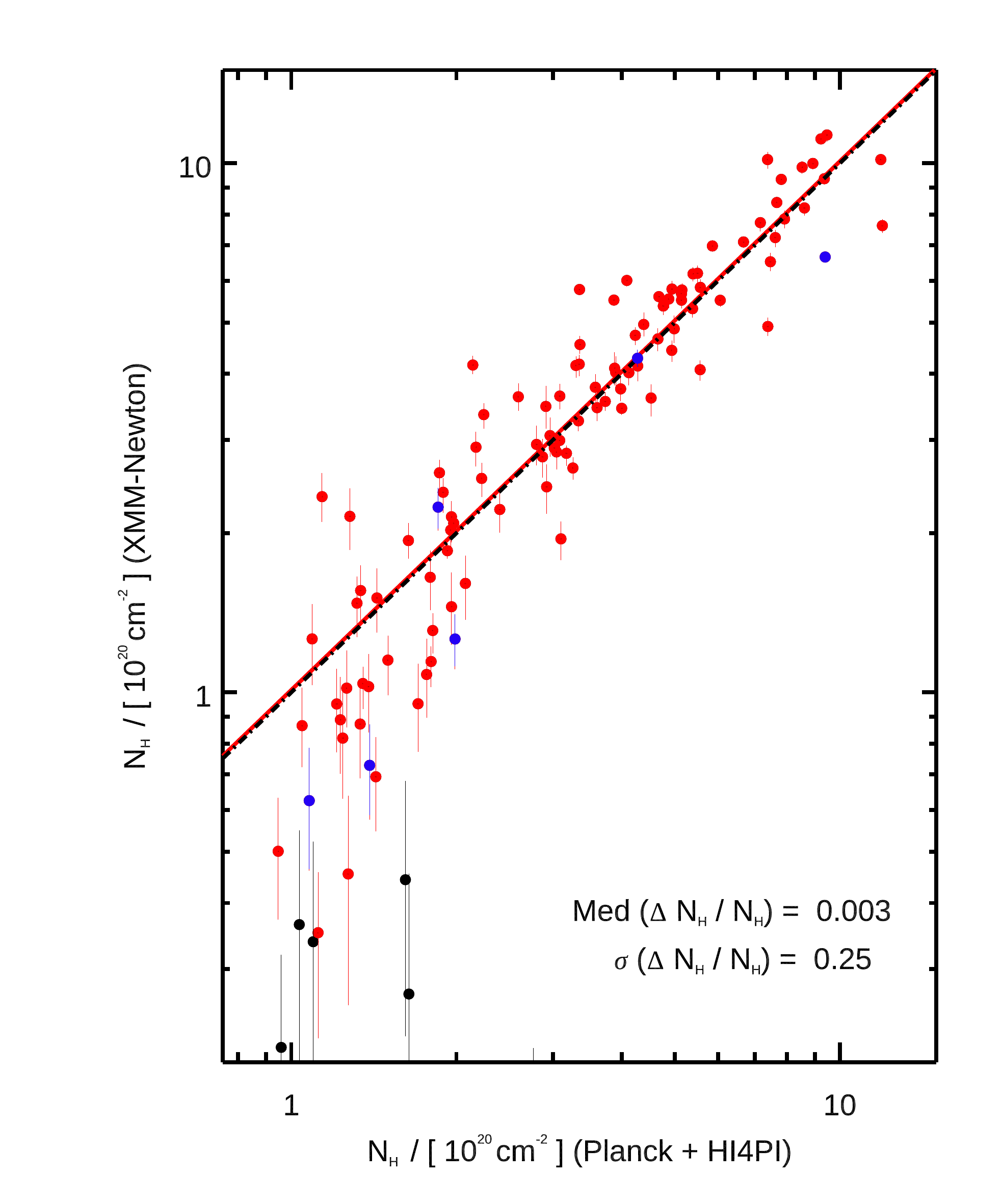}
    \caption{Estimates of the hydrogen column density derived from X-ray spectroscopy and considering the dust emission excess from \textit{Planck}+HI4PI (Bourdin et al. in prep.) toward the CHEX-MATE galaxy clusters. The six clusters analysed in this work are depicted as blue points in the figure.}
    \label{fig:NHXvsNHP}
\end{figure}

X-ray absorption and its relationship with the total (atomic+molecular) Galactic hydrogen density column toward CHEX-MATE galaxy clusters have been investigated in a companion analysis (Bourdin et al. in prep.). Briefly, a joint-analysis of \textit{Planck} and HI4PI data allowed us to estimate the mass fraction of molecular gas across the lines of sight toward CHEX-MATE galaxy clusters, $\rm f=2N_{H_2}/N_H$, by looking for thermal dust emission excesses with respect to the neutral atomic hydrogen density column, $\rm N_{H_I}$. We found that the CHEX-MATE cluster catalogue can be divided into three categories: 40\% are clusters located behind low $\rm N_{H_I}$ regions ($\rm N_{H_I}<2\,10^{20} \rm cm^{-2}$) where the molecular mass fraction is negligible, 40\% are clusters located behind intermediate $\rm N_{H_I}$ regions ($2\,10^{20} {\rm cm^{-2}} < {\rm N_H} < 5\,10^{20} {\rm cm^{-2}}$) where the molecular gas fraction is $\sim10\%$ on average, and the remaining 20\% of the cluster sample are located behind high $\rm N_{H_I}$ regions where higher molecular gas fractions could locally affect the analyses of a few observations. A comparison of $N_H$ estimates obtained from X-ray spectroscopy (XMM-\textit{Newton}) and dust emission excesses (\textit{Planck}+HI4PI) with respect to $H_I$ is shown in Fig.~\ref{fig:NHXvsNHP}. Given the scatter of 25\% that separates these $\rm N_{H}$ estimates, the regression line shows that both of them are compatible with one another. The present work focuses on a subsample of the CHEX-MATE cluster catalogue made of six clusters, listed in Tab.~\ref{table:sample} and depicted using blue points in Fig.~\ref{fig:NHXvsNHP}. Among these CHEX-MATE+SPT clusters, four clusters belong to the low $\rm N_{H_I}$ category, a regime in which molecular gas fractions do not significantly affect temperature measurements. The remaining two clusters are PSZ2G271.18-30.95 and PSZ2G263.68-22.55, which belong to the intermediate and high $\rm N_{H_I}$ categories, respectively. For each CHEX-MATE+SPT cluster excepting PSZ2G63.68-22.55, the relative difference between XMM and \textit{Planck}+HI4PI inferences of the $\rm N_{H}$ values is lower than the characteristic dispersion of 25\% measured for the overall CHEX-MATE sample. XMM and \textit{Planck}+HI4PI inferences of the $\rm N_{H}$ values differ instead by a relative amount as high as $dN_H/N_H=70\%$ in the case of PSZ2G63.68-22.55, which makes this cluster an outlier of the relationship observed between XMM and \textit{Planck}+HI4PI inferences for the complete CHEX-MATE sample. Furthermore, the observed spectrum in the soft X-ray band of this galaxy cluster shows an overestimation of the Galactic absorption when the molecular fraction is considered. These characteristics of the CHEX-MATE+SPT subsample led us to adopt the \textit{Planck}+HI4PI inferences of the $\rm N_{H}$ values for all clusters except PSZ2G63.68-22.55. In the case of PSZ2G63.68-22.55, the $\rm N_{H}$ value has been inferred from X-ray spectroscopy as a result of a joint-fit with projected temperature and metallicities of the ICM in annular regions delimited by a cluster-centric radii range of $[0.15,0.6]R_{500}$. The resulting value of $\rm N_H=(6.88^{+0.12}_{-0.13})\,10^{20} \rm cm^{-2}$ ($\chi^2_\nu=1.09$) yields a molecular gas fraction of $f=0.21$, which is significantly positive but twice lower than expected from \textit{Planck}+HI4PI data.

\begin{figure*}
    \centering
	 \begin{subfigure}[a]{0.495\textwidth}
	\includegraphics[width=\textwidth]{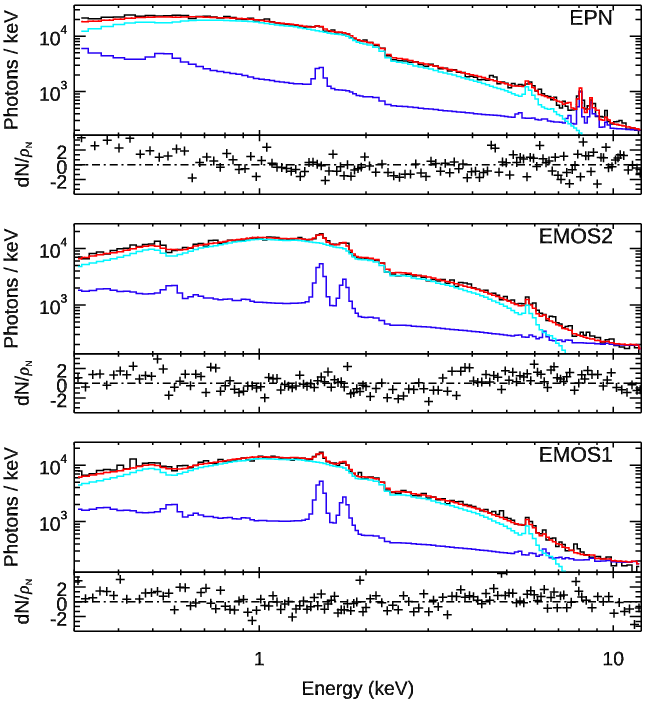}
	\end{subfigure}
	 \begin{subfigure}[a]{0.495\textwidth}
	\includegraphics[width=\textwidth]{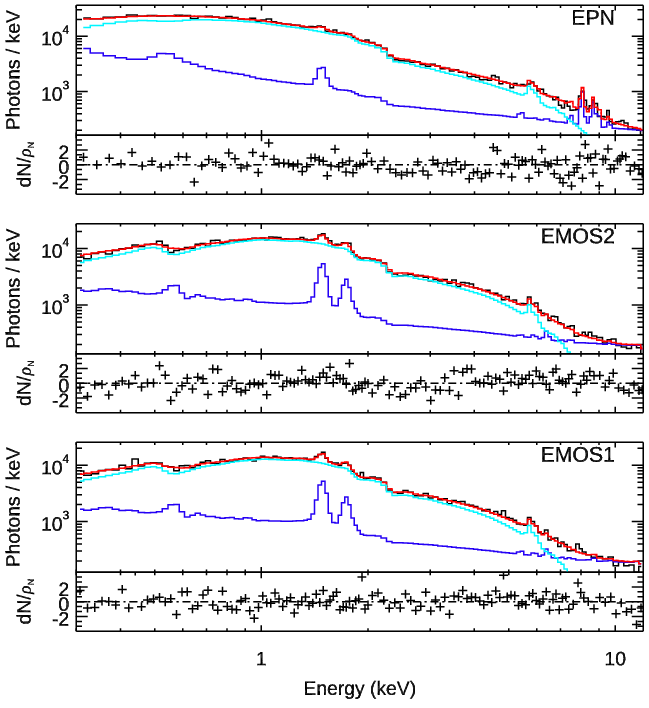}
	\end{subfigure}
    \caption{Fit and residuals of the X-ray spectrum of PSZ2G263.68-22.55 when we assume 
    the molecular hydrogen column density from Bourdin et al. (in prep.) (left panel) or when $N_H$ is free to vary in the fit (right). For the three cameras (EPN, MOS1, and MOS2) of the XMM-\textit{Newton} telescope, the light blue curve is the fitted model of the cluster spectrum, and the dark blue line represents the total contribution from the background or foreground components (CXB plus QPB, SP, and the Galactic emission, see Sec.~\ref{sec:Xray bck}), the red curve is the sum of these two components, and the black curve is the observed spectrum.}
    \label{fig:psz2g263_nh_hbnh}
\end{figure*}

We show the effects of the different $\rm N_H$ values in the X-ray cluster spectrum in Fig.~\ref{fig:psz2g263_nh_hbnh}. The black curve in each panel of Fig.~\ref{fig:psz2g263_nh_hbnh} represents the observed spectrum, as seen from the three cameras (MOS1, MOS2, EPN) of the XMM-\textit{Newton} telescope, while the light and dark blue curves are, respectively, the fitted cluster spectrum and the background plus foreground model in the energy range ($[0.3, 12.1]$ keV) in interest. In the two panels of Fig.~\ref{fig:psz2g263_nh_hbnh} we show the residuals of the fits, for the two values of the $\rm N_H$ examined in this work. In particular, we show the spectrum with the \textit{Planck}+HI4PI $\rm N_H$ hydrogen column density ($\chi^2_\nu=1.48$), and the results from the X-ray spectroscopy in the left and right panels, respectively. When the molecular contribution is considered in the fit, we observe an overestimate of the Galactic absorption (below $0.6$ keV), while the observed spectrum of the cluster is more accurately modelled fitting the hydrogen column density ($\chi^2_\nu=1.09$). We also test the case where no molecular contribution is assumed in the X-ray spectral modelling, for which we observe a systematic underestimate of the Galactic absorption in the soft X-ray band, with a reduced $\chi^2_\nu$ of $\chi^2_\nu=1.18$.

\end{appendix}

\end{document}